\documentclass[twocolumn,showpacs,preprintnumbers,amsmath,amssymb]{revtex4}
\usepackage{epsfig}
\usepackage{dcolumn}% Align table columns on decimal point
\usepackage{bm}% bold math
\usepackage{color}
\newcommand{\beq}{\begin{equation}}
\newcommand{\eeq}{\end{equation}}
\newcommand{\beqa}{\begin{eqnarray}}
\newcommand{\eeqa}{\end{eqnarray}}
\newcommand{\bea}{\begin{array}}
\newcommand{\ea}{\end{array}}

\newcommand{\dd}{\mathrm{d}}

\newcommand{\lag}{\langle}
\newcommand{\rag}{\rangle}

\newcommand{\ii}{{\rm i}}
\newcommand{\inta}{\int_{-\ii\infty}^{+\ii\infty}}

\newcommand{\ve}{{\bf e}}

\newcommand{\vv}{{\bf v}}

\newcommand{\vx}{{\bf x}}
\newcommand{\vk}{{\bf k}}
\newcommand{\vq}{{\bf q}}

\newcommand{\vPsi}{{\bf \Psi}}

\newcommand{\tdelta}{{\tilde{\delta}}}
\newcommand{\tu}{{\tilde{u}}}
\newcommand{\tW}{{\tilde{W}}}

\newcommand{\cF}{{\cal F}}

\newcommand{\cP}{{\cal P}}

\newcommand{\rhob}{\overline{\rho}}

\newcommand{\Pa}{P^{\parallel}}
\newcommand{\Psc}{P_{\rm c.w.}}

\begin{document}

\title{Matter power spectrum from a Lagrangian-space regularization of perturbation theory}

\author{Patrick Valageas}
\affiliation{Institut de Physique Th\'eorique,\\
CEA, IPhT, F-91191 Gif-sur-Yvette, C\'edex, France\\
CNRS, URA 2306, F-91191 Gif-sur-Yvette, C\'edex, France}
\author{Takahiro Nishimichi}
\affiliation{Kavli Institute for the Physics and Mathematics of the Universe,
Todai Institutes for Advanced Study, The University of Tokyo,
Kashiwa, Chiba 277-8583, Japan (Kavli IPMU, WPI)}
\author{Atsushi Taruya}
\affiliation{Research Center for the Early Universe, School of Science,
The University of Tokyo, Bunkyo-ku, Tokyo 113-0033, Japan\\
Kavli Institute for the Physics and Mathematics of the Universe,
Todai Institutes for Advanced Study, The University of Tokyo,
Kashiwa, Chiba 277-8583, Japan (Kavli IPMU, WPI)}
\vspace{.2 cm}

\date{\today}
\vspace{.2 cm}

\begin{abstract}
We present a new approach to computing the matter density power spectrum, from
large linear scales to small highly nonlinear scales.
Instead of explicitly computing a partial series of high-order diagrams, as in perturbative 
resummation schemes, we embed the standard perturbation theory within a realistic
nonlinear Lagrangian-space ansatz. 
We also point out that an ``adhesion-like'' regularization of the shell-crossing regime
is more realistic than a ``Zel'dovich-like'' behavior, where particles freely escape to infinity.
This provides a ``cosmic web'' power spectrum with good small-scale properties that
provide a good matching with a halo model on mildly nonlinear scales.
We obtain a good agreement with numerical simulations on large scales,
better than $3\%$ for $k\leq 1 h$Mpc$^{-1}$, and on small scales, better than $10\%$ for
$k \leq 10 h$Mpc$^{-1}$, at $z \geq 0.35$, which improves over previous methods.

\end{abstract}

 \vskip2pc

\maketitle

\section{Introduction}
\label{Introduction}

The description of gravitational clustering can follow two different frameworks: the
Eulerian approach, where one studies density and velocity fields on a fixed grid,
and the Lagrangian approach, where one studies particle trajectories themselves
\cite{Bernardeau2002}.

The Eulerian approach is more convenient for some practical purposes, because we observe
density fields and not trajectories (and it is also better suited to baryonic physics, which
involves pressure and temperature fields). 
However, it has the theoretical disadvantage that one should in principle work with the
phase-space distribution function $f(\vx,\vv;t)$, which involves seven variables and
is very heavy for numerical and analytical computations.
Then in practice, most analytical approaches are based on the fluid approximation,
which replaces the Vlasov equation with hydrodynamical equations for the density and
velocity fields $\rho(\vx,t)$ and $\vv(\vx,t)$, which are much easier to handle.
However, this is only exact in the single-stream regime and these equations of motion
themselves break down after shell crossing. This makes the matching between the
perturbative and nonperturbative regimes rather difficult. More precisely, because the
dynamics is not well defined (these equations of motion are not sufficient to determine
the evolution after shell crossing), the asymptotic regime at high wave numbers of the
power spectrum obtained within this framework is not obvious and not clearly related
to physical behaviors.

In contrast, in the Lagrangian approach, the trajectories $\vx(\vq,t)$, where particles are
identified by their initial position $\vq$, are always well defined and the behavior
after shell crossing is (at least implicitly) defined within each peculiar scheme.
This offers the prospect of a more convenient matching to the highly nonlinear regime.
Moreover, in contrast to the Eulerian case the Lagrangian approach is not sensitive to
the ``sweeping effect''  associated with almost uniform translations by long wavelengths
of the velocity field \cite{Valageas2007a,Bernardeau2008a,Bernardeau2012}
(i.e., this can be automatically removed as a random uniform shift),
and it is a more direct probe of the structures of the density field.
From an observational point of view, this framework is also more convenient for including
redshift-space distortions. However, to compare theoretical predictions with observations,
one typically needs to compute again the density fields, and the derivation of the latter from
the particle trajectories usually introduces additional approximations and problems.

This latter disadvantage explains why most perturbative schemes that have been 
recently developed follow the Eulerian framework
\cite{Crocce2006a,Crocce2006b,Valageas2007,Taruya2008,Pietroni2008,Bernardeau2008,Valageas2011d,Bernardeau2013,Anselmi2012,Taruya2012}.
They provide a good match with numerical simulations on quasilinear scales by using
resummation schemes that are exact up to the one- or two-loop order and improve over the
standard perturbative approach by including partial resummations of higher-order
diagrams.
However, it has been difficult so far to obtain a good model up to the highly nonlinear
scales (by embedding the perturbative predictions within a phenomenological halo model),
as one usually underestimates the power spectrum on transition scales
\cite{Valageas2011d,Valageas2011e}.
This discrepancy is even worse for Lagrangian schemes, which usually give rise to
a perturbative power spectrum that decays faster than the linear power spectrum at
high $k$ \cite{Matsubara2008,Valageas2011d,Okamura2011}, whereas the Eulerian
schemes can show a variety of behaviors, a similar fast
decay, a convergence back to the linear power, or a higher tail.

In this paper, we follow two main motivations:

(1) Develop a Lagrangian-based approach.

(2) Obtain a good matching with the highly nonlinear regime.

Indeed, a Lagrangian-based approach is expected to provide a better starting point
for eventually tackling redshift-space distortions and biasing (although we do not address
these points here). 
Moreover, it is not sensitive to the ``sweeping effect'', so that
the modeling can be directly focused on density structures, that is, relative displacements.
Next, because our final goal is to obtain a unified description from linear to highly
nonlinear scales, it is necessary to pay attention to transition scales.
Moreover, the requirement of a good matching to nonlinear
scales is likely to shed light on the asymptotic behavior that should be satisfied by
good perturbative schemes. 

To reach these goals, we advocate a new perspective while drawing on some previous
works. Our main new ideas are the following:

(a) Instead of building a perturbative resummation scheme, which explicitly computes
a partial series of high-order diagrams,
we build a self-consistent ansatz that we next make consistent with standard
perturbation theory up to some order (one-loop order in this paper).

(b) Instead of the usual ``Zel'dovich-like'' behavior in the shell-crossing regime, we impose
an ``adhesion-like'' behavior.

These two strategies are motivated by the wish to obtain a good
behavior up to nonlinear scales.

First, most perturbative schemes are not guaranteed to provide at each order
self-consistent
predictions, whether they involve a sharp truncation at a finite order (as in standard
perturbation theory) or include partial resummations of high-order terms.
In other words, they do not guarantee
that their predictions can be realized by a physical density field (such that the density
is real and positive), and one may encounter unphysical behaviors at high $k$
(such as a negative power spectrum).
This is not necessarily a problem, if one restricts the focus to the range of validity of 
the perturbative scheme. However, since we intend to build a unified model that applies
up to nonlinear scales, we need to obtain as a building block a perturbative power spectrum
that remains well behaved in all regimes. A second reason is that
physical self-consistency (at least to some degree) can be expected to ensure
good convergence properties or at least a reasonable high-$k$ tail.
For instance, this clearly rules out the bad behavior of the standard Eulerian perturbation
theory, where higher-order contributions grow increasingly fast on small scales and 
lead to power spectra that can change sign with the truncation order.
Then, we can hope that a better-controlled high-$k$ behavior and a better convergence
can in turn improve the accuracy on quasilinear scales.

The method that we develop to achieve this goal is first to build an
ansatz for the power spectrum that is always well behaved and next to tune
its parameters (in our case, the skewness of the longitudinal displacement) so that its
perturbative expansion matches the standard perturbative expansion
up to the required order (here up to the second order over $P_L$).

Second, the ``Zel'dovich-like'' behavior of usual Lagrangian perturbative schemes,
where particles escape to infinity after shell crossing, leads to a fast decay of the
power spectrum on nonlinear scales. 
This is due to the erasing of intermediate-scale structures such as pancakes soon after
their formation, because particles do not remain trapped within potential wells.
As recalled above, this may be the source of a significant underestimation of the
power spectrum on transition scales. The ``adhesion model'' introduced in 
Ref.\cite{Gurbatov1989}, where particles stick together after shell crossing, seems a more
realistic approximation. In particular, this provides a good description of the cosmic web.
Because it is not easy to implement the exact adhesion model within a model for the power 
spectrum, we use a simplified ansatz, inspired from Ref.\cite{Valageas2011a}, to
include some form of ``sticking'' of particle pairs after shell crossing.
This models the formation of pancakes and yields additional power at high $k$,
which should be more realistic and more accurate than previous methods.

These ingredients allow us to build a model for the ``cosmic web'' power spectrum,
associated with large and intermediate scale structures (bulk flows, voids, and 
pancackes). Next, following Refs.\cite{Valageas2011d,Valageas2011e}, we combine
these results with a phenomenological halo model to extend our model up to highly
nonlinear scales, associated with inner halo regions. 

This paper is organized as follows.
After recalling the expression of the matter density power spectrum in a
Lagrangian-space framework in Sec.~\ref{Lagrangian}, 
we describe in Sec.~\ref{large-scale-power} our model for the ``cosmic web'' power
spectrum. This corresponds to both the large-scale perturbative regime and the
nonperturbative intermediate-scale regime associated with pancakes.
Next, we explain in Sec.~\ref{Combining} how we combine this cosmic web power
spectrum with a halo model to extend our model to highly nonlinear scales, associated
with inner halo regions.
Then, we compare our model with numerical simulations in Sec.~\ref{results},
and we conclude in Sec.~\ref{Conclusion}.

The reader who is only interested in our final numerical results may directly go to
Sec.~\ref{results}.

In this paper, we mainly focus on a flat $\Lambda$CDM cosmology derived from
the five-year observation by the WMAP satellite \cite{Komatsu2009}. We measure some of the ingredients in our
model from five out of 60 realizations of N-body simulations performed in \cite{Taruya2012}, adopting 
this cosmological model.
The numerical data for the power spectrum at large scales ($k<1 h$ Mpc$^{-1}$) is taken from the full 60 simulations
in that paper, while we refer to a higher-resolution simulation by \cite{Valageas2011d} for 
smaller scales.
Finally, we show how well our model can reproduce the power spectrum from N-body simulations 
done in two other cosmologies.

\section{Matter density power spectrum in a Lagrangian framework}
\label{Lagrangian}

In a Lagrangian framework one considers the trajectories $\vx(\vq,t)$
of all particles, of initial Lagrangian coordinates $\vq$ and Eulerian coordinates
$\vx$ at 
time $t$. In particular, at any given time $t$, this defines a mapping, $\vq\mapsto\vx$, 
from Lagrangian to Eulerian space, which fully determines the Eulerian density field
$\rho(\vx)$ through the conservation of matter,
\beq
\rho(\vx) \, \dd \vx = \rhob \, \dd\vq ,
\label{continuity}
\eeq
where $\rhob$ is the mean comoving matter density of the Universe and we work
in comoving coordinates.
Then, defining the density contrast as
\beq
\delta(\vx,t) = \frac{\rho(\vx,t)-\rhob}{\rhob} ,
\label{deltadef}
\eeq
and its Fourier transform as 
\beq
\tdelta(\vk) = \int\frac{\dd\vx}{(2\pi)^3} \, e^{-\ii\vk\cdot\vx} \, \delta(\vx) ,
\label{tdelta}
\eeq
one obtains from Eq.(\ref{continuity})
\beq
\tdelta(\vk) = \int\frac{\dd\vq}{(2\pi)^3} \, \left( e^{-\ii\vk\cdot\vx(\vq)} 
- e^{-\ii\vk\cdot\vq} \right) .
\label{tdelta-q}
\eeq
Next, defining the density power spectrum as
\beq
\lag \tdelta(\vk_1) \tdelta(\vk_2) \rag = \delta_D(\vk_1+\vk_2) P(k_1) ,
\label{Pkdef}
\eeq
we obtain from Eq.(\ref{tdelta-q}), using statistical homogeneity
\cite{Schneider1995,Taylor1996},
\beq
P(k) = \int\frac{\dd\Delta\vq}{(2\pi)^3} \, \lag e^{\ii \vk \cdot \Delta\vx} - 
e^{\ii\vk\cdot\Delta\vq} \rag ,
\label{Pkxq}
\eeq
where we introduced the Lagrangian-space and Eulerian-space separations
$\Delta\vq$ and $\Delta\vx$, 
\beq
\Delta\vq = \vq_2 - \vq_1 , \;\;\; \Delta\vx = \vx_2 - \vx_1 ,
\label{Delta-x}
\eeq
between two particles $\vq_1$ and $\vq_2$.
The expression (\ref{Pkxq}) is fully general since it is a simple consequence
of the matter conservation [Eq.(\ref{continuity})] and of statistical homogeneity.
In particular, it holds for any dynamics, such as the one associated with the
Zel'dovich approximation \cite{Zeldovich1970}, where the mapping $\vx(\vq)$
is given by the linear displacement field.

\section{Large-scale power spectrum (cosmic web)}
\label{large-scale-power}

Our final goal is to build a unified model for the power spectrum that applies from
linear to highly nonlinear scales, which requires at some level a phenomenological
model to describe small scales. As described in Sec.~\ref{Combining} below, 
following \cite{Valageas2011d,Valageas2011e} we use a halo model to combine 
large-scale perturbative schemes with phenomenological approaches.
However, we first focus on our perturbative approach, where we neglect the formation
of virialized halos.
Thus, we consider in this section the power spectrum associated with large-scale bulk flows,
that is, the formation of the cosmic web, disregarding small-scale structures such as
inner halo regions.
This actually includes two regimes: i) the very large scales where shell crossing can be 
neglected, and ii) the intermediate scales, associated with pancakes or filaments,
where shell crossing comes into play to shape the cosmic web, but where the
internal structure of pancakes and filaments can be neglected.
The first regime can be described by perturbation theory, and it is considered in
Secs.~\ref{Zeldovich} and \ref{Perturbative}.
Next, we tackle the second regime in Sec.~\ref{shell-crossing}.

\subsection{Zel'dovich power spectrum}
\label{Zeldovich}

If all particle pairs can be described by the same perturbative framework, the power
spectrum [Eq.(\ref{Pkxq})] can be written as
\beqa
P(k) & = & \int\frac{\dd\Delta\vq}{(2\pi)^3} \, \lag e^{\ii \vk \cdot \Delta\vx} \rag 
\label{Pk-def} \\
& = & \int\frac{\dd\Delta\vq}{(2\pi)^3} \, \exp \left[ \sum_{n=1}^{\infty} 
\frac{\lag (\ii \vk \cdot \Delta\vx)^n \rag_c}{n!} \right] .
\label{Pk-cum}
\eeqa
In the first line we used the fact that the integral of the last term in Eq.(\ref{Pkxq}) gives
a Dirac factor $\delta_D(\vk)$ that can be discarded for $k>0$, and in the second line
we used the usual expansion over cumulants.

In the well-known Zel'dovich approximation \cite{Zeldovich1970}, we use the linear
prediction for the Eulerian separation $\Delta \vx$. Then, for Gaussian initial conditions,
only the first- and second-order cumulants are nonzero. Introducing the displacement field
$\vPsi$, $\vx(\vq,t)= \vq+\vPsi(\vq,t)$, at linear order we have the longitudinal and
transverse variances (with respect to the direction defined by the initial Lagrangian
separation $\Delta\vq$):
\beqa
\sigma^2_{\parallel}(\Delta q) & = & 2 \int \dd\vk \, [1-\cos(k_1 \Delta q)] \, 
\frac{k_1^2}{k^4} \, P_L(k) ,
\label{sig2-par} \\
\sigma^2_{\perp}(\Delta q) & = & 2 \int \dd\vk \, [1-\cos(k_1 \Delta q)] \, 
\frac{k_2^2}{k^4} \, P_L(k) .
\label{sig2-perp}
\eeqa
Here we took $\Delta \vq = (\Delta q) \ve_1$ along the first axes, $\ve_1$, and we labeled
the two transverse axis as $\ve_2$ and $\ve_3$. Thus, 
$\sigma^2_{\parallel}=\lag (\Delta \Psi_{L\parallel})^2 \rag = \lag (\Delta \Psi_{L1})^2 \rag$ and
$\sigma^2_{\perp}=\lag (\Delta \Psi_{L2})^2 \rag = \lag (\Delta \Psi_{L3})^2 \rag$.
Introducing the integrals
\beq
I_{\ell}(q) = \frac{4\pi}{3} \int_0^{\infty} \dd k \; P_L(k) \; j_{\ell}(q k) ,
\label{Il-def}
\eeq
and the variance of the linear one-point displacement along one dimension,
\beq
\sigma_v^2 =  \frac{1}{3} \lag |\vPsi_L(\vq)|^2 \rag = I_0(0) ,
\label{sigv-def}
\eeq
we have:
\beqa
\sigma^2_{\parallel}(\Delta q) & = & 2 \sigma_v^2 - 2 I_0(\Delta q) + 4 I_2(\Delta q) , 
\label{sig-par-I}
\\
 \sigma^2_{\perp}(\Delta q) & = & 2 \sigma_v^2 - 2 I_0(\Delta q) -2 I_2(\Delta q) .
\label{sig-perp-I}
\eeqa
Then, in the Zel'dovich approximation, where $\vx = \vq+\vPsi_L$, Eq.(\ref{Pk-cum})
reads as
\beq
P^{\rm Z}(k) = \int\frac{\dd\Delta\vq}{(2\pi)^3} \; e^{\ii k \mu \Delta q
-\frac{1}{2}k^2\mu^2\sigma_{\parallel}^2 -\frac{1}{2}k^2(1-\mu^2)\sigma_{\perp}^2} ,
\label{PZ-def}
\eeq
where $\mu=(\vk\cdot\Delta\vq)/(k\Delta q)$.
(By symmetry, we always have $\lag\Delta\vPsi\rag=0$.)
At large distances, the linear displacements $\vPsi(\vq_1)$ and $\vPsi(\vq_2)$ of the two
particles become independent and $\sigma_{\parallel}^2$ and $\sigma_{\perp}^2$ converge
to $2\sigma_v^2$, in agreement with Eqs.(\ref{sig-par-I})-(\ref{sig-perp-I}).
Then, Eq.(\ref{PZ-def}) is formally divergent at large $q$, as it contains a Dirac factor
$\delta_D(\vk)$, and for numerical computations it is convenient to compute in a
separate fashion the low-order terms.

\begin{figure}
\begin{center}
\epsfxsize=8.5 cm \epsfysize=6.5 cm {\epsfbox{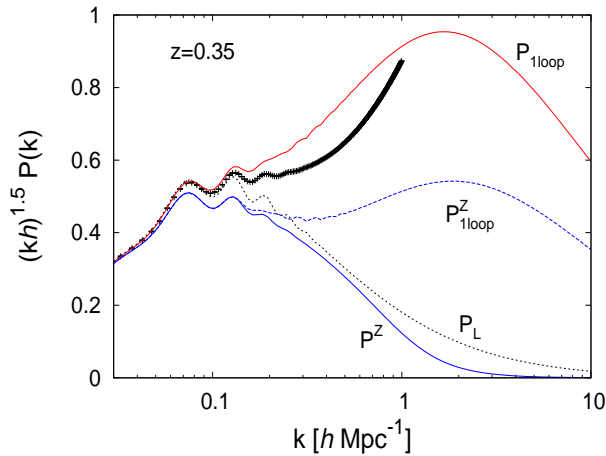}}
\end{center}
\caption{Matter density power spectrum at $z=0.35$ (multiplied by a factor $(k h)^{1.5}$
to decrease the vertical range). We show the results from N-body simulations (data points),
the linear power spectrum $P_L$, the standard one-loop prediction $P_{\rm 1loop}$,
the one-loop prediction $P_{\rm 1loop}^{\rm Z}$ (\ref{P1loop-Z}) within the Zel'dovich
approximation, and the full nonlinear Zel'dovich power spectrum $P^{\rm Z}$ (\ref{PZ-def}).}
\label{fig-Pk_Z1loop_z0.35}
\end{figure}

By expanding the exponential (\ref{Pk-cum}) over powers of $P_L$, one recovers the
standard Eulerian perturbation theory. Within the Zel'dovich approximation, we obtain, for
instance, up to second order \cite{Valageas2008},
\beq
P^{\rm Z}(k) = P_L(k) + P^{\rm Z}_{\rm 1loop}(k) + ...
\eeq
with
\beqa
P^{\rm Z}_{\rm 1loop}(k) & = & - k^2\sigma_v^2 P_L(k) + \int\dd\vk_1\dd\vk_2 \;
\delta_D(\vk_1+\vk_2-\vk) \nonumber \\
&& \times \frac{(\vk\cdot\vk_1)^2(\vk\cdot\vk_2)^2}{2k_1^4k_2^4} P_L(k_1) P_L(k_2) .
\label{P1loop-Z}
\eeqa
This is different from the exact one-loop contribution $P_{\rm 1loop}$ generated by the actual
gravitational dynamics, because at this order we should include the third-order
cumulants and loop corrections to the second-order cumulants
(the cumulant of order $n$ scales as $\lag (\Delta\vPsi)^n\rag_c \propto P_L^{n-1}$
at leading order).

We compare in Fig.~\ref{fig-Pk_Z1loop_z0.35} the power spectra obtained from the
true gravitational dynamics (at linear and one-loop order, as well as the full nonlinear
power spectrum from N-body simulations) with those associated with the Zel'dovich
approximation (at one-loop order and for the full nonlinear power spectrum).
As is well known, the standard Eulerian one-loop prediction $P_{\rm 1loop}$ improves the
agreement with simulations as compared with linear theory on large scales, but quickly
gives too much power and is badly behaved at high $k$.
In contrast, the nonlinear Zel'dovich power spectrum $P^{\rm Z}$ decays faster than the
linear power at high $k$ \cite{Schneider1995,Taylor1996}. For instance, for a power-law
linear power spectrum $P_L(k) \propto k^n$ with $-3<n<-1$, we have
$P^{\rm Z}(k) \sim k^{-3+3(n+3)/(n+1)}$ at high $k$ \cite{Valageas2007a}.
This is due to the fact that within the approximation of linear trajectories, particles keep
moving along straight lines after shell crossing, so that pancakes and filaments quickly
fatten and dissolve after their formation.
This also leads to a one-loop Zel'dovich prediction (\ref{P1loop-Z}) that is smaller than the
linear power spectrum on weakly nonlinear scales, where it is meaningful and follows the
full nonlinear power $P^{\rm Z}$. Therefore, at one-loop order, the Zel'dovich approximation
is actually worse than the simple linear approximation.

Figure~\ref{fig-Pk_Z1loop_z0.35} teaches us two things if we wish to describe
the power spectrum in a Lagrangian framework:

(1) We must explicitly modify the Zel'dovich approximation at one-loop order to (partly) cure the
underestimation of the power spectrum on weakly nonlinear scales.

(2) We must modify the behavior after shell crossing to cure the high-$k$ damping.

\subsection{Perturbative Lagrangian ansatz}
\label{Perturbative}

\subsubsection{Factorized ansatz}
\label{Factorized}

Before tackling the high-$k$ regime in Sec.~\ref{shell-crossing} (point 2 above),
we address the weakly nonlinear regime in this section (point 1), where we disregard
shell crossing.
Thus, we look for a Lagrangian ansatz that improves over (\ref{PZ-def}) on weakly
nonlinear scales and shows a perturbative expansion over integer powers of $P_L$
as in the true gravitational dynamics.

By symmetry, the means $\lag (\Delta\Psi_{\parallel})^2\Delta\vPsi_{\perp}\rag_c$ 
and $\lag (\Delta\vPsi_{\perp})^3\rag_c$ are zero, and at order $P_L^2$
we are left with the two third-order
cumulants, $\lag (\Delta\Psi_{\parallel})^3\rag_c$ and 
$\lag \Delta\Psi_{\parallel}(\Delta\Psi_{\perp})^2\rag_c$.
In particular, the latter average shows that beyond the Gaussian order (\ref{PZ-def}), the
longitudinal and transverse displacements are generically correlated.
Nevertheless, because our goal is to build a simple ansatz for Eq.(\ref{Pk-cum}), we consider
a simplified model where the longitudinal and transverse displacements are uncorrelated.
Then, since $\lag \Delta\Psi_{\perp}^4\rag_c$ is of order $P_L^3$, we keep the linear Gaussian
for the transverse part and look for an ansatz of the form
\beq
\Pa(k) = \int \frac{\dd\Delta\vq}{(2\pi)^3} \; \lag e^{\ii k \mu \Delta x_{\parallel}} \rag_{\parallel} \;
e^{-\frac{1}{2} k^2 (1-\mu^2) \sigma_{\perp}^2} ,
\label{Pa-def}
\eeq
where $\Delta x_{\parallel}=\Delta q+\Delta\Psi_{\parallel}$ is the longitudinal Eulerian
separation and we need to specify a model for the mean $\lag .. \rag_{\parallel}$.
Thus, the difference from the Zel'dovich approximation (\ref{PZ-def}) is that we go beyond
the Gaussian for the longitudinal part.
Apart from simplicity and ordering in perturbation theory, the reason why we focus on the
longitudinal part is that it should be more sensitive than the transverse part to the
progress of nonlinear clustering, even at a qualitative level.
Indeed, by symmetry, the probability distribution of the relative displacement along
any transverse direction, $\cP(\Delta\Psi_{\perp})$, remains even and peaks at
$\Delta\Psi_{\perp}=0$, so that on a qualitative level its shape remains similar to a Gaussian.
In contrast, the probability distribution of the longitudinal Eulerian separation,
$\cP(\Delta x_{\parallel})$, which in the linear regime is a Gaussian centered on
$\Delta x_{\parallel}=\Delta q$, develops
an asymmetry (skewness) that is characteristic of the gravitational dynamics.
Moreover, following \cite{Valageas2011a}, negative values of $\Delta x_{\parallel}$ are
expected to be closely associated with shell crossing, so that keeping track of
$\Delta x_{\parallel}$ will be useful to handle shell-crossing effects,
as explained in Sec.~\ref{shell-crossing} below.

\subsubsection{One-loop order}
\label{One-loop}

It is convenient to define the dimensionless Eulerian longitudinal separation
$\kappa_{\parallel}$ by
\beq
\kappa_{\parallel} = \frac{\Delta x_{\parallel}}{\Delta q} = 1 + 
\frac{\Delta\Psi_{\parallel}}{\Delta q} ,
\label{kappa-def}
\eeq
and its linear variance and its skewness by
\beq
\sigma_{\kappa_{\parallel}}^2= \frac{\sigma_{\parallel}^2}{(\Delta q)^2} ,  \hspace{0.7cm}
S^{\kappa_{\parallel}}_3 = 
\frac{\lag \kappa_{\parallel}^3\rag_c}{\lag \kappa_{\parallel}^2\rag_c^2} = (\Delta q) \; 
\frac{\lag (\Delta\Psi_{\parallel})^3\rag_c}{\lag (\Delta\Psi_{\parallel})^2\rag_c^2} .
\label{sig-kappa-def}
\eeq

\begin{figure}
\begin{center}
\epsfxsize=8.5 cm \epsfysize=6.5 cm {\epsfbox{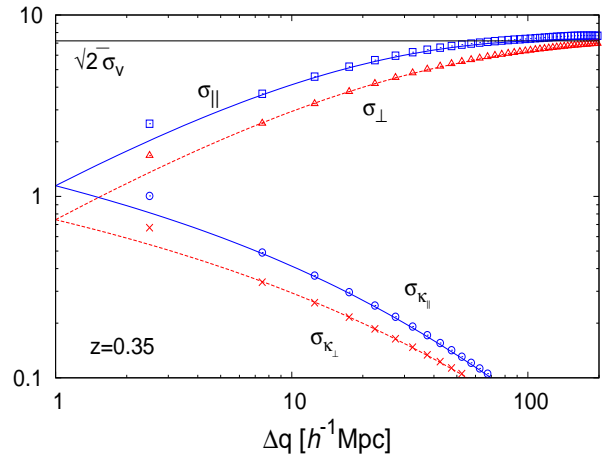}}
\end{center}
\caption{The linear variances $\sigma_{\parallel}$ and $\sigma_{\perp}$
(in units of $h^{-1}$Mpc), and $\sigma_{\kappa}$, at $z=0.35$. The lines are the
linear theory predictions (\ref{sig2-par})-(\ref{sig2-perp}), and the points are the results
from N-body simulations.}
\label{fig-sig2v_z0.35}
\end{figure}

We show the linear variances $\sigma_{\parallel}$ and $\sigma_{\kappa_{\parallel}}$
[as well as the transverse counterparts $\sigma_{\perp}$ and
$\sigma_{\kappa_{\perp}}=\sigma_{\perp}/(\Delta q)$] in Fig.~\ref{fig-sig2v_z0.35}.
We can check that above $3 h^{-1}$Mpc, at $z=0.35$, linear theory gives a good estimate
of these variances.
On smaller scales, nonlinearities and nonperturbative effects (virialized motions in the
shell-crossing regime), which we do not consider in this section, increase the amplitude
of these rms displacements. 
As noticed above, at large distances $\sigma_{\parallel}$ and $\sigma_{\perp}$ converge to
$\sqrt{2}\sigma_v$, because the motions of the two particles become independent.
This implies that $\sigma_{\kappa}$ decreases as $1/(\Delta q)$ on large scales.
On small separations, for CDM power spectra that decrease faster than $k^{-3}$ at high $k$,
$\sigma_{\parallel}$ and $\sigma_{\perp}$ decrease as $\Delta q$, and $\sigma_{\kappa}$
converges to a constant. However, we can see that these asymptotic regimes are only
reached on very large or very small scales.
The value of $\sigma_{\kappa}$ is a better measure of nonlinearity than the rms 
displacements $\sigma_{\parallel}$ and $\sigma_{\perp}$, and $\sigma_{\kappa} \sim 1$
marks the transition to the nonlinear regime. From the definition (\ref{kappa-def}),
in a spherical dynamics $\kappa_{\parallel} \sim (1+\delta)^{-1/3}$, and deviations of order
unity
of $\kappa_{\parallel}$ from its mean are associated with density contrasts of order unity.
In particular, $\kappa_{\parallel}-1=-1$ corresponds to shell crossing and infinite density.
In CDM cosmologies, smaller scales turn nonlinear first, and we can check in
Fig.~\ref{fig-sig2v_z0.35} that $\sigma_{\kappa}$ increases on smaller scales.

To further simplify our ansatz, we keep the linear theory prediction
$\sigma_{\kappa_{\parallel}}^2$ for the variance $\lag \kappa_{\parallel}^2\rag_c$.
Then, the only difference at order $P_L^2$ of the power spectrum (\ref{Pa-def}) from the
Zel'dovich approximation (\ref{PZ-def}) arises from the skewness $S_3^{\kappa_{\parallel}}$:
\beq
\Pa_{\rm 1loop}(k) = P^{\rm Z}_{\rm 1loop}(k) + P_{S_3}(k) ,
\label{Pa-1loop}
\eeq
with
\beq
P_{S_3}(k) = \int\frac{\dd\Delta\vq}{(2\pi)^3} \; \sin(k\mu\Delta q) \;
(k\mu\Delta q)^3 \; \frac{S_3^{\kappa_{\parallel}} \sigma_{\kappa_{\parallel}}^4}{6} .
\label{P-S3-def}
\eeq
The integration over angles gives
\beqa
P_{S_3}(k) & = & \int_0^{\infty} \frac{\dd\Delta q}{2\pi^2} \; (\Delta q)^2 \;
\frac{S_3^{\kappa_{\parallel}}\sigma_{\kappa_{\parallel}}^4}6 \; 
\biggl \lbrace [6-(k\Delta q)^2]  \nonumber \\
&& \hspace{-0.8cm} \times \cos(k\Delta q) + [(k\Delta q)^2-2] \, 
3 \frac{\sin(k\Delta q)}{k\Delta q} \biggl \rbrace .
\label{PS3-1}
\eeqa
Then, to complete our model at one-loop order, we must set the skewness
$S_3^{\kappa_{\parallel}}$.
A first choice would be to use the prediction from lowest-order perturbation theory.
Then, going up to second order in Lagrangian perturbation theory
(``2LPT'', see \cite{Buchert1994,Bouchet1995,Scoccimarro1998}), we obtain
\beqa
S_3^{\kappa_{\parallel},{\rm P.T.}}(\Delta q) & = & \frac{18}{7\sigma_{\kappa_{\parallel}}^4} 
\int\dd\vk_1\dd\vk_2 \; P_L(k_1) P_L(k_2) \nonumber \\
&& \hspace{-2cm} \times \; \frac{k_{1\parallel} k_{2\parallel} (k_{1\parallel}+k_{2\parallel})}
{k_1^2 k_2^2 \, |\vk_1+\vk_2|^2 (\Delta q)^3} 
\left[ 1 - \frac{(\vk_1\cdot\vk_2)^2}{k_1^2k_2^2} \right] \nonumber \\
&& \hspace{-2cm} \times \{ \sin[(k_{1\parallel}+k_{2\parallel})\Delta q] 
- \sin(k_{1\parallel}\Delta q) - \sin(k_{2\parallel}\Delta q) \} . \nonumber \\
&& \label{psi3-tree}
\eeqa
However, this choice implies that the power spectrum $\Pa(k)$ only agrees with perturbation
theory at the linear level.
Indeed, it misses the contribution from the cross-correlation
$\lag\Delta\Psi_{\parallel}(\Delta\vPsi_{\perp})^2\rag_c$ and one-loop contributions to the
variances $\lag(\Delta\Psi_{\parallel})^2\rag_c$ and $\lag(\Delta\vPsi_{\perp})^2\rag_c$. 
In principle, this could be included by building an ansatz for the bivariate distribution
$\cP(\Delta\Psi_{\parallel},\Delta\vPsi_{\perp})$ that does not factorize and is parameterized
by the exact second and third cumulants (up to one-loop order).
In this spirit, one may include perturbative results up to any finite order, at the price of
increasingly complex models for $\cP(\Delta\Psi_{\parallel},\Delta\vPsi_{\perp})$.

\begin{figure}
\begin{center}
\epsfxsize=8.5 cm \epsfysize=6.5 cm {\epsfbox{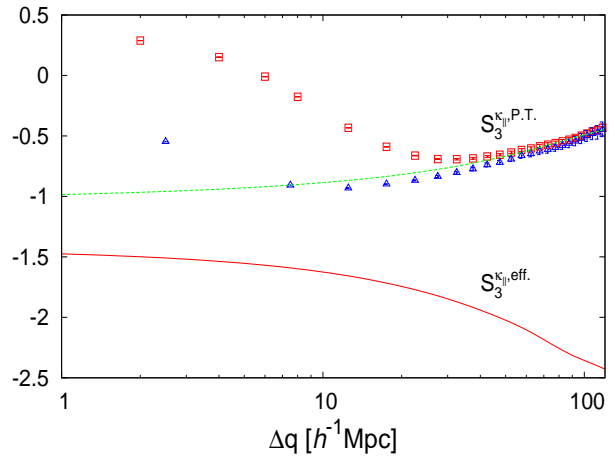}}
\end{center}
\caption{The skewness $S_3^{\kappa_{\parallel}}$ given by lowest-order perturbation theory
[Eq.(\ref{psi3-tree}), upper dashed line] and by the one-loop power-spectrum matching
[Eq.(\ref{S3-1loop}), lower solid line].
The points are the results from N-body simulations at $z=0.35$ (squares) and
$z=2$ (triangles)}
\label{fig-S3}
\end{figure}

Our approach in this paper follows a different strategy. We consider
Eq.(\ref{Pa-def}) as a qualitative ansatz that allows us to build a simple model that is 
physically consistent and provides a good behavior in all regimes, as we explain below.
Then, we treat the parameters of this ansatz as free parameters that we set so as to
match the exact perturbative expansion up to the required order.
In this paper we only go up to one-loop order (i.e., $P_L^2$), which means that we choose
$S_3^{\kappa_{\parallel}}$ so that $\Pa_{\rm 1loop}(k)=P_{\rm 1loop}(k)$, that is,
\beq
P_{S_3}(k) = P_{\rm 1loop}(k) - P^{\rm Z}_{\rm 1loop}(k) ,
\label{P1loop-P1loopZ}
\eeq
where $P_{\rm 1loop}$ is the exact one-loop power spectrum given by standard perturbation
theory.
Inverting Eq.(\ref{PS3-1}) this gives the effective skewness
\beqa
S_3^{\kappa_{\parallel},{\rm eff.}}(\Delta q) & = & - \frac{24\pi}{\sigma_{\kappa_{\parallel}}^4} 
\int_0^{\infty} \dd k \; \frac{P_{\rm 1loop}(k) - P^{\rm Z}_{\rm 1loop}(k)}{(\Delta q)^4 k^2} 
\nonumber \\
&& \times \left[ 2 + \cos(k \Delta q) - 3 \frac{\sin(k\Delta q)}{k\Delta q} \right] .
\label{S3-1loop}
\eeqa
The explicit expression (\ref{S3-1loop}) allows us to recover the exact one-loop power
spectrum
for any cosmology. To be meaningful, such an approach requires that we start from an
ansatz that is not too far from the exact dynamics. This is indeed the case on a qualitative
level,
as Eqs.(\ref{psi3-tree}) and (\ref{S3-1loop}) show similar functional forms: in both cases
$(S_3^{\kappa_{\parallel}}\sigma_{\kappa_{\parallel}}^4)$ is a quadratic integral over
$P_L$, and $S_3^{\kappa_{\parallel}}$ is independent of redshift and of the amplitude of the
linear power spectrum.
On a quantitative level, the comparison in Fig.~\ref{fig-S3} shows that the effective
skewness is greater (in amplitude) than the perturbative skewness by a factor of about
2 and that it has the same sign.

Of course, if one is interested in the statistics of $\kappa_{\parallel}$ for its own sake,
one should use the perturbative prediction (\ref{psi3-tree}) rather than the effective value
(\ref{S3-1loop}). In particular, Fig.~\ref{fig-S3} shows that the skewness is well described
by lowest-order perturbation theory above $15 h^{-1}$Mpc, at $z=0.35$.
(On small scales, the skewness measured in the simulations increases
and becomes positive because of nonlinearities and nonperturbative effects.
As expected, this upturn shifts to larger scales at lower redshift.)

\begin{figure}
\begin{center}
\epsfxsize=8.5 cm \epsfysize=6.5 cm {\epsfbox{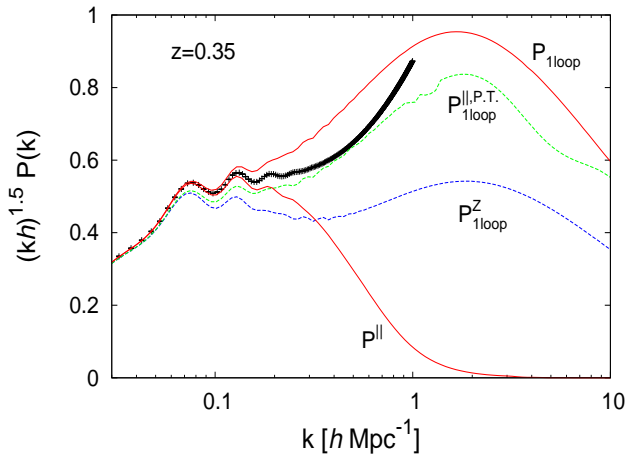}}
\end{center}
\caption{Matter density power spectrum at $z=0.35$, as in Fig.~\ref{fig-Pk_Z1loop_z0.35}.
We show the results from N-body simulations (data points),
the standard one-loop prediction $P_{\rm 1loop}$,
the one-loop prediction $P_{\rm 1loop}^{\rm Z}$ (\ref{P1loop-Z}) within the Zel'dovich
approximation, the one-loop prediction $P^{\rm \parallel,P.T.}_{\rm 1loop}$ (\ref{Pa-1loop})
using the perturbative skewness (\ref{psi3-tree}), and the full nonlinear prediction $\Pa$
of Eq.(\ref{Pa-1}) using the effective skewness (\ref{S3-1loop}).}
\label{fig-Pk_S3_z0.35}
\end{figure}

The effective value (\ref{S3-1loop}) only makes sense as an ingredient for a model for
the power spectrum.
Thus, we also show in Fig.~\ref{fig-Pk_S3_z0.35} the one-loop power spectrum
(\ref{Pa-1loop})
that would be obtained using the perturbative skewness (\ref{psi3-tree}).
This yields a power spectrum $P^{\parallel,\rm P.T.}_{\rm 1loop}$
that is halfway between the Zel'dovich and gravitational one-loop
power spectra and much closer to the N-body data.
Therefore, it already provides a more realistic starting point than the Zel'dovich
approximation. Then, modifying the skewness as in Eq.(\ref{S3-1loop}) to recover the
exact one-loop power spectrum is not a very large modification, and we can hope that the
ansatz determined by Eqs.(\ref{Pa-def}) and (\ref{S3-1loop}) remains sufficiently close
to the exact dynamics to be useful.

\subsubsection{Resummed ansatz}
\label{resummed}

The one-loop power spectrum (\ref{Pa-1loop}) was obtained by expanding the exponential
(\ref{Pk-cum}) up to order $P_L^2$, which only involves the second and third cumulants.
However, one of the main goals of this paper is precisely not to expand the exponential
(\ref{Pk-cum}). Indeed, it is well known that no probability distribution exists such that all
its cumulants vanish beyond some order $n>3$ (i.e., the only distribution with a finite set
of nonzero cumulants is the Gaussian). In particular, approximations such as the
Gram-Charlier or Edgeworth expansions, where we expand up to some finite order
over cumulants, lead to functions that are not guaranteed to be positive and may not be
valid probability distributions.
Of course, this is not necessarily a problem, if one restricts oneself to the range of validity of
such approximations.
However, the spirit of this paper is to avoid as far as possible such inconsistencies and
to develop an approach that remains well behaved in all regimes.

Indeed, we can hope that by satisfying such constraints, the convergence of our scheme
will be improved and the matching to the highly nonlinear regime will be smoother.
In particular, this should avoid the bad behavior encountered in the standard Eulerian 
perturbative expansion, where higher orders improve the accuracy on very large scales
but yield increasingly divergent quantities on small scales.
Thus, the rise of $\Pa_{\rm 1loop}$ at high $k$ in Fig.~\ref{fig-Pk_S3_z0.35}, around
$k > 0.5 h/$Mpc, is an artifact due to the truncation at one-loop order. As for the Eulerian
perturbative expansion, higher orders will also give large contributions on these scales,
and the much smaller full nonlinear power spectrum will result from a compensation
between these much larger terms. [Indeed, as explained in point 2 in Sec.~\ref{Zeldovich},
we know that the logarithmic nonlinear power spectrum $k^3 P(k)$ should not grow
much beyond the nonlinear scale as long as the shell-crossing regime has not been
drastically modified from the Zel'dovich dynamics by trapping particles in small-scale
structures.]
This means that the perturbative expansion has a very slow convergence at best
(provided it is convergent, which is not always the case \cite{Valageas2007a}), and
that including an increasing number of higher-order terms is not sufficient to build
a useful model for our purposes.

To resum all cumulants, it is convenient to define the cumulant-generating function
$\varphi_{\parallel}(y)$ of $\kappa_{\parallel}$ as
\beq
e^{-\varphi_{\parallel}(y)/\sigma_{\kappa_{\parallel}}^2} = \lag e^{-y\kappa_{\parallel}/\sigma^2_{\kappa_{\parallel}}} 
\rag_{\parallel} ,
\label{phi-def}
\eeq
which can be expanded at $y=0$ as
\beq
\varphi_{\parallel}(y) = - \sum_{n=1}^{\infty} \frac{S^{\kappa_{\parallel}}_n}{n!} \; (-y)^n , 
\;\;\; S^{\kappa_{\parallel}}_n = 
\frac{\lag\kappa_{\parallel}^n\rag_c}{\sigma_{\kappa_{\parallel}}^{2(n-1)}} .
\label{phi-cum}
\eeq
(This function is always well defined by the average (\ref{phi-def}), at least along
the whole imaginary axis, even when the series (\ref{phi-cum}) is divergent or cumulants
beyond a finite order do not exist.)
The choice of the generating function $\varphi_{\parallel}$ will define our ansatz (\ref{Pa-def}).
From the Taylor expansion (\ref{phi-cum}), we have the behavior at the origin
\beq
y \rightarrow 0 : \;\; \varphi_{\parallel}(y) = y - \frac{y^2}{2} + S^{\kappa_{\parallel}}_{3} \, 
\frac{y^3}{6} + ...
\label{phi-y-0}
\eeq
Defining the probability distribution $\cP_{\parallel}(\kappa_{\parallel})$ through its
cumulant-generating function $\varphi_{\parallel}(y)$ shows several advantages.
First, imposing the expansion (\ref{phi-y-0}) up to order $y^3$ at the origin automatically
ensures
that the probability distribution is normalized to unity, its mean is 
$\lag\kappa_{\parallel}\rag=1$, its variance $\sigma_{\kappa_{\parallel}}^2$, and its skewness
$S^{\kappa_{\parallel}}_3$.
Second, choosing coefficients $S^{\kappa_{\parallel}}_n$ 
(i.e., a function $\varphi_{\parallel}$) that do not
depend on time or the amplitude of the linear power spectrum automatically ensures the
scaling over $P_L$ given by perturbation theory (at leading order).
Indeed, as we have already noticed above, in perturbation theory
$\lag\kappa_{\parallel}^n\rag_c$ scales
as $P_L^{n-1}$. This ensures that the power spectrum (\ref{Pa-def}) can be expanded over
integer powers of $P_L$ as in the exact perturbation theory
(but of course, the exact value of each term will not be reproduced by our simple model).
Third, from Eq.(\ref{phi-def}), we can see that the average $\lag..\rag_{\parallel}$ that
enters Eq.(\ref{Pa-def}) has an explicit expression in terms of $\varphi_{\parallel}$.
This avoids performing an additional integration over the probability distribution
$\cP_{\parallel}(\kappa_{\parallel})$, which is convenient for numerical purposes.
In this paper, we choose the following ansatz for $\varphi_{\parallel}(y)$,
\beq
\varphi_{\parallel}(y) = \frac{1-\alpha}{\alpha} \left(1+\frac{y}{1-\alpha}\right)^{\alpha}
- \frac{1-\alpha}{\alpha} ,
\label{phi-alpha-def}
\eeq
where the scale-dependent parameter $\alpha(\Delta q)$ is set from Eq.(\ref{phi-y-0}),
\beq
S^{\kappa_{\parallel}}_3 = \frac{\alpha-2}{\alpha-1} .
\label{S3-alpha}
\eeq
This gives the dependence of $\varphi_{\parallel}$ on the scale $\Delta q$
through Eq.(\ref{S3-1loop}).
As seen in Fig.~\ref{fig-S3}, the skewness $S_3^{\kappa_{\parallel},{\rm eff.}}$ is negative, 
hence the parameter $\alpha$ falls in the range
\beq
1< \alpha \leq 2 \; .
\label{alpha-1-2}
\eeq
The case of a zero skewness corresponds to $\alpha=2$, where the generating function
(\ref{phi-alpha-def}) becomes the quadratic polynomial $\varphi_{\parallel}(y)=y-y^2/2$,
and we recover the Gaussian case (i.e., the Zel'dovich approximation (\ref{PZ-def})).

\begin{figure}
\begin{center}
\epsfxsize=8.5 cm \epsfysize=6.5 cm {\epsfbox{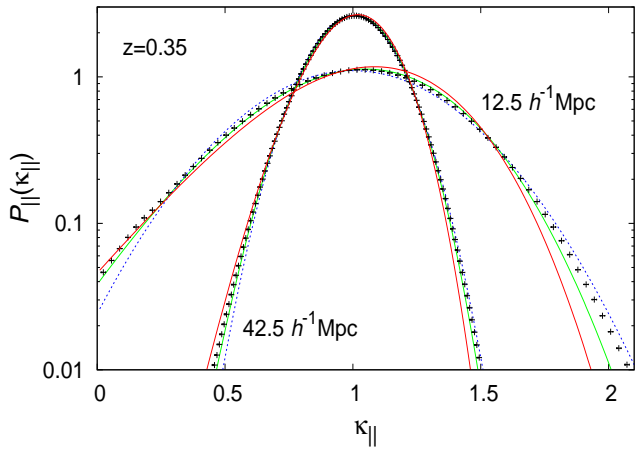}}
\end{center}
\caption{The probability distribution of the longitudinal relative displacement,
$\kappa_{\parallel}=\Delta x_{\parallel}/\Delta q$, at $z=0.35$, for the Lagrangian separations
$\Delta q=12.5 h^{-1}$Mpc and $\Delta q=42.5 h^{-1}$Mpc.
We show the linear-theory  Gaussian (dotted line) and our model (\ref{Pkappa-def})
(solid lines), using for the skewness either Eq.(\ref{psi3-tree}) (intermediate curve) or (\ref{S3-1loop})
(largest departure from the Gaussian).
The points are the results from N-body simulations.}
\label{fig-Pkappa_z0.35}
\end{figure}

The power law (\ref{phi-alpha-def}) is the simplest function that obeys the constraint
(\ref{phi-y-0}) while being a valid cumulant-generating function. [This requires, for
instance, that $-\varphi_{\parallel}(y)$ be convex.]
Thus, it provides a simple continuation of the series of cumulants: instead of truncating at
third order (i.e., at the skewness), we automatically generate all higher-order cumulants
in a manner that provides a meaningful probability distribution 
$\cP_{\parallel}(\kappa_{\parallel})$ in all regimes (i.e., $\cP_{\parallel}(\kappa_{\parallel})$ 
is always positive, normalized to unity, and with a mean $\lag\kappa_{\parallel}\rag=1$). 
From Eq.(\ref{phi-def}), this ``perturbative'' probability distribution 
$\cP_{\parallel}(\kappa_{\parallel})$ is given by the inverse Laplace transform
\beq
\cP_{\parallel}(\kappa_{\parallel}) = \inta \frac{\dd y}{2\pi\ii\sigma^2_{\kappa_{\parallel}}} \; 
e^{[\kappa y -\varphi_{\parallel}(y)]/\sigma^2_{\kappa_{\parallel}}} .
\label{Pkappa-def}
\eeq
Because the skewness $S^{\kappa_{\parallel}}_3$ is negative, the right tail is sharper than the
left tail, as can be checked in Fig.~\ref{fig-Pkappa_z0.35}. 
In agreement with Fig.~\ref{fig-S3}, Fig.~\ref{fig-Pkappa_z0.35} shows that using the
perturbative prediction (\ref{psi3-tree}), with the ansatz (\ref{phi-alpha-def}), provides a
good match to
the full probability distribution measured on large scales in N-body simulations. As expected, 
using the larger value (\ref{S3-1loop}) for the skewness amplifies the deviation from the
Gaussian and yields a distribution that is somewhat too skewed towards the left.
However, the point of Fig.~\ref{fig-Pkappa_z0.35} is only to check that our model,
defined by Eqs.(\ref{S3-1loop}) and (\ref{phi-alpha-def}), provides a physically consistent
and realistic ansatz. This should be understood as a simpler effective system that is not
expected to simultaneously reproduce with the same accuracy all statistical properties of
the dynamics.
Thus, depending on whether one is interested in $\cP_{\parallel}(\kappa_{\parallel})$ or in
$P(k)$, one should use either Eq.(\ref{psi3-tree}) or (\ref{S3-1loop}).

In both cases, the skewness is negative, which
can be understood from the effects of gravity, as compared with
the Zel'dovich dynamics (associated with the linear theory Gaussian).
Indeed (considering for instance a spherical configuration), particles that move outward
($\kappa_{\parallel}>1$) will be slowed down by gravity until they turn around and eventually
merge into a single halo. This leads to an Eulerian separation that is smaller than the one 
predicted by linear theory, whence a sharper cutoff of the probability distribution
$\cP_{\parallel}(\kappa_{\parallel})$ for large
positive $\kappa_{\parallel}$. In contrast, particles that move inward will be accelerated by
gravity (because of the $1/r^2$ behavior of the three-dimensional gravity). This leads to a
smaller value of the Eulerian separation than the linear prediction, which now gives a
more extended tail towards low values of $\kappa_{\parallel}$, below the static value
$\kappa_{\parallel,\rm static}=1$.
Of course, this argument does not extend to $\kappa_{\parallel}<0$, where shell crossing and 
changes of direction within virialized halos are expected.

The spirit of our approach is different from resummation schemes, where one
performs partial resummations of higher-order terms by computing a subset of higher-order
Feynman diagrams \cite{Crocce2006a,Crocce2006b,Matarrese2007,Bernardeau2008}
or expanding over some auxiliary parameter \cite{Valageas2007,Valageas2007a}
(such as the dimension of space $d$ or the number of field components $N$).
Here we do not explicitly compute approximations to $S_n^{\kappa_{\parallel}}$ for $n\geq 4$
from a subset of diagrams,
as these coefficients are automatically generated by the functional form (\ref{phi-alpha-def})
from the low-order ones. (Here we only take $S_3^{\kappa_{\parallel}}$ as input, but as in
resummed perturbative approaches, we could exactly include all terms up to order $n$
and generate approximations for higher orders by using an ansatz for $\varphi_{\parallel}$
that is parameterized by the set $\{S_3^{\kappa_{\parallel}},..,S_n^{\kappa_{\parallel}}\}$.)
The advantage of our approach is that it ensures a well-behaved ``resummed''
(or rather ``regularized'') probability
distribution $\cP_{\parallel}(\kappa_{\parallel})$, while explicit perturbative approaches do not
always guarantee that their resummation is physically consistent.

As explained in the Introduction, we can hope that by ensuring physical consistency,
our method will be better behaved and show a faster convergence.
In this approach we only ensure self-consistency up to a partial degree, in terms of the
probability distribution $\cP_{\parallel}(\Delta \Psi_{\parallel})$, but we do not show that
physical density fields can exactly realize such a probability distribution, which is a more
difficult problem.
Nevertheless, this is already a significant improvement over previous Lagrangian-space
methods that expand the exponential (\ref{Pk-cum}) and truncate at some finite
order \cite{Matsubara2008,Okamura2011,Carlson2012}.

Then, Eqs.(\ref{phi-alpha-def}), (\ref{S3-alpha}), and (\ref{S3-1loop}) fully define the
power spectrum (\ref{Pa-def}). Using Eq.(\ref{phi-def}), this yields
\beq
\Pa(k) = \int \frac{\dd\Delta\vq}{(2\pi)^3} \; e^{-\varphi_{\parallel}(-y)/\sigma_{\kappa_{\parallel}}^2} \;
e^{-\frac{1}{2} k^2 (1-\mu^2) \sigma_{\perp}^2} ,
\label{Pa-1}
\eeq
where we denote
\beq
y= \ii \, k \mu \Delta q \, \sigma_{\kappa_{\parallel}}^2 .
\label{y-def}
\eeq
We recover the Zel'dovich approximation (\ref{PZ-def}) for $\varphi_{\parallel}(y)=y-y^2/2$,
which corresponds to $\alpha=2$ and $S_3^{\kappa_{\parallel}}=0$ in Eqs.(\ref{phi-alpha-def})
and (\ref{S3-alpha}).
Thus, we can see that our ansatz (\ref{Pa-1}) is a continuous generalization of the
Zel'dovich power spectrum, parameterized by the skewness $S_3^{\kappa_{\parallel}}$.
In agreement with a well-known mathematical result, it is only in the Gaussian case
(i.e., the Zel'dovich approximation) that the cumulant-generating function is a finite-order
polynomial.

As seen in Fig.~\ref{fig-Pk_S3_z0.35}, the power spectrum (\ref{Pa-1}) follows the
one-loop power (\ref{Pa-1loop}) on large scales, which is identical to the exact
one-loop power
spectrum because we use the skewness (\ref{S3-1loop}), and it improves over the Zel'dovich
approximation (compare with the Zel'dovich one-loop power $P_{\rm 1loop}^{\rm Z}$
or with Fig.~\ref{fig-Pk_Z1loop_z0.35}).
At high $k$, $\Pa(k)$ converges back to a behavior similar to the Zel'dovich power
spectrum, due to the fact that particles still escape to infinity after shell crossing.
Although we will modify this behavior in the next section, to include the building
of pancakes in a simplified form, this shows the main property that we looked for 
in this section: the power spectrum remains well behaved at high $k$, with a
universal behavior that does not change with the order of the perturbation theory
up to which one requires consistency.

The perturbative expansion of the power spectrum (\ref{Pa-1})
over powers of the linear power spectrum $P_L$ is likely to be divergent for $k>k^{\max}$,
with
\beq
k^{\max} = \min_{\Delta q} \left[ \frac{\alpha - 1}{\Delta q \sigma^2_{\kappa_{\parallel}}} \right] , 
\label{kmax}
\eeq
because of the finite radius of convergence of the generating function
$\varphi_{\parallel}(y)$ around $y=0$. 
This gives $k^{\max} \sim 0.2 h$Mpc$^{-1}$ at $z=0.35$ (and $k^{\max}$ decreases
with time as $1/\sigma_{\kappa_{\parallel}}^2$).
This does not necessarily apply to the actual power spectrum built by the gravitational
dynamics, and these values would be modified by using different generating functions
than (\ref{phi-alpha-def}).
However, it explicitly shows a possible limitation of some perturbative approaches
and the importance of choosing appropriate expansions or resummations.
This is independent of the limitation due to shell crossing and only due to the fast
growth of high-order cumulants shown by our ansatz.

The power spectrum (\ref{Pa-1}) is closely related to
Lagrangian perturbation theory. Indeed, it can be expanded over integer powers of
$P_L$, and it is based on a partial implicit resummation of higher-order cumulants
$\lag(\Delta\Psi_{\parallel})^n\rag_c$, which scale at leading order over $P_L$ as in
perturbation theory. For simplicity, we have only included the linear theory variances
$\lag(\Delta\Psi_{\parallel})^2\rag$ and $\lag(\Delta\vPsi_{\perp})^2\rag$
and the third-order cumulant $\lag(\Delta\Psi_{\parallel})^3\rag_c$, which is set by the
matching with the exact one-loop contribution to the power spectrum.
However, this approach could be made exact up to an arbitrary order of
perturbation theory by including loop contributions to these quantities and higher-order
cumulants, such as $\lag(\Delta\Psi_{\parallel})^4\rag_c$, and cross-correlations, such
as $\lag\Delta\Psi_{\parallel}(\Delta\Psi_{\perp})^2\rag_c$.
As explained above, in contrast to the standard Eulerian perturbation theory,
the resulting power spectrum would remain well behaved at high $k$,
with a Zel'dovich-like damping.

An advantage over the Eulerian-based approaches is that we are not sensitive
to the ``sweeping effect''  \cite{Valageas2007a,Valageas2008,Bernardeau2008a}
associated with random advection by long wavelengths of the velocity field,
which only move structures by a random uniform shift, because we directly work with
relative displacements.
As compared with most Eulerian perturbative resummation schemes, this means that we
do not need to build a model (from a phenomenological approach or a partial resummation
of perturbative diagrams) for different-time Eulerian response functions or propagators,
which are governed by the one-point velocity distribution
\cite{Bernardeau2010b,Bernardeau2012}. 
This should make such Lagrangian-space approaches more robust because they are
not sensitive to such quantities that introduce additional approximations.

\subsection{Adhesion-like shell-crossing continuation}
\label{shell-crossing}

Whatever the order up to which standard perturbation theory is exactly
recovered by the power spectrum $\Pa(k)$, following the procedure described in the
previous section, one ingredient is still missing: the nonperturbative shell-crossing regime.
In contrast to the standard Eulerian perturbation theory that becomes
meaningless after shell crossing (the equations of motion themselves no longer make
sense in the multistreaming regime), the Lagrangian approach, which is based on particle
trajectories, still makes sense after shell crossing. For instance, in the Zel'dovich
approximation, the particles keep following straight lines after shell crossing,
$\vx(\vq,t) = \vq+\vPsi_L(\vq,t)$. A similar behavior is implicitly included in the
Lagrangian ansatz (\ref{Pa-1}). Shell crossing does not appear in this
formalism and particles eventually escape to infinity as in the Zel'dovich
approximation: the distributions $\cP_{\parallel}(\Delta x_{\parallel})$ and 
$\cP_{\perp}(\Delta\vx_{\perp})$ widen with time.

This implicitly provides a regularization (or completion) of the Eulerian hydrodynamical
equations of motion. Indeed, as noticed above, the latter do not provide a complete
description of the dynamics because they break down after shell crossing and one
should explicitly state how particles behave afterwards. For instance, the
Zel'dovich dynamics can also be written in terms of the Eulerian equations of motion
\cite{Gurbatov1989,Vergassola1994,Valageas2007a}. As compared with the gravitational
case, one replaces the gravitational potential (which is coupled to the density field
through the Poisson equation) with the velocity potential (this is exact at linear order).
This yields an Euler equation for the velocity field that decouples from the density field
and gives back the trajectories predicted by Lagrangian linear theory.
As in the gravitational case, this Euler equation breaks down at shell crossing (when
multistreaming appears), and to continue the dynamics at later times one needs to
go back to the Lagrangian interpretation in terms of trajectories and state that particles
keep moving along their initial direction forever.
However, this is not the only possible continuation.
For instance, following the ``adhesion model'' introduced in \cite{Gurbatov1989},
one can choose to make particles stick together after collisions in order to mimic
the trapping in gravitational potential wells. 
In fact, in dimensions greater than 1, several continuations are possible. In the
most convenient geometrical formulation in terms of convex hulls and Legendre transforms,
one obtains an intricate process of halo mergings and fragmentations 
\cite{Bernardeau2010c,Valageas2011b}, but it is also possible to use another continuation
(which has no simple geometrical interpretation), where halos cannot fragment but show
complex motions along the shock manifold \cite{Bogaevsky2006}.
These two continuations and the Zel'dovich dynamics are identical before shell crossing
and have the same perturbative expansion (they obey the same fluid equations in the
single-stream regime); they only differ after shell crossing, and this only appears
through nonperturbative terms. 

This simple example shows that we could imagine different continuations of the
power spectrum (\ref{Pa-1}) into the shell-crossing regime.
Thus, Eq.(\ref{Pa-1}) implicitly includes a ``Zel'dovich-like'' continuation, but one of the
main ideas of this paper is that this is not the best choice.
Indeed, even though in this section we neglect halo formation, or more precisely the
inner halo regions, we would like to obtain a reasonable description of the cosmic
web. In terms of the halo model that we use in Sec.~\ref{Combining} below, the
one-halo term describes the inner halo shells, and the two-halo term, which is based
on the large-scale power spectrum that we consider here, describes the cosmic web
with its pancakes and filaments.
As is well-known, such intermediate-scale structures are erased in the Zel'dovich
approximation
as particles escape to infinity \cite{Bouchet1995}. This has led to the ``truncated Zel'dovich
approximation'' \cite{Coles1993,Sathyaprakash1995}, where one removes the initial power
at high $k$ to suppress the damping of small-scale structures.
More generally, a Zel'dovich-like continuation, such as (\ref{Pa-1}), leads to a falloff of the
density power spectrum at high $k$, due to this erasing of small-scale structures,
and truncating the initial power cannot provide more than a $k^{-3}$ tail 
\cite{Schneider1995,Taylor1996,Valageas2007a} (e.g., for a power-law linear power
spectrum $P_L(k) \propto k^n$ with $-3<n<-1$, we have
$P^{\rm Z}(k) \sim k^{-3+3(n+3)/(n+1)}$ at high $k$).

This means that Zel'dovich-like continuations are not sufficient to describe the cosmic
web. Indeed, they miss the formation of pancakes and filaments that give rise to
additional power at high $k$. For instance, pancakes, or more precisely, extended sheets
of zero thickness, would give a contribution $P(k) \sim k^{-2}$ whereas infinitesimally thin
filaments would give $P(k) \sim k^{-1}$.
In the actual case of the gravitational dynamics with a hierarchical CDM power spectrum,
we do not have such universal and singular structures (pancakes and filaments
can show holes of various sizes), and we would rather have a (multi)fractal density
field with a typical scaling $P(k) \sim k^{-1.2}$ [associated with a correlation function
$\xi(r) \sim r^{-1.8}$ \cite{Balian1989,Balian1989a,Peebles1980}].
Another modification is the finite width of the pancakes and filaments, which can be
associated with the inner halo structure and is not our concern in this section.

On a quantitative level, as recalled in the Introduction, previous approaches that try to
match perturbation theory with a halo model give too little power on intermediate scales,
where $\Delta^2(k) \sim 1$. This could be traced back to the fact that most resummation
schemes converge back to the linear power at high $k$ or even show a sharper cutoff.
From the discussion above, one explanation is that to match the very large-scale perturbative
regime, where shell crossing is truly negligible, to the high-$k$ regime associated with
inner halo regions, we need to take into account intermediate-scale structures
such as pancakes that give a non-negligible contribution on transition scales.
This means that an ``adhesion-like'' continuation should be more efficient than the
Zel'dovich-like continuation.

As recalled above, in dimensions greater than 1, shell crossing is a difficult problem.
In fact, in contrast to the one-dimensional case, defining shell crossing itself is not obvious,
as two particles coming from opposite sides may join circular orbits in a potential well
without physically crossing each other.
In this paper, we take the simple criterion that was used in \cite{Valageas2011a} to
evaluate the impact of shell crossing on the power spectrum.
If the Lagrangian-space to Eulerian-space mapping is potential, that is,
$\vx(\vq,t) = \partial\Phi/\partial\vq$, as in the Zel'dovich and adhesion models, then 
shell crossing is associated with the loss of convexity of the potential $\Phi(\vq,t)$.
[In the Zel'dovich dynamics, $\Phi^{\rm Z}$ is of the form $|\vq|^2/2+\phi_L$, where
$\phi_L$ is stochastic and given by linear theory, and $\Phi^{\rm Z}$ loses its initial convexity
as $\phi_L$ grows with time, while in the adhesion model we take its convex hull,
${\rm conv}(\Phi^{\rm Z})$, which prevents shell crossing but gives rise to shocks
\cite{Vergassola1994,Bernardeau2010c,Valageas2011b}.]
Then, convexity of $\Phi(\vq,t)$ implies that $\Delta x_{\parallel} \geq 0$
(i.e., the longitudinal Eulerian separation does not change sign, for any pair of particles)
\cite{Noullez1994,Brenier2003}.
Then we can choose $\Delta x_{\parallel}<0$ as a criterion of shell crossing 
\cite{Valageas2011a}.
This is obvious in one dimension and provides the adequate generalization to higher
dimensions for potential mappings. In the case of the gravitational dynamics, the mapping
is only potential up to the second order of perturbation theory \cite{Buchert1994a}, but we can
still expect that it gives a useful criterion for the formation of the first large-scale
structures.

\begin{figure}
\begin{center}
\epsfxsize=8.5 cm \epsfysize=6.5 cm {\epsfbox{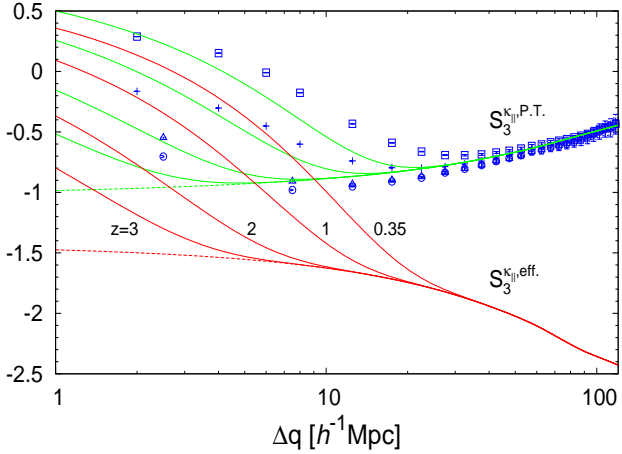}}
\end{center}
\caption{Skewness $S_3^{\kappa_{\parallel}}$ given by low-order perturbation theory
(\ref{psi3-tree}) (upper dashed line) and by the effective value (\ref{S3-1loop}) 
(lower dashed line).
The solid lines that arise from these two large-scale asymptotes are the predictions at
redshifts $z=0.35, 1, 2$ and $3$ (from right to left) obtained from the nonperturbative
adhesion-like probability distribution (\ref{Pkappa-sc}), determined by the corresponding
perturbative part $\cP_{\parallel}$. 
The points are the results from N-body simulations at $z=0.35$ (squares), $z=1$ 
(diamonds), $z=2$ (triangles), and $z=3$ (circles).}
\label{fig-S3_sc}
\end{figure}

Therefore, we modify the ``perturbative'' longitudinal probability distribution 
(\ref{Pkappa-def}) by setting $\Delta x_{\parallel}=0$ to all pairs that had
$\Delta x_{\parallel}<0$. More precisely, we define the ``adhesion-like'' longitudinal probability
distribution as
\beq
\cP_{\parallel}^{\rm ad.}(\kappa_{\parallel}) =  a_1 \, \Theta(\kappa_{\parallel}>0)
\cP_{\parallel}(\kappa_{\parallel}) + a_0 \, \delta_D(\kappa_{\parallel}) ,
\label{Pkappa-sc}
\eeq
where $\Theta(\kappa_{\parallel}>0)$ is the Heaviside function ($1$ for
$\kappa_{\parallel}>0$ and 0 for $\kappa_{\parallel}<0$).
The parameters $a_0$ and $a_1$ are set by the constraints $\lag 1\rag=1$
(i.e., the probability distribution is normalized to unity) and $\lag\kappa_{\parallel}\rag=1$.
From the expression (\ref{Pkappa-def}) we obtain
\beq
a_1 = (1+A_1)^{-1} \;\;\; \mbox{and} \;\;\; a_0= 1 - a_1 + a_1 A_0 ,
\label{a0-a1-def}
\eeq
where we introduce
\beq
A_n = \int_{0^+-\ii\infty}^{0^++\ii\infty} \frac{\dd y}{2\pi\ii\sigma_{\kappa_{\parallel}}^2}
\; \left( \frac{\sigma^2_{\kappa_{\parallel}}}{y} \right)^{n+1} \; 
e^{-\varphi_{\parallel}(y)/\sigma^2_{\kappa_{\parallel}}} ,
\label{An-def}
\eeq
where the contour over $y$ runs to the right of the pole at the origin and to the left of
the branch cut at $y_s=\alpha-1$.
The coefficients $A_n$ are nonperturbative and scale as
$e^{-(\alpha-1)/(\alpha\sigma_{\kappa_{\parallel}}^2)}$, which gives
\beq
(a_1 - 1) \sim a_0 \sim e^{-(\alpha-1)/(\alpha\sigma_{\kappa_{\parallel}}^2)}  .
\label{ao-a1-scale}
\eeq
Therefore, the ``perturbative'' distribution $\cP_{\parallel}$ obtained in Eq.(\ref{Pkappa-def})
in Sec.~\ref{resummed} and its ``adhesion-like'' modification (\ref{Pkappa-sc}) are identical
to all orders of perturbation theory.

We show in Fig.~\ref{fig-S3_sc} the skewness $S^{\kappa_{\parallel}}_3$ defined by this
new probability distribution (\ref{Pkappa-sc}), using either the perturbative skewness
(\ref{psi3-tree}) or the effective value (\ref{S3-1loop}) for the perturbative part
$\cP_{\parallel}$.
On large scales, we recover the skewness associated with the regular distribution
(\ref{Pkappa-def}), with a very fast convergence because the deviation decays as
$\sim e^{-(\Delta q)^2/\sigma_v^2}$.
On small scales, the skewness increases and becomes positive in a fashion similar to the
behavior measured in the simulations. As for the negative sign in the perturbative regime
that we explained in Sec.~\ref{Perturbative}, this can be understood from the 
dynamics. Indeed, because of the ``sticking'' of particle pairs at $\Delta x_{\parallel}=0$,
in the nonlinear regime (i.e., on small scales), which becomes sensitive to this modification,
the low-$\kappa_{\parallel}$ tail becomes very sharp (it is zero for $\kappa_{\parallel}<0$), 
whereas the high-$\kappa_{\parallel}$ tail still extends to infinity, albeit with an
exponential-like decay.
Therefore, the global shape of the probability distribution now shows a broader right tail,
in contrast with the perturbative regime shown in Fig.~\ref{fig-Pkappa_z0.35},
which now leads to a positive skewness.
In the actual gravitational case, there is no such exact ``sticking'' to $\Delta x_{\parallel}=0$,
but there is a trapping within small virialized halos. Thus, collapsed pairs remain bound
with a separation set by the typical size $x_{\rm halo}$ of virialized halos. This plays the
role of the left-tail cutoff, which is no longer sharp at $\Delta x_{\parallel}=0$ but decays
on a scale of order $x_{\rm halo}$, whereas the right tail still extends to infinity and is not
strongly affected by smaller-scale virialization (it corresponds to rare voids).
Of course, we cannot expect the simple model (\ref{Pkappa-sc}) to provide an accurate
prediction for $S_3^{\kappa_{\parallel}}$, even when we use the correct perturbative
limit (\ref{psi3-tree}) on large scales.
However, we can check in Fig.~\ref{fig-S3_sc}
that it already provides a good qualitative description. In particular, it captures the
dependence on redshift of the upturn of $S^{\kappa_{\parallel}}_3$, due to these 
nonperturbative effects that occur after shell crossing.

\begin{figure}
\begin{center}
\epsfxsize=8.5 cm \epsfysize=6.5 cm {\epsfbox{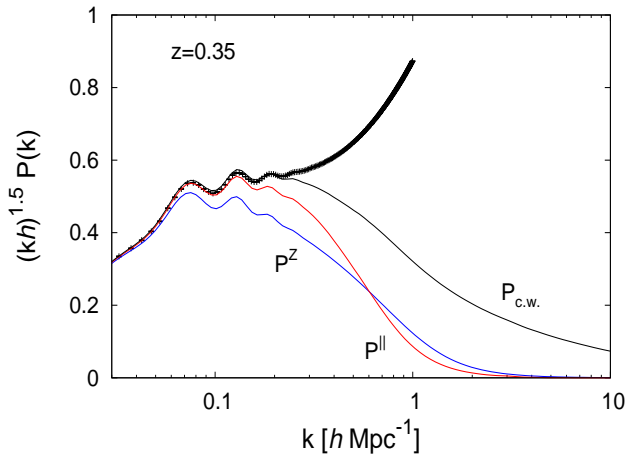}}
\end{center}
\caption{Matter density power spectrum at $z=0.35$, as in Figs.~\ref{fig-Pk_Z1loop_z0.35}
and \ref{fig-Pk_S3_z0.35}.
We show the results from N-body simulations (data points),
the nonlinear Zel'dovich power spectrum
$P^{\rm Z}$ (\ref{PZ-def}), the nonlinear ansatz $\Pa$ (\ref{Pa-1}), and the
``adhesion-like'' continuation $\Psc$ (\ref{Psc-def}) for the ``cosmic web''.}
\label{fig-Pk_sc_z0.35}
\end{figure}

Then, we modify the perturbative power spectrum
(\ref{Pa-1}) by replacing the perturbative probability distribution (\ref{Pkappa-def})
with its adhesion-like extension (\ref{Pkappa-sc}).
Substituting into Eq.(\ref{Pa-def}) gives the ``cosmic web'' power spectrum
\beqa
\Psc(k) & = & \int \frac{\dd\Delta\vq}{(2\pi)^3} \; \frac{1}{1+A_1}
\biggl \lbrace e^{-\varphi_{\parallel}(-\ii k\mu\Delta q\,\sigma_{\kappa_{\parallel}}^2)/\sigma_{\kappa_{\parallel}}^2}
+ A_1 \nonumber \\
&& \hspace{-1.3cm} + \int_{0^+-\ii\infty}^{0^++\ii\infty} \frac{\dd y}{2\pi\ii} \; 
e^{-\varphi_{\parallel}(y)/\sigma^2_{\kappa_{\parallel}}} 
\left[ \frac{1}{y} - \frac{1}{y\! + \! \ii k\mu\Delta q \, \sigma^2_{\kappa_{\parallel}}} \right]
\biggl \rbrace \nonumber \\
&& \hspace{-1.3cm} \times \; e^{-\frac{1}{2} k^2 (1-\mu^2) \sigma_{\perp}^2} ,
\label{Psc-def}
\eeqa
where the contour over $y$ again crosses the real axis between
$0<y<\alpha-1$.
As in \cite{Valageas2011a}, this is a ``sticky model'' that can be seen as
a very simplified version of the full 3D adhesion model. Since we do not modify the transverse
motion, the ``adhesion'' only takes place along the longitudinal direction, which also serves
as the signal of shell crossing. Therefore, we only include planar structures (thin pancakes).
To describe filaments we should also include some sticking along one transverse direction,
but this would require additional ingredients and free parameters (e.g., to set the relative
mass between filaments and pancakes). Hence, for simplicity we only take into account
pancakes as in Eq.(\ref{Psc-def}), which fits in a natural fashion in our framework where we
have already kept track of the longitudinal separation.

As explained above, the shell-crossing correction is nonperturbative, and the power spectrum
(\ref{Psc-def}) is identical to the power spectrum (\ref{Pa-1}) to all orders over
$P_L$.
 
We compare in Fig.~\ref{fig-Pk_sc_z0.35} the three Lagrangian-space power spectra
that we have obtained, the usual Zel'dovich approximation $P^{\rm Z}$, 
our nonlinear ansatz $\Pa$, and its ``adhesion-like'' continuation $\Psc$.
On large scales, the nonperturbative correction is negligible, and we
recover the perturbative power spectrum $\Pa$, which is somewhat larger than the
Zel'dovich power spectrum because of the nonzero skewness $S^{\kappa_{\parallel}}_3$
that allows us to ensure consistency with standard perturbation theory up to one-loop order.
On small scales, the nonperturbative correction becomes dominant
and gives some extra power, associated with the formation of pancakes, with
a high-$k$ tail $\sim k^{-2}$ that decreases more slowly than the Zel'dovich-like
tails (that are steeper than $k^{-3}$).
The nonlinear power spectrum $\Psc$ is the Lagrangian ansatz that we
use in this paper to describe the ``cosmic web''.

\section{Combining the ``cosmic web power spectrum'' with the halo model}
\label{Combining}

\subsection{The halo model from a Lagrangian point of view}
\label{Halo-model}

Because our goal is to build a unified model for the power spectrum that applies from
linear to highly nonlinear scales, we must combine the perturbative approach described 
in the previous section (which is systematic and accurate but only applies to large scales)
with phenomenological models (that are not systematic and only show an accuracy of
$10\%$ but can be applied to small scales). 
Following \cite{Valageas2011d,Valageas2011e}, we consider the halo model
from a Lagrangian point of view instead of the usual Eulerian framework \cite{Cooray2002}.
This provides a convenient framework to include our Lagrangian perturbative approach
within the halo model (the latter being mostly used to describe small,
highly nonlinear scales). In particular, the transition from the perturbative large scales,
driven by bulk flows, to the inner halo regions, driven by virialization, takes place in a
gradual fashion while satisfying matter conservation (i.e., without double counting).

Following \cite{Valageas2011d}, we split the average in Eq.(\ref{Pkxq}) over two terms,
$P_{\rm 1H}$ and $P_{\rm 2H}$, associated with pairs $\{\vq_1,\vq_2\}$ that belong either
to a single halo or to two different halos,
\beq
P(k) = P_{\rm 1H}(k) + P_{\rm 2H}(k) ,
\label{Pk-halos}
\eeq
with
\beq
P_{\rm 1H}(k) = \int\frac{\dd\Delta\vq}{(2\pi)^3} \, F_{\rm 1H}(\Delta q) \, 
\lag e^{\ii \vk \cdot \Delta\vx} - e^{\ii\vk\cdot\Delta\vq} \rag_{\rm 1H}
\label{Pkxq-1H}
\eeq
and
\beq
P_{\rm 2H}(k) = \int\frac{\dd\Delta\vq}{(2\pi)^3} \, F_{\rm 2H}(\Delta q) \, 
\lag e^{\ii \vk \cdot \Delta\vx} - e^{\ii\vk\cdot\Delta\vq} \rag_{\rm 2H} .
\label{Pkxq-2H}
\eeq
Here the averages $\lag ... \rag_{1\rm H}$ and $\lag ... \rag_{2\rm H}$ are the conditional 
averages, knowing that the pair of length $\Delta q$ belongs to a single halo or to two halos,
while $F_{\rm 1H}$ and $F_{\rm 2H}$ are the
associated probabilities. The approximation of spherical halos in Lagrangian space
gives, for the probability that the pair belongs to a single halo \cite{Valageas2011d},
\beq
F_{\rm 1H}(\Delta q) = \int_{\nu_{\Delta q/2}}^{\infty} \frac{\dd\nu}{\nu} \, f(\nu) \,
\frac{(2 q_M-\Delta q)^2 (4 q_M+\Delta q)}{16 q_M^3} ,
\label{F1H-def}
\eeq
where $q_M$ is the Lagrangian radius associated with the mass $M$,
$M=4\pi \rhob q_M^3/3$, and $f(\nu)$ is the scaling function that defines the halo mass
function,
\beq
n(M) \frac{\dd M}{M} = \frac{\rhob}{M} \, f(\nu) \frac{\dd\nu}{\nu} , \;\;
\mbox{with} \;\; \nu = \frac{\delta_L}{\sigma(M)} .
\eeq
Here $\sigma(M)$ is the rms linear density contrast at scale $M$, and 
$\delta_L=\cF^{-1}(200)$
is the linear density contrast associated with the nonlinear density threshold that defines
collapsed halos, which we choose to equal $200$. In numerical computations, we use the
fit to the halo mass function from \cite{Valageas2009}, which has been shown to match
numerical simulations  while obeying the asymptotic large-mass tail 
$f(\nu) \sim e^{-\nu^2/2}$ \cite{Valageas2002b}.
The probability that the pair belongs to two different halos reads as
\beq
F_{\rm 2H}(\Delta q) =  1 - F_{\rm 1H}(\Delta q) .
\label{F2H-F1H}
\eeq
This automatically avoids any double counting in the decomposition (\ref{Pk-halos}),
and it ensures that we count all matter once.

\subsection{One-halo term}
\label{One-halo}

The decomposition (\ref{Pk-halos}) also corresponds
to the one-halo and two-halo terms of the usual halo model \citep{Cooray2002}.
In particular, Eqs.(\ref{Pkxq-1H}) and (\ref{F1H-def}) yield
\cite{Valageas2011d,Valageas2011e}
\beq
P_{\rm 1H}(k) = \int_0^{\infty} \frac{\dd\nu}{\nu} f(\nu) \frac{M}{\rhob (2\pi)^3}
\left(  \tu_M(k) - \tW(k q_M) \right)^2 .
\label{Pk-1H}
\eeq
Here $\tu_M(k)$ is the normalized Fourier transform of the halo radial profile,
\beq
\tu_M(k) = \frac{\int\dd\vx \, e^{-\ii\vk\cdot\vx} \rho_M(x)}{\int\dd\vx \, \rho_M(x)}
= \frac{1}{M} \int\dd\vx \, e^{-\ii\vk\cdot\vx} \rho_M(x) ,
\label{uM-k-def}
\eeq
where $\rho_M(x)$ is the halo density profile for a halo of mass $M$, 
and $\tW(k q)$ is the normalized Fourier transform of the top hat of radius $q$,
\beq
\tW(k q) = \int_V \frac{\dd \vq}{V} \; e^{\ii \vk\cdot\vq} = 3 
\frac{\sin(kq)-kq \cos(kq)}{(k q)^3} .
\label{W-def}
\eeq
To derive Eq.(\ref{Pk-1H}) we used the approximation of fully virialized halos: the two
particles $\vq_1$ and $\vq_2$ have lost all memory of their initial locations and are
independently located at random within the halo.
As in \cite{Valageas2011d}, in numerical computations we use the usual NFW halo profile
\cite{Navarro1997}.

The counterterm $\tW$ in Eq.(\ref{Pk-1H}) arises from the counterterm
$e^{\ii\vk\cdot\Delta\vq}$ of Eq.(\ref{Pkxq-1H}) (this subtracts the contribution from the
mean density). 
The computation in \cite{Valageas2011d} would rather give a factor $(\tu_M^2-\tW^2)$, but we
prefer to use the factor $(\tu_M-\tW)^2$ that readily extends to higher-order multispectra as
seen in \cite{Valageas2011e}.
Moreover, this ensures that the one-halo contribution to the matter power spectrum
will decay as $k^4$ at low $k$, as implied by the conservation of matter and momentum
for small-scale redistributions of matter \cite{Peebles1974}
[whereas the factor $(\tu_M^2-\tW^2)$ only ensures a $k^2$ tail, which is consistent with
the conservation of matter but not of momentum].
We show in Fig.~\ref{fig-dDk_k4} in the Appendix the impact of this low-$k$ tail of the
one-halo term on the power spectrum. 
We find that using a factor $(\tu_M^2-\tW^2)$, which gives a slower falloff at low $k$, can
overestimate the power spectrum by about $5\%$ on transition scales.
Indeed, at higher $k$, in the highly nonlinear regime, the counterterm $\tW$ is negligible, whereas
at lower $k$ the one-halo term itself is negligible.
This shows that on transition scales, where $\Delta^2(k) \sim 1$, the power spectrum is sensitive
to details of the halo model if we require an accuracy of a few percent.
Fortunately, this only appears on a limited range of scales and has no impact on the
perturbative scales. (However, it is important to ensure that the one-halo term decays at least
as $k^2$ at low $k$ rather than converging to a constant as in the usual prescription without any
counterterm.)

\subsection{Two-halo term}
\label{Two-halo}

At a perturbative level, $F_{\rm 1H}$ and $P_{\rm 1H}$ are identically zero while
$F_{\rm 2H}$ is unity, because of the exponential decay of the halo mass function,
of the form $e^{-1/\sigma^2(M)}$, in the rare event limit.
Then, as noticed in \cite{Valageas2011d}, the power spectrum given by perturbation theory
is included in the two-halo contribution (\ref{Pkxq-2H}). More precisely, the expansion
over powers of $P_L$ of $P_{\rm 2H}$ must recover the standard perturbative expansion.
In \cite{Valageas2011d}, we used the simple approximation $P_{\rm 2H}(k) \simeq
F_{\rm 2H}(1/k) P_{\rm pert}(k)$, where $P_{\rm pert}(k)$ is the power spectrum given
by the Eulerian ``steepest-descent" resummation scheme developed 
in \cite{Valageas2007,Valageas2008}. 
Any other Eulerian or Lagrangian resummation scheme could be used, provided it
is well behaved at high $k$, where it becomes subdominant with respect to the
one-halo term. (This excludes the standard perturbation theory, which grows too fast
at high $k$, unless one adds an extra high-$k$ cutoff.)

However, the approaches investigated in \cite{Valageas2011d} did not manage to
provide a fully satisfactory matching to the highly nonlinear regime, because they predicted
too little power on intermediate scales (where $\Delta^2(k) \sim 1 - 10$).
This can be traced to the fact that they usually go back to the linear power at most at high
$k$, which leads to insufficient power on transition scales where $P_{\rm 2H}$ and
$P_{\rm 1H}$ are of the same order.
Another problem was that Lagrangian perturbative schemes, which would be more
convenient to embed within Eq.(\ref{Pkxq-2H}) and to extend to redshift space, made this
lack of power even worse, because they usually display a strong cutoff at high $k$.
In this paper, our goal is to improve over \cite{Valageas2011d} by implementing a
Lagrangian perturbative scheme that is free of this problem and provides reasonably
accurate predictions up to the transition scales.
Thus, we use the ``cosmic web'' power spectrum developed in Sec.~\ref{large-scale-power}
as a basis of our two-halo term, which we write as
\beqa
P_{\rm 2H}(k) & = & \int \frac{\dd\Delta\vq}{(2\pi)^3} \; F_{\rm 2H}(\Delta q) \;
\lag e^{\ii\vk\cdot\Delta\vx} \rag^{\rm vir}_{\Delta q} \;
\frac{1}{1+A_1} \nonumber \\
&& \hspace{-1.4cm} \times \; e^{-\frac{1}{2} k^2 (1-\mu^2) \sigma_{\perp}^2} \;
\biggl \lbrace e^{-\varphi_{\parallel}(-\ii k\mu\Delta q \, \sigma^2_{\kappa_{\parallel}})
/\sigma_{\kappa_{\parallel}}^2} + A_1 \nonumber \\
&& \hspace{-1.4cm} \!\! + \!\! \int_{0^+-\ii\infty}^{0^++\ii\infty} \frac{\dd y}{2\pi\ii} \; 
e^{-\varphi_{\parallel}(y)/\sigma^2_{\kappa_{\parallel}}} 
\left( \! \frac{1}{y} - \frac{1}{y\! + \! \ii k\mu\Delta q \, \sigma^2_{\kappa_{\parallel}}} \! \right)
\!\! \biggl \rbrace  . \nonumber \\
&& 
\label{Pk-2H-1}
\eeqa
We recognize the cosmic web power spectrum given by Eq.(\ref{Psc-def}),
to which we added
the factors $F_{\rm 2H}$ and $\lag e^{\ii\vk\cdot\Delta\vx} \rag^{\rm vir}_{\Delta q}$.
As explained in Eqs.(\ref{Pk-halos})-(\ref{Pkxq-2H}), the factor $F_{\rm 2H}$,
given by Eqs.(\ref{F2H-F1H}) and (\ref{F1H-def}), ensures that mass is conserved:
there is no double counting of particle pairs.
The factor $\lag e^{\ii\vk\cdot\Delta\vx} \rag^{\rm vir}_{\Delta q}$ is associated with small-scale
motions.

Indeed, as seen from the derivation of Eq.(\ref{Psc-def}) in
Sec.~\ref{large-scale-power}, the cosmic web power spectrum $\Psc$
arises from the statistical average of $e^{\ii\vk\cdot\Delta\vx}$ due to large-scale motions,
associated with the bulk flows described by perturbation theory that we regularize
by an adhesion-like continuation. As noticed in Sec.~\ref{shell-crossing}, this simple
procedure takes into account the formation of idealized, infinitesimally thin pancakes.
In the actual gravitational process, pancakes and filaments have a finite width, as
particles keep moving for some time after shell crossing instead of instantaneously
sticking together, and have finite-size turnaround radii and virialized orbits.
More generally, we can split the motion of particles into two components, a first one
associated with large-scale bulk flows, which was considered in Sec.~\ref{large-scale-power}
and corresponds to the ``skeleton'' of the large-scale structures, and a second one
associated with small-scale virialized motions, which gives some thickness to this skeleton.
In particular, for hierarchical linear power spectra with a high-$k$ tail that does not
decrease faster than $k^{-3}$, all particles are expected to belong to collapsed objects
(this may not be the case for linear power spectra with less power at high $k$).
Then, making the approximation that these small-scale and large-scale motions are
decorrelated, we write
\beq
\lag e^{\ii\vk\cdot\Delta\vx}\rag = \lag e^{\ii\vk\cdot\Delta\vx}\rag^{\rm bulk} \; 
\lag e^{\ii\vk\cdot\Delta\vx}\rag^{\rm vir} .
\label{bulk-vir}
\eeq
The first part was the focus of Sec.~\ref{large-scale-power} and corresponds to the
cosmic web power spectrum $\Psc$.
Assuming, as in the derivation of the one-halo term (\ref{Pk-1H}), full virialization
\cite{Valageas2011d}, that is, that the two particles $\vq_1$ and $\vq_2$ are
independently located at random in their host halo, we write
\beqa
\lag e^{\ii\vk\cdot\Delta\vx}\rag^{\rm vir}_{\Delta q} & = & \!\! 
\frac{ \int \frac{\dd\nu_1}{\nu_1} f(\nu_1) 
\lag e^{\ii\vk\cdot\vx_1}\rag_{\!M_1} } { \int \frac{\dd\nu_1}{\nu_1} f(\nu_1) }
\frac{ \int \frac{\dd\nu_2}{\nu_2} f(\nu_2) 
\lag e^{\ii\vk\cdot\vx_2}\rag_{\!M_2} } { \int \frac{\dd\nu_2}{\nu_2} f(\nu_2) }
\nonumber \\
& = & \left[ \frac{ \int_0^{\nu_{\Delta q/2}} \frac{\dd\nu}{\nu} f(\nu) \, \tu_M(k) }
{  \int_0^{\nu_{\Delta q/2}} \frac{\dd\nu}{\nu} f(\nu) } \right]^2 .
\label{vir-1}
\eeqa
In contrast with the one-halo case (\ref{Pk-1H}) where the two particles belong to the same
halo, which gave rise to the factor $\tu_M^2$, here the two particles belong to two different
halos, whence the integration over two halo mass functions, each one with its factor $\tu_M$.
In Eq.(\ref{vir-1}), we only integrate over halos of radius smaller than $\Delta q/2$,
to take into account in an approximate fashion that if a pair of separation $\Delta q$ belongs
to two different halos, these halos are unlikely to have a radius much greater than
$\Delta q/2$.
Thus, massive halos only contribute to the one-halo term (\ref{Pk-1H}) and to the two-halo
term (\ref{Pk-2H-1}) at large distances.

We show in Fig.~\ref{fig-dDk_k4} in the Appendix the impact of this ``virial damping''
factor on the matter power spectrum, by plotting the deviation that would be obtained
by neglecting this term.
Because of the upper bound on halo mass in Eq.(\ref{vir-1}), this factor does not induce
a significant damping of the power spectrum (\ref{Pk-2H-1}) on large scales.
Indeed, in the weakly
nonlinear regime where the two-halo term is dominant, power at wavenumber $k$ typically
comes from pairs of initial separation $\Delta q\sim 1/k$, which selects halo radii $q_M$ that
are smaller than $1/k$ for which the factor $\tu_M(k)$ is close to unity. 
On small, highly nonlinear scales it yields a greater damping as $1/k$ becomes smaller than the
typical size of the halos.
However, on these scales, the matter power spectrum given by our approach is not accurate to
better than $10\%$ because of the one-halo term itself, which involves the halo profiles and their
concentration parameters that are not modeled to a very high accuracy.
Nevertheless, we include this factor in our computations because it naturally arises 
in our framework.

\subsection{Nonperturbative effects on large scales}
\label{nonperturbative-large-scales}

Thus, in our approach we include nonperturbative and shell-crossing effects on the 
large-scale power spectrum through the factors $F_{\rm 2H}$ and
$\lag e^{\ii\vk\cdot\Delta\vx} \rag^{\rm vir}_{\Delta q}$, associated with halo formation and
virialized motions, and through the adhesion-like continuation described in
Sec.~\ref{shell-crossing} (in addition to the one-halo term itself).
This is different from recent Eulerian-space works \cite{Pietroni2012,Carrasco2012}
that propose to include the impact of such 
nonperturbative effects on large scales through additional terms in the hydrodynamical
equations of motion of the fluid approximation, that may be obtained by a combination of
coarse-graining (that sets the form of these terms) and phenomenology (they include
coefficients that are measured in simulations).
Indeed, in our approach the hydrodynamical equations of motion enter through the
perturbative expansion that they imply for the power spectrum, which is
not affected by these nonperturbative effects.
Then, the impact of a small-scale velocity dispersion, due to virialized motions, is  
described by the factor $\lag e^{\ii\vk\cdot\Delta\vx} \rag^{\rm vir}_{\Delta q}$
in Eq.(\ref{bulk-vir}). This would play the role of some pressurelike terms included
in \cite{Pietroni2012,Carrasco2012}.
The trapping within potential wells, that is described in a simplified manner by the
nonperturbative correction (\ref{Psc-def}) associated with the adhesion-like continuation,
may also be described within such methods through a pressure or viscosity term, as in the
original adhesion model \cite{Gurbatov1989}, where the pressureless Euler equation is
replaced with the Burgers equation.

As seen in Sec.~\ref{shell-crossing}, the ``sticky model'' that we use for the cosmic web
power spectrum corresponds to a one-dimensional adhesion model in the inviscid limit,
that is; when the viscosity parameter $\nu$ is sent to $0^+$. The associated Euler equation
is also known as the Burgers equation \cite{Burgers1974}, introduced as a model for
turbulence. Within a specific geometrical formulation of the inviscid limit, the equations of
motion can be explicitly solved in any dimension, through Legendre transforms
and convex hulls \cite{Vergassola1994,Valageas2011b}. However, it remains very difficult
to obtain explicit results for the statistical properties of the associated density and velocity
fields, except for a few cases in one dimension (for power-law initial conditions with
$n=-2$ \cite{Bertoin1998,Valageas2009a} or $n=0$ \cite{Frachebourg2000,Valageas2009b}).
If we keep a finite viscosity parameter $\nu$, the solution is more regular (there are no shocks),
but there are no explicit results for statistical properties of the displacement field either.
This is why we use the simple model of Sec.~\ref{shell-crossing}, which provides
the simple explicit expression (\ref{Psc-def}), whereas a fully three-dimensional adhesion
model with finite viscosity would yield more intricate path integrals.
A second reason for our choice is to avoid introducing additional free parameters.
Indeed, by introducing a finite viscosity parameter, which may
also depend on the local density and on time, as well as other nonlinear terms in the
equations of motion, one needs to build a model for these new parameters or functions.
Although these parameters may be estimated from simulations, we prefer in this paper
to stick to the simplest possible modeling that captures some of the nonperturbative
shell-crossing effects. It may be possible to improve over this first step by considering
more realistic equations of motion (e.g., see \cite{Dominguez2000,Buchert2005}
for earlier works), but we leave this for future studies.

Another approach to take into account such effects would be to go back to the Vlasov
equation of motion \cite{Valageas2001,Valageas2004,Tassev2011}.
In principle, this could provide systematic schemes, but the methods that have been
proposed so far lead to heavier computations, and no fast and accurate method has been
presented yet.

As noticed in the Introduction, the advantage of the Lagrangian-space framework 
over the Eulerian-space framework is that particle trajectories can describe both the
single-stream and multistream regimes, which allows us to include these nonperturbative
effects in the simple fashion described in the previous sections. However, this approach
is not fully systematic, and it involves some phenomenological insight.

\section{Numerical results}
\label{results}

\subsection{WMAP5 cosmology}
\label{WMAP5}

We first compare our model with numerical simulations for a WMAP5 cosmology,
that is, with the set of cosmological parameters $h=0.701$, $\Omega_{\rm m}=0.279$,
$\sigma_8=0.8159$, $n_s=0.96$, and $w_{\rm de}=-1$.
While we use the N-body data of \cite{Taruya2012} on large scales ($k<0.3 h$ Mpc$^{-1}$),
we use the measurements of \cite{Valageas2011d} on smaller scales.
This is because the former one is tuned to give an accurate power spectrum on BAO scales
and has a larger total volume [$60\times(2048\,h^{-1}$ Mpc$)^3\sim515\,h^{-3}$ Gpc$^3$], while the latter gives 
more reliable results over a wide range in $k$ by carefully combining the measurements from some simulations with 
different resolutions
[the highest-resolution simulation employs $2048^3$ particles in a $(512\,h^{-1}$ Mpc$)^3$ cube].
Similarly, we use the simulation results from these papers for the two-point correlation function:
data points from \cite{Taruya2012} are shown on BAO scales (from $70$ to $140\,h^{-1}$ Mpc), 
while those from \cite{Valageas2011d} are depicted on smaller separations.

We show our results for the redshifts $z=3$, $1$, and $0.35$, which was already used in
the previous sections to illustrate the building blocks of our model.
The redshift $z=0.35$ is the lowest available redshift for this set of simulations (but we
also consider $z=0$ in another set of simulations in Sec.~\ref{WMAP3}).
However, it is also relevant from an observational point of view, as recent or upcoming
redshift surveys aiming at measuring cosmic expansion or structure growth via
baryon acoustic oscillations and redshift-space distortions probe similar redshifts
(e.g., the SDSS LRG sample \cite{Eisenstein2005,Kazin2010}) or higher
redshifts (e.g., $z\sim 0.57$ for BOSS DR9 \cite{Anderson2013}).

\subsubsection{Power spectrum}
\label{power-spectrum}

\begin{figure*}
\begin{center}
\epsfxsize=6.35 cm \epsfysize=6.5 cm {\epsfbox{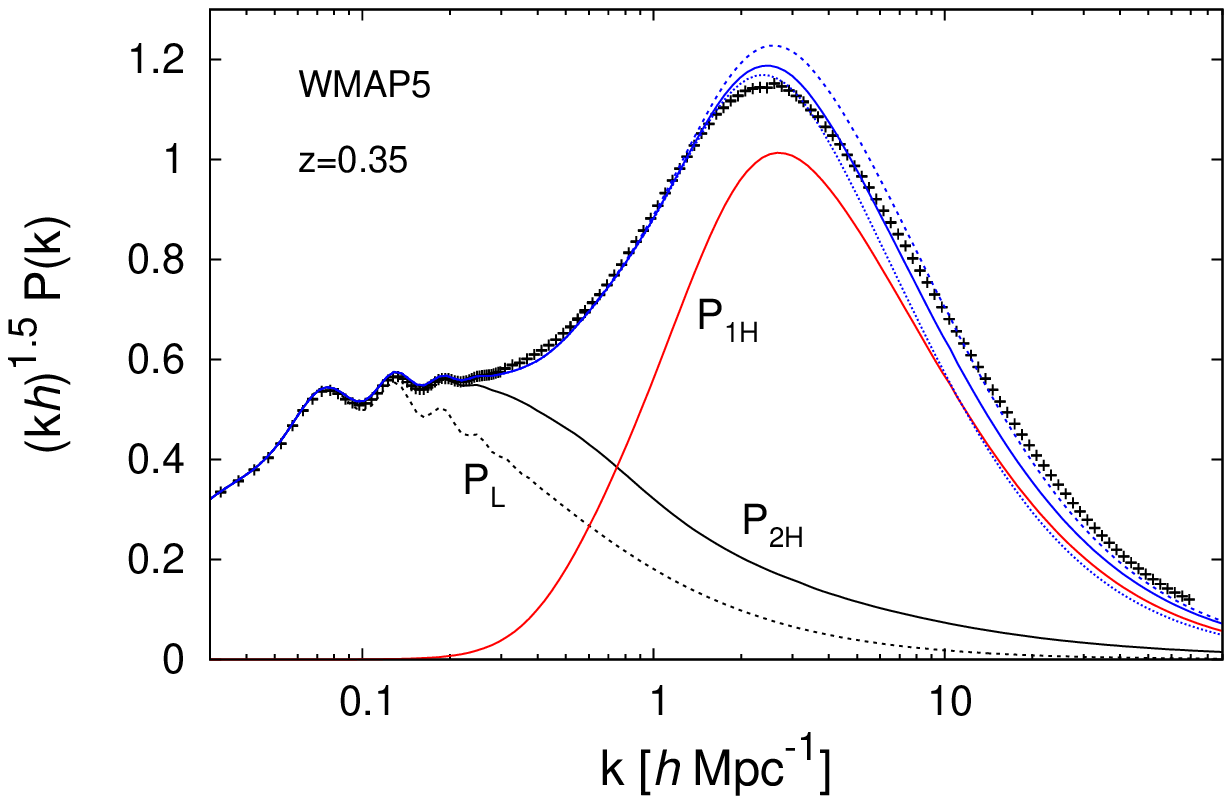}}
\epsfxsize=5.65 cm \epsfysize=6.5 cm {\epsfbox{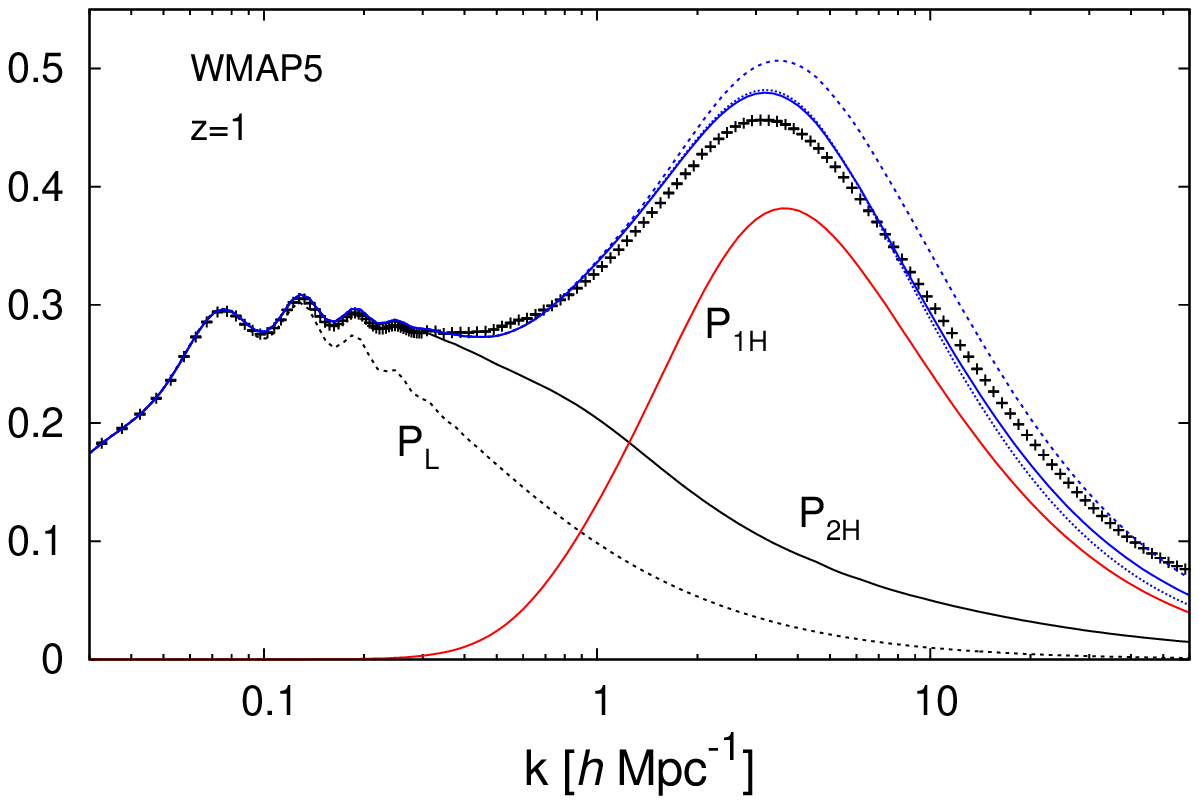}}
\epsfxsize=5.65 cm \epsfysize=6.5 cm {\epsfbox{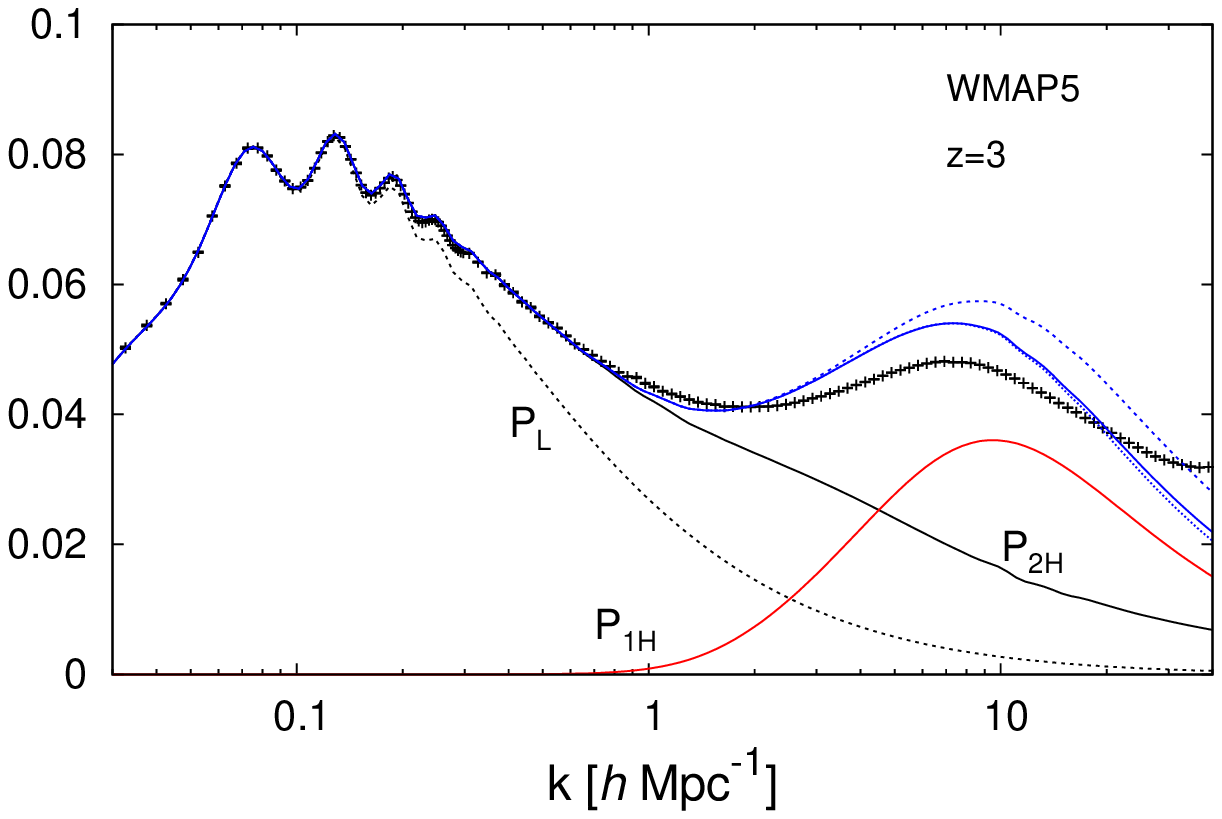}}\\
\epsfxsize=6.35 cm \epsfysize=6.5 cm {\epsfbox{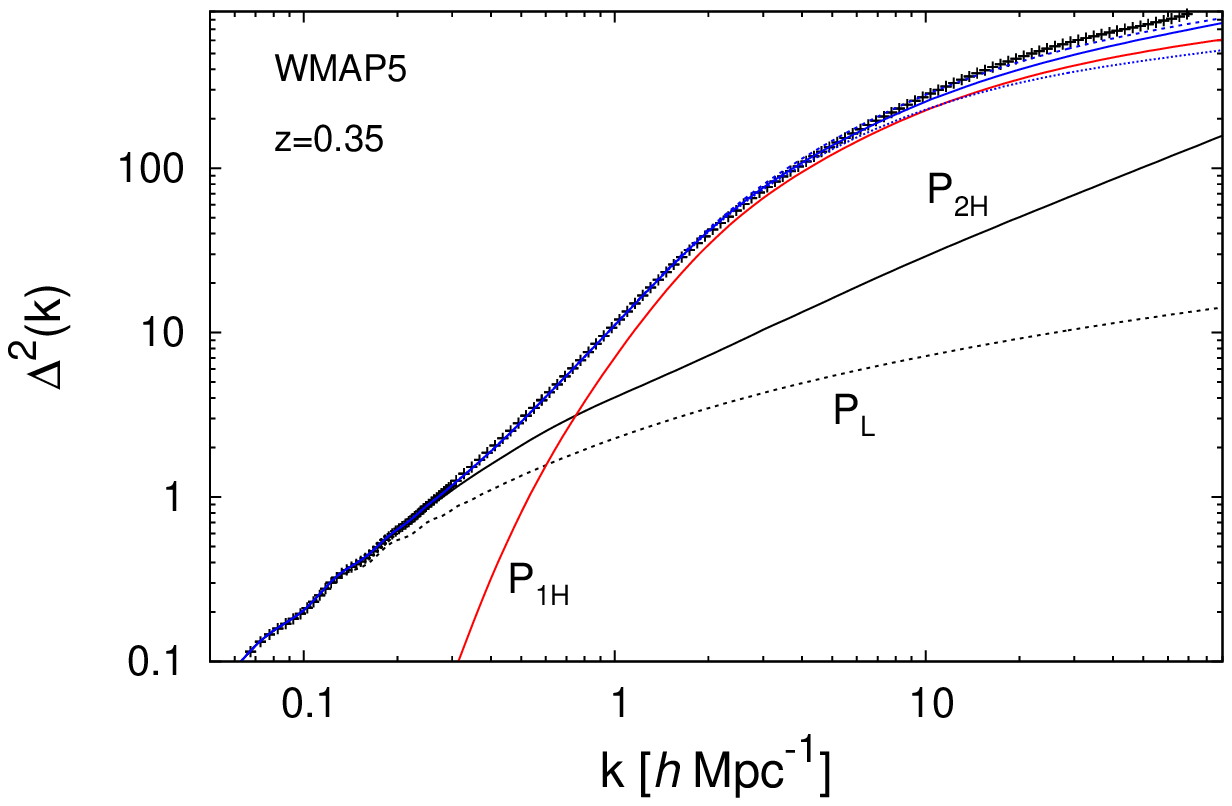}}
\epsfxsize=5.65 cm \epsfysize=6.5 cm {\epsfbox{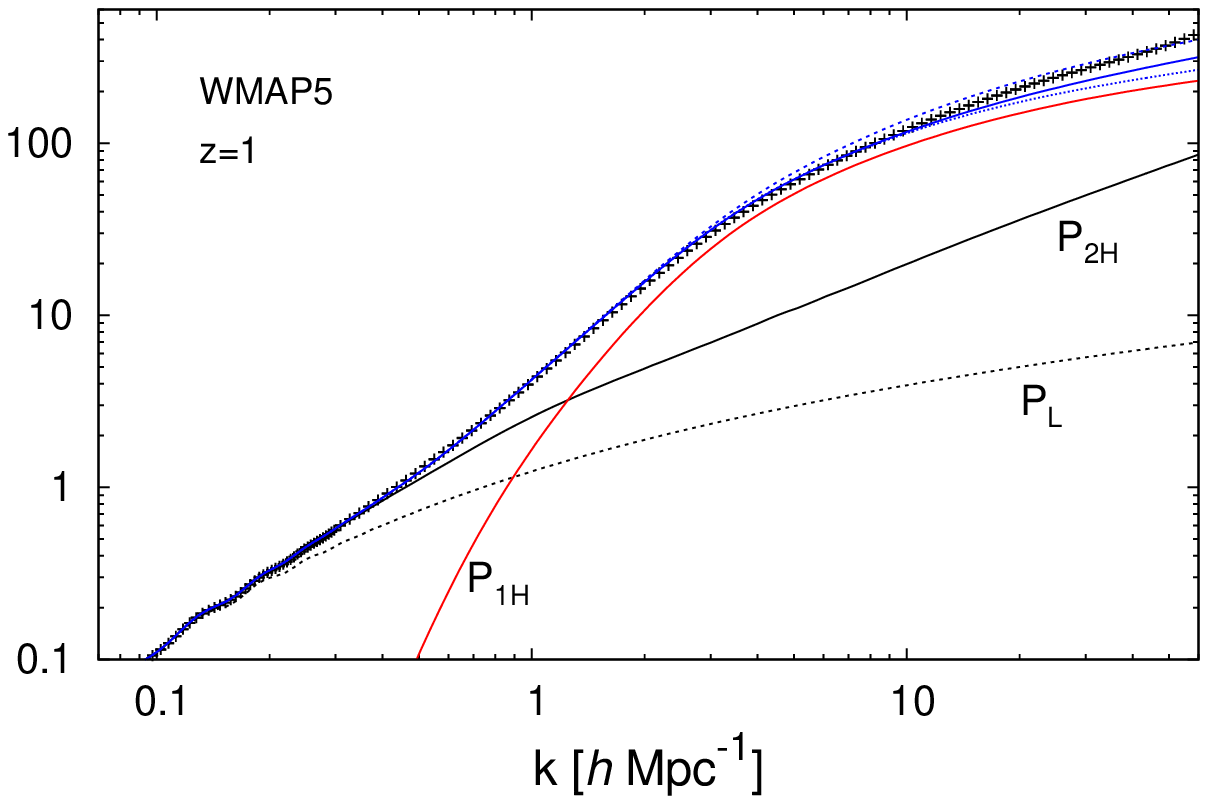}}
\epsfxsize=5.65 cm \epsfysize=6.5 cm {\epsfbox{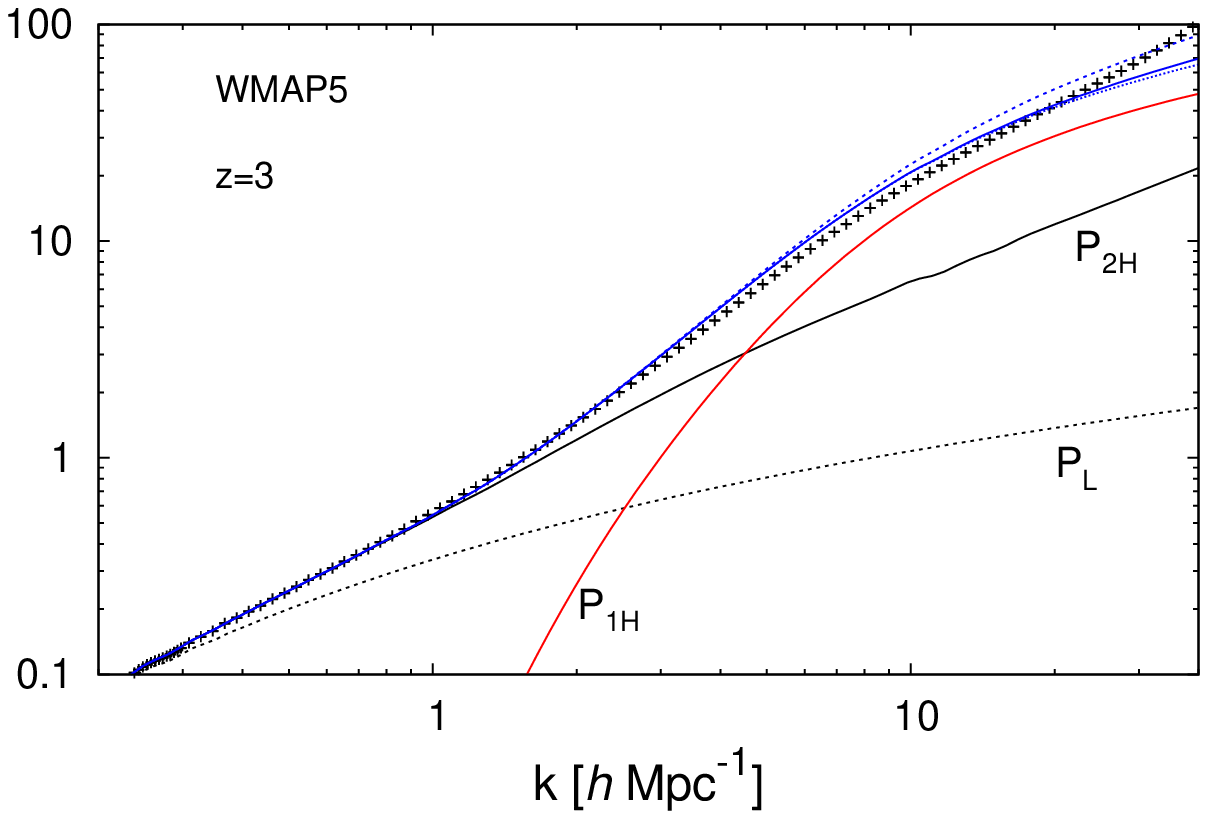}}
\end{center}
\caption{{\it Upper row:} matter density power spectrum $P(k)$ at $z=0.35$, $1$, and 
$3$, from left to right, multiplied by a factor $(k h)^{1.5}$ as in Fig.~\ref{fig-Pk_sc_z0.35}.
{\it Lower row:} logarithmic matter power spectrum, $\Delta^2(k)=4\pi k^3 P(k)$.
We show the linear power spectrum, the two-halo and one-halo contributions
$P_{\rm 2H}$ (\ref{Pk-2H-1}) and $P_{\rm 1H}$ (\ref{Pk-1H}), the full nonlinear power
spectrum (\ref{Pk-halos}), and the results from numerical simulations.
For the full nonlinear power spectrum (\ref{Pk-halos}), in addition to the result obtained
with the mass-concentration relation (\ref{c-def}) (solid line) we also show the results
obtained using the mass-concentration relations from \cite{Duffy2008} and
\cite{Bhattacharya2013} (dashed lines).}
\label{fig-Pk_Halo}
\end{figure*}

We show in Fig.~\ref{fig-Pk_Halo} the final power spectrum obtained by our model,
combining the cosmic web power spectrum obtained in Sec.~\ref{large-scale-power}
with the halo model.
We clearly see how the two-halo and one-halo contributions are dominant on large and
small scales, respectively.
The transition takes place around $k \sim 1 h$Mpc$^{-1}$ at low redshift, in a gradual
manner.
In contrast with previous studies \cite{Valageas2011d,Valageas2011e}, we no longer
underestimate the power spectrum on these transition scales, and we obtain a good
agreement up to the highly nonlinear regime. 
This is because our two-halo contribution is no longer based on a perturbative power
spectrum that goes back to the linear power at high $k$ but on the ``cosmic web''
power spectrum of Sec.~\ref{large-scale-power}, which shows more power on nonlinear
scales because it takes into account intermediate-scale structures such as pancakes.
This gives a more realistic basis that appears indeed to provide a better match to the
N-body results.

Although our simple modeling takes into account the formation of pancakes, 
as described in Sec.~\ref{large-scale-power}, it does not include filaments.
Indeed, this would require going beyond the linear regime in more than one dimension,
to model the trapping of particles within filaments. Then, we may expect an
underestimation of the power spectrum on transition scales. We do not try to
remedy for this effect in this paper, and Fig.~\ref{fig-Pk_Halo} shows that the impact
on our results is not very large. In fact, it is likely that this loss of power is hidden
within the limited accuracy of our model (due, for instance, to the ambiguities associated
with the splitting over one-halo and two-halo terms, while the actual cosmic web
is clearly more complex, as pancakes, filaments, and virialized objects are connected with
each other without sharp symmetric boundaries.).

The upper row in Fig.~\ref{fig-Pk_Halo} clearly shows how the slope of the power spectrum
evolves with redshift. For CDM power spectra, the transition scale shifts to higher $k$ at
higher redshift, which leads to a ``redder'' power spectrum (the local slope $n$ of the
linear power spectrum decreases and typically shifts from $-1.5$ to $-2.5$).
In turn, this yields a two-halo contribution that becomes larger with respect to the one-halo
contribution on highly nonlinear scales. In retrospect, this explains why the underestimation
of the power spectrum on transition scales was more severe at higher redshift in
\cite{Valageas2011d,Valageas2011e}.
Although using the ``cosmic web'' power spectrum of Sec.~\ref{large-scale-power}
is a significant improvement over these previous works, our two-halo contribution is
unlikely to be very accurate at high $k$ and this is probably one of the reasons for the
discrepancy found in the rightmost panels of Fig.~\ref{fig-Pk_Halo} at
$k > 3 h$Mpc$^{-1}$. On these small scales, we may overestimate the two-halo
contribution.

Another source of inaccuracy in this regime is the uncertainty of the halo-profile parameters
themselves, in particular the mass-concentration relation $c(M)$.
We show in Fig.~\ref{fig-Pk_Halo} the results obtained by using for $c(M)$ the fits to
N-body simulations from \cite{Duffy2008} and \cite{Bhattacharya2013}, as well as the
formula
\beq
c(M) = 11 \left(\frac{M}{2 \times 10^{12} h^{-1} M_{\odot}}\right)^{-0.1}
(1+z)^{-1.5} ,
\label{c-def}
\eeq
for $0.35 \leq z \leq 3$, where halos are defined by a density contrast of 200 with respect to
the mean density of the Universe.
This gives a power spectrum that is in between those obtained with the fits from 
\cite{Duffy2008} and \cite{Bhattacharya2013} and gives a slightly better match
to our simulations at high $k$. In principle, the relation $c(M)$ used for the power
spectrum is expected to be slightly different from the one measured in simulations because
of its finite scatter, as inclusion in the matter power spectrum is associated with an implicit
weight that may differ from the averaging procedure used to measure the relation $c(M)$
in the simulations. In any case, we can see in Fig.~\ref{fig-Pk_Halo} that the discrepancy
between our model and the power spectrum measured in the N-body simulation is of the
same order as the difference between the predictions that use either \cite{Duffy2008} or 
\cite{Bhattacharya2013} for $c(M)$.
This means that the power spectrum at these high wave numbers still shows an uncertainty
on the order of $10\%$, whether it is computed through the halo model or directly from
the N-body simulations.
In particular, shot noise certainly explains part of the rise of the power spectrum measured
in the N-body simulations at $z=3$ for $k>20 h$Mpc$^{-1}$ above our predictions
(lower-right panel in Fig.~\ref{fig-Pk_Halo}).
This is clearly seen from the comparison with the panels in the lower row obtained at lower
redshift, which probe deeper into the nonlinear regime before being affected by shot noise,
where we can see that the logarithmic power spectrum follows the shape predicted from the
halo model. In particular, at $z=0.35$ we clearly see the slowing down of the growth of
$\Delta^2(k)=4\pi k^3 P(k)$ in the highly nonlinear regime, and we would expect a
similar behavior at $z=3$. (Actually, we would expect a slightly faster slowdown because
the local slope $n$ of the linear power spectrum is redder.)
Therefore, in this regime it seems that semianalytical approaches 
like ours, based on the halo model, are competitive with direct N-body 
simulations.
(At very high $k$, the semianalytic approaches are expected
to remain reasonable, because they are based on a physically reasonable 
ansatz and/or assumption, whereas the direct results from simulations 
suffer from shot noise.)

\begin{figure}
\begin{center}
\epsfxsize=8.5 cm \epsfysize=6.5 cm {\epsfbox{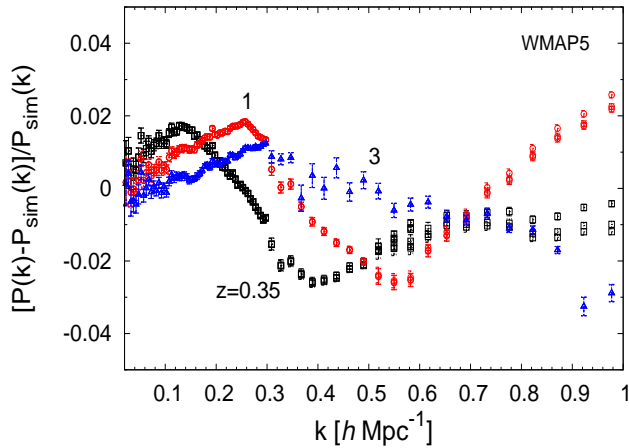}}
\end{center}
\caption{Relative deviation between the density power spectrum predicted by our model
and the result from numerical simulations, at $z=0.35$ (black squares), $1$ (red circles),
and $3$ (blue triangles).
We show the results (which almost coincide) obtained using the mass-concentration
relation (\ref{c-def}) and the ones from \cite{Duffy2008} and \cite{Bhattacharya2013}.}
\label{fig-dPk}
\end{figure}

\begin{figure}
\begin{center}
\epsfxsize=8.5 cm \epsfysize=6.5 cm {\epsfbox{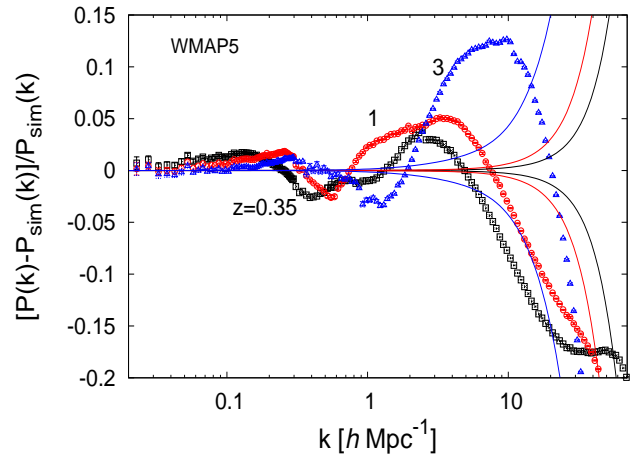}}
\end{center}
\caption{Relative deviation between the density power spectrum predicted by our model
and the result from numerical simulations, at $z=0.35$ (squares), $1$ (circles), and $3$
(triangles).
The solid lines are estimates of the effect of the shot noise in numerical simulations
(smaller impact at lower redshift).}
\label{fig-dDk}
\end{figure}

We show in Figs.~\ref{fig-dPk} and \ref{fig-dDk} the relative deviation between the power 
spectrum predicted by our model and the N-body measurements.
We obtain an accuracy of about $2\%$ up to $k \sim 0.3 h$Mpc$^{-1}$, and
$5\%$ up to $k \sim 3 h$Mpc$^{-1}$, for $z \geq 0.35$.
The small underestimation of the power spectrum on the transition scales
($k \sim 0.5 h$Mpc$^{-1}$ at $z \leq $1) may be due to the fact that filaments are not
explicitly included in our model of the cosmic web.
The accuracy degrades rapidly in the highly nonlinear regime, because of the uncertainty
of the halo model and of the N-body simulations themselves (in particular because of shot
noise at very high $k$).
Fortunately, as shown in Fig.~\ref{fig-dPk}, the uncertainties of the halo model
(i.e., the parameters of halo profiles) do not contaminate the predictions for the power
spectrum on large scales, $k< 1 h$Mpc$^{-1}$. Therefore, these large scales remain a
robust probe of cosmology.
This is also one interest of such analytical approaches that are complementary to numerical
simulations: they allow us to estimate the impact of different processes on the final
power spectrum and to estimate the range of wave numbers that are not affected by
small-scale uncertainties and can be safely used to constrain cosmology up to a good
accuracy.

\subsubsection{Two-point correlation function}
\label{correlation}

\begin{figure*}
\begin{center}
\epsfxsize=6.35 cm \epsfysize=6.5 cm {\epsfbox{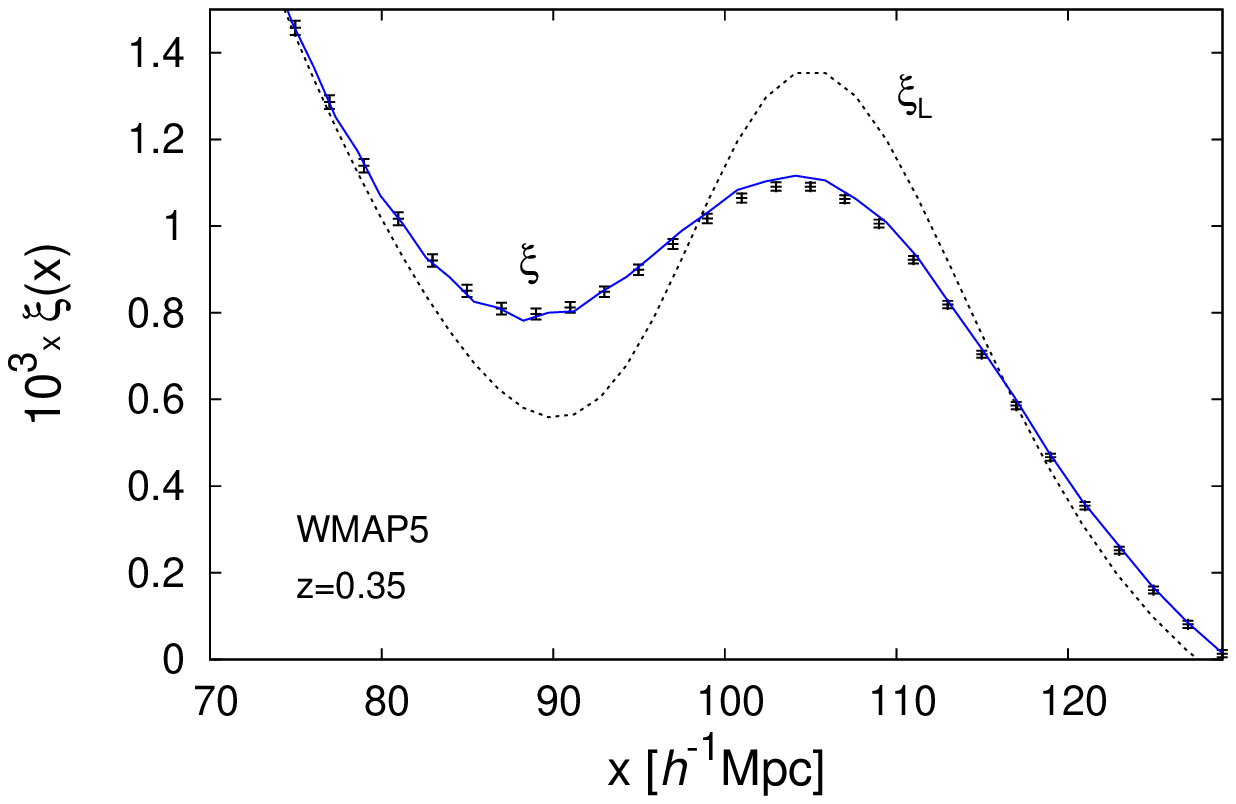}}
\epsfxsize=5.65 cm \epsfysize=6.5 cm {\epsfbox{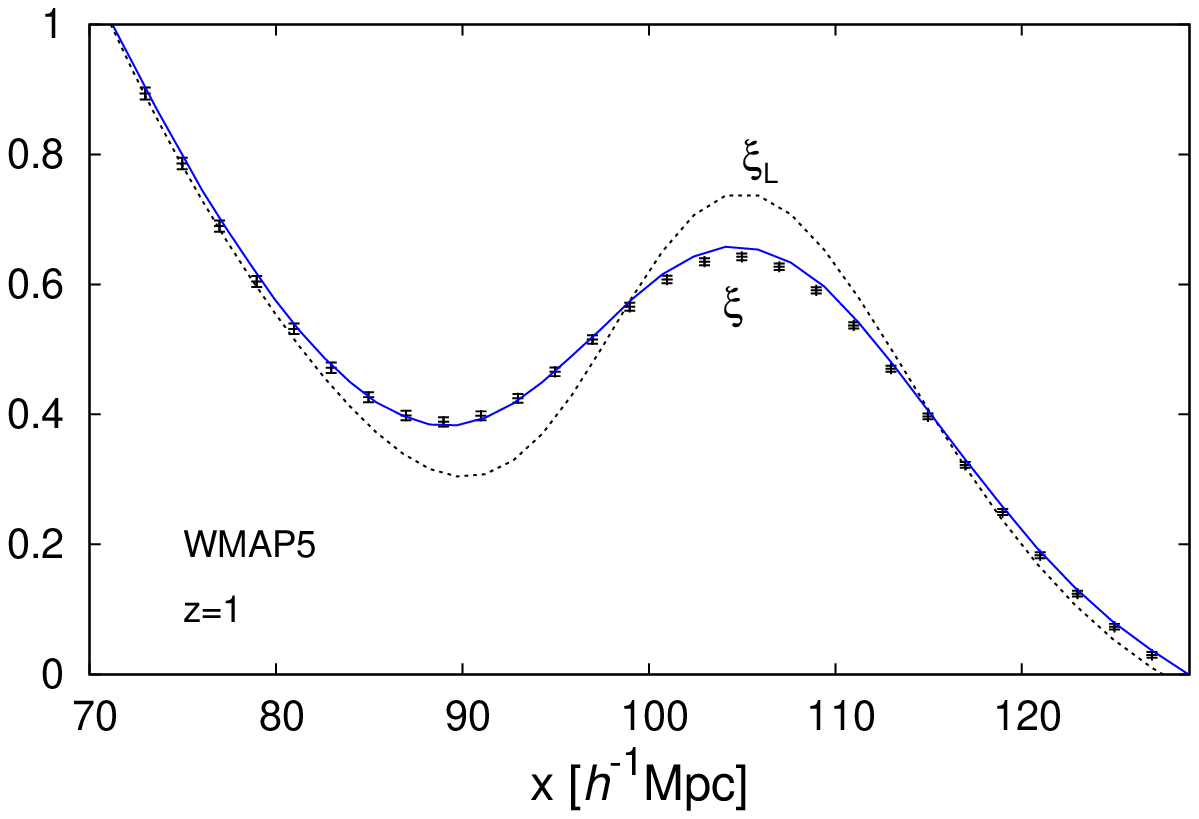}}
\epsfxsize=5.65 cm \epsfysize=6.5 cm {\epsfbox{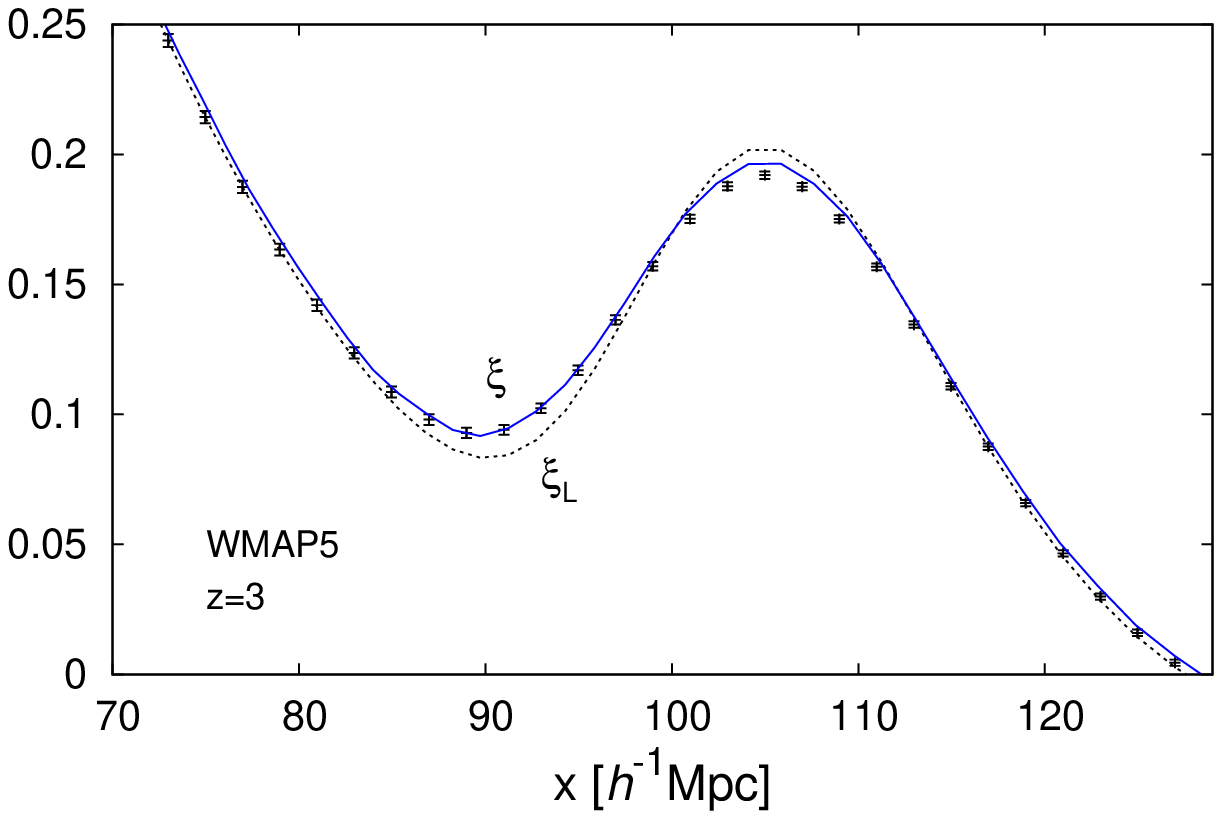}}\\
\epsfxsize=6.35 cm \epsfysize=6.5 cm {\epsfbox{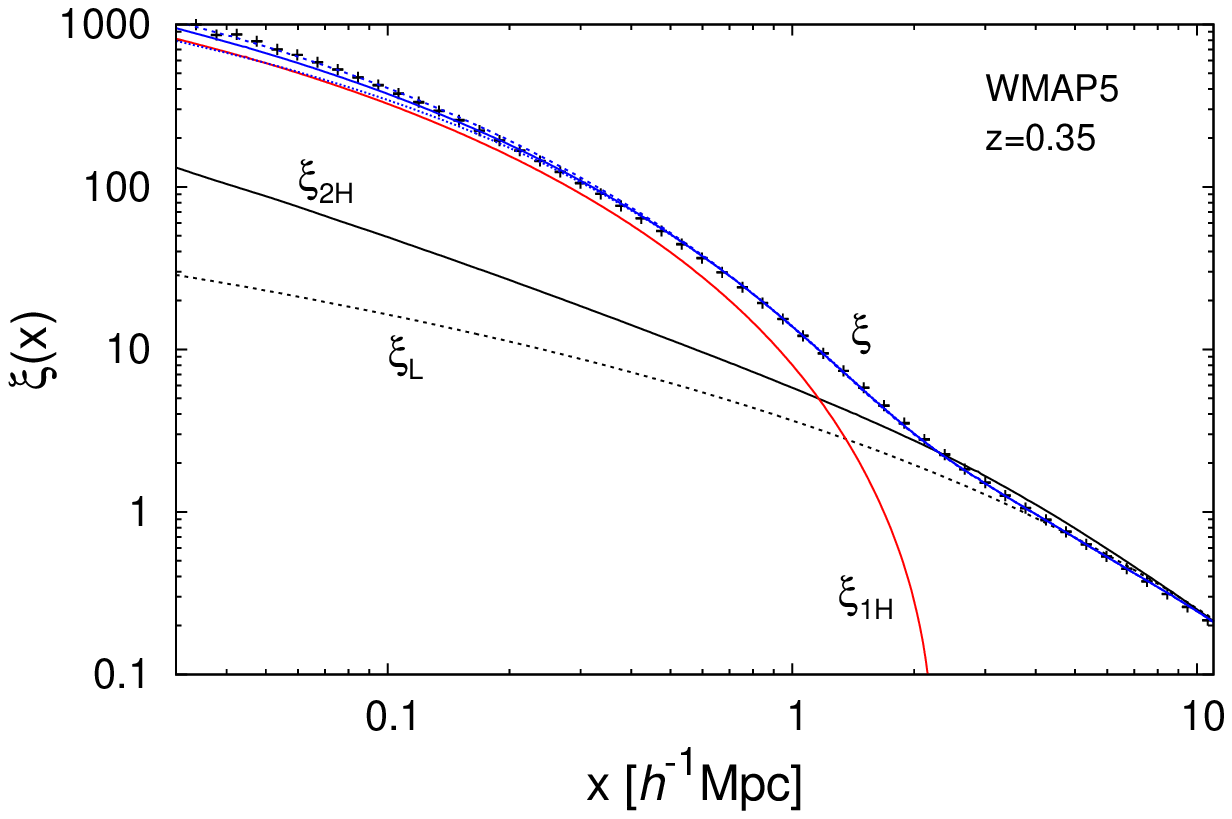}}
\epsfxsize=5.65 cm \epsfysize=6.5 cm {\epsfbox{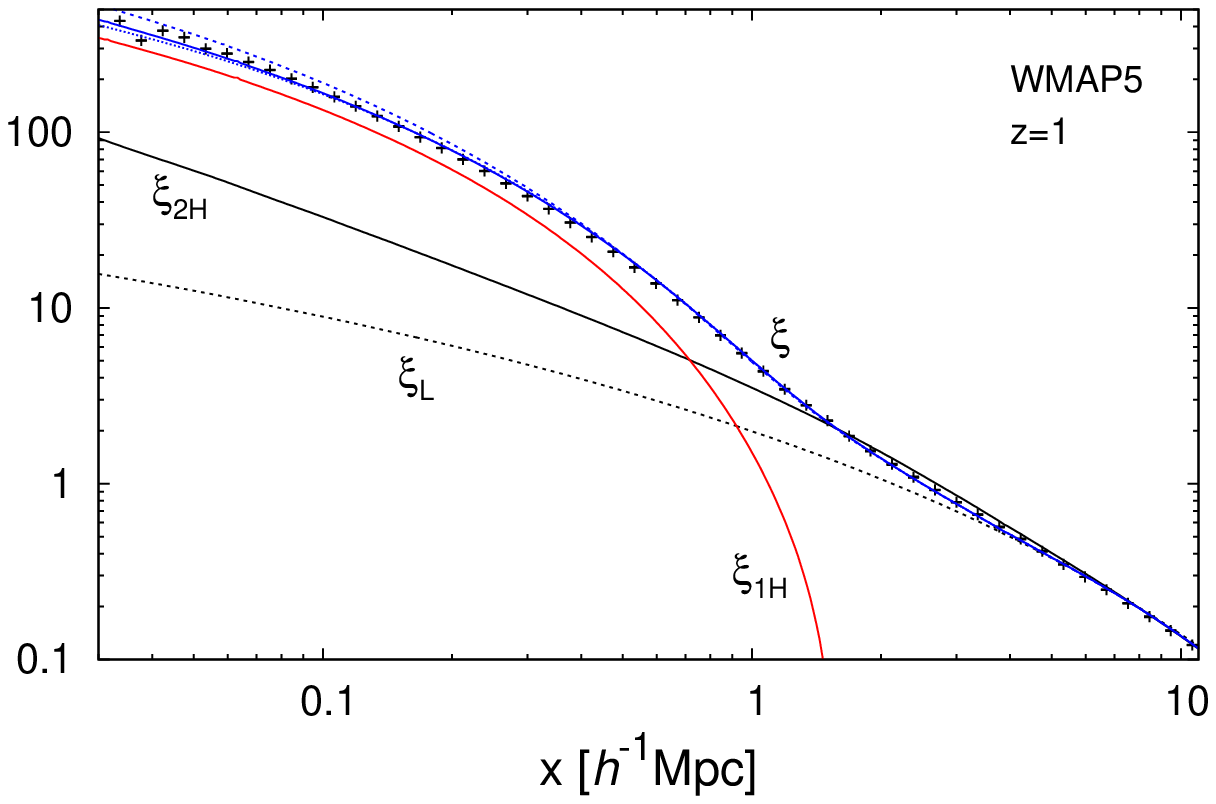}}
\epsfxsize=5.65 cm \epsfysize=6.5 cm {\epsfbox{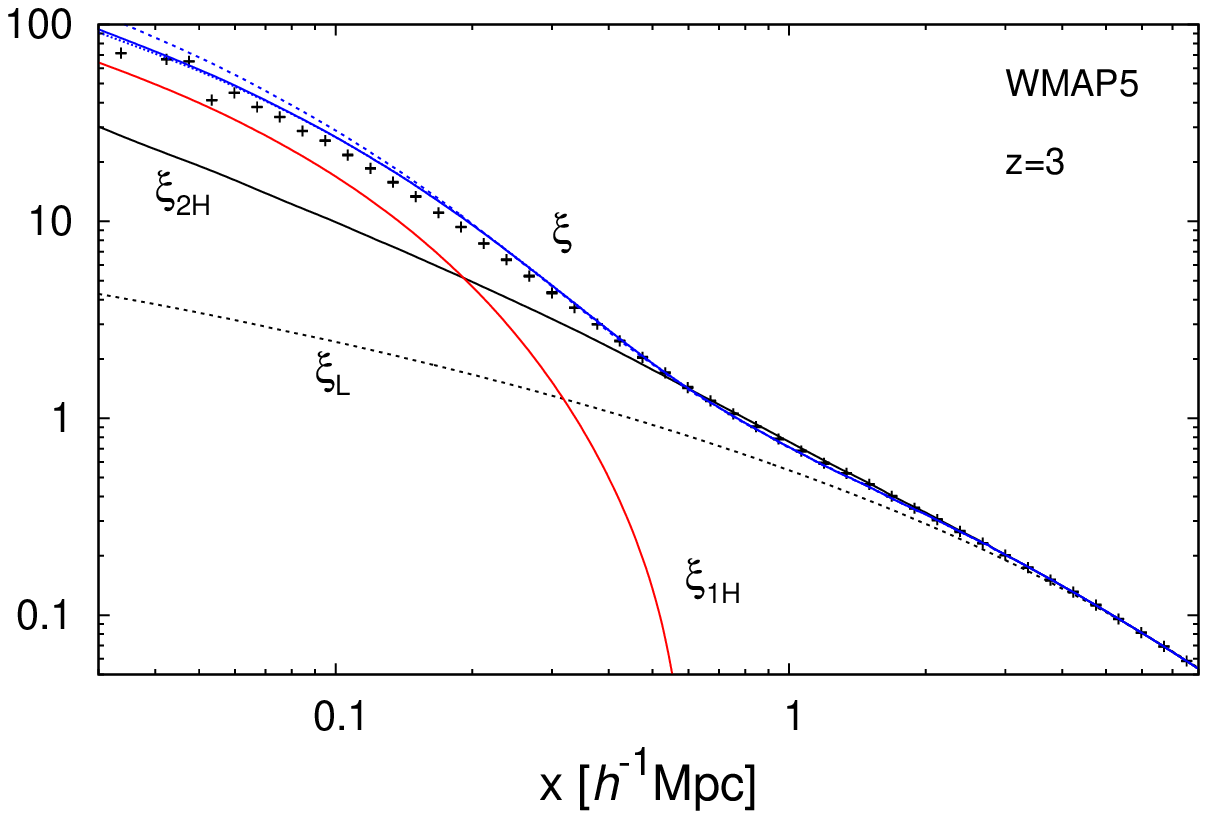}}
\end{center}
\caption{{\it Upper row:} matter density correlation function $\xi(x)$ at $z=0.35$, $1$, and $3$,
from left to right, multiplied by a factor $10^3$, on BAO scales.
{\it Lower row:} correlation function $\xi(x)$ on smaller scales.
We show the linear correlation $\xi_L$, the two-halo and one-halo contributions
$\xi_{\rm 2H}$ and $\xi_{\rm 1H}$ from Eqs.(\ref{Pk-2H-1}) and (\ref{Pk-1H}), and the full
nonlinear correlation $\xi$ from Eq.(\ref{Pk-halos}) (using the mass-concentration relation
(\ref{c-def}) (solid line) and the fits from \cite{Duffy2008} and
\cite{Bhattacharya2013} (dashed lines)).
The points are the results from numerical simulations.}
\label{fig-xi}
\end{figure*}

We show in Fig.~\ref{fig-xi} our results for the matter density two-point correlation $\xi(x)$,
given by
\beq
\xi(x) = 4\pi \int_0^{\infty} \dd k \; k^2 P(k) \; \frac{\sin(k x)}{k x} .
\label{xi-def}
\eeq
As in previous works \cite{Crocce2008,Taruya2009,Valageas2011d,Valageas2011e}, 
we can see that using a 
well-behaved perturbative contribution that includes one-loop contributions provides
a good accuracy on baryon acoustic oscillation (BAO) scales.
In particular, we reproduce the well-known damping of the baryonic peak at
$\sim 105 h^{-1}$Mpc as compared with linear theory.
We can see in the lower panel the transition between the two-halo and one-halo
contributions, at $x \sim 1 h^{-1}$Mpc.
As for the power spectrum, we obtain a significant improvement over previous studies
\cite{Valageas2011d,Valageas2011e}, as we no longer underestimate the two-point
correlation on these transition scales.
In particular, we recover the shape of the two-point correlation from linear to highly
nonlinear scales.
Again, the two-halo contribution becomes significantly greater than the linear 
correlation (or the linear power spectrum in Fig.~\ref{fig-Pk_Halo}) on small scales because
of the adhesion-like continuation explained in Sec.~\ref{shell-crossing}.
This is more important at higher redshift because the local slope of the linear
power spectrum on transition scales becomes redder while our ``cosmic web''
power spectrum shows a universal $k^{-2}$ tail.  
The good match with simulations on transition scales suggests that using such a two-halo
contribution that is greater than the linear one on nonlinear scales is an
important ingredient for recovering the full nonlinear power spectrum.

On small scales, the discrepancy between our model and the numerical simulations is
again on the order of the scatter due to the uncertainty of the halo-model
mass-concentration relation. Fortunately, this only affects very small scales, where
the physics of baryons should also be included (for instance, through the halo-model
parameters) \cite{Guillet2010}.

\begin{figure}
\begin{center}
\epsfxsize=8.5 cm \epsfysize=6.5 cm {\epsfbox{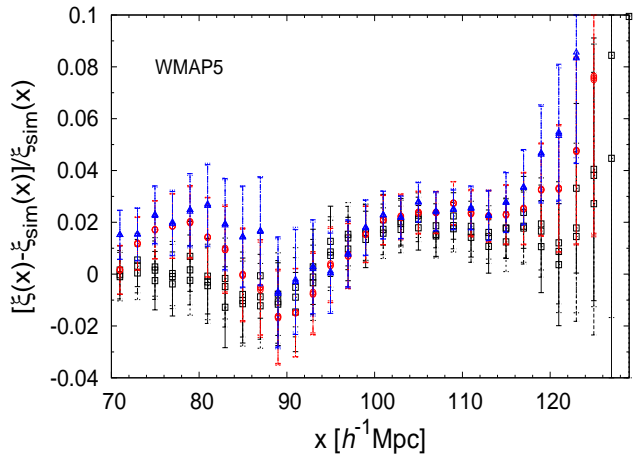}}
\end{center}
\caption{Relative deviation between the two-point correlation $\xi(x)$ predicted by our model
and the result from numerical simulations, at $z=0.35$ (squares), $1$ (circles), and $3$
(triangles).
We show the results (which almost coincide) obtained using the mass-concentration
relation (\ref{c-def}) and the ones from \cite{Duffy2008} and \cite{Bhattacharya2013}.
The error bars are the statistical error bars of the N-body simulations.}
\label{fig-dxi}
\end{figure}

\begin{figure}
\begin{center}
\epsfxsize=8.5 cm \epsfysize=6.5 cm {\epsfbox{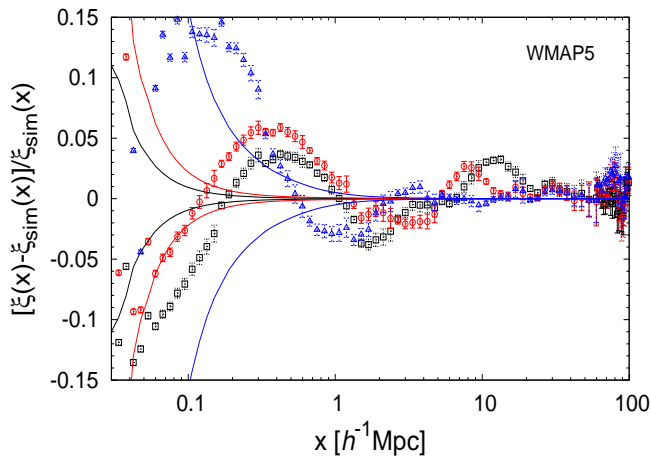}}
\end{center}
\caption{Relative deviation between the two-point correlation $\xi(x)$ predicted by our model
and the result from numerical simulations, at $z=0.35$ (squares), $1$ (circles), and $3$
(triangles).
The error bars are the statistical error bars of the N-body simulations. 
The solid lines are estimates of the effect of the shot noise in numerical simulations
(smaller impact at lower redshift).}
\label{fig-dlxi}
\end{figure}

We show in Figs.~\ref{fig-dxi} and \ref{fig-dlxi} the relative deviation between the two-point
correlation predicted by our model and the one measured in the N-body simulations.
We obtain an accuracy of about $2\%$ on the BAO scales and of about $5\%$ down to
$0.5 h^{-1}$Mpc.
Again, the accuracy degrades on smaller scales because of the uncertainties of the halo-model
parameters and of the finite resolution of the N-body simulations themselves.
On very large scales, the statistical error bars of the simulations grow because of the finite box
size and become much larger than the inaccuracy of the analytical model.
In fact, it seems that we predict more power than is measured in the simulations by about
$2\%$ on large scales. Part of this offset may be due to a lack of large-scale power in the 
N-body simulations because of their finite size. Indeed, we obtain a better agreement
on intermediate scales, $20 < x < 60 h^{-1}$Mpc.
Moreover, the two-point correlation function changes sign and vanishes at
$x_0 \sim 130 h^{-1}$Mpc, so that small absolute deviations are amplified after we take the
ratio to $\xi_{\rm sim}(x)$, and the relative deviation becomes infinite at $x_0$ because
of the finite accuracy of models and simulations.
Thus, as for the power spectrum, semianalytical approaches like ours appear competitive
with direct N-body simulations.

In this paper, we have not considered the impact of baryon physics on the matter density
power spectrum and correlation function. Hydrodynamic simulations that include a
strong AGN feedback (to cure the overcooling problem and to reproduce X-ray data)
suggest that such effects can modify the power spectrum by $10\%$ at
$k \sim 1 h$Mpc$^{-1}$ at $z=0$ \cite{van-Daalen2011}. 
Fortunately, most of these effects may be described through modifications to the halo
model \cite{Semboloni2011a} (e.g., by distinguishing the stellar, gas, and dark matter
profiles around and within virialized halos). This may also be incorporated within our
framework.
In this respect, semianalytic approaches like ours can be used to investigate the impact
of such modifications of the halo model parameters onto integrated quantities, such
as the matter power spectrum or correlation function, and in particular how they affect
different scales.

\subsubsection{Baryon acoustic oscillations}
\label{BAO}

\begin{figure}
\begin{center}
\epsfxsize=8.5 cm \epsfysize=6.5 cm {\epsfbox{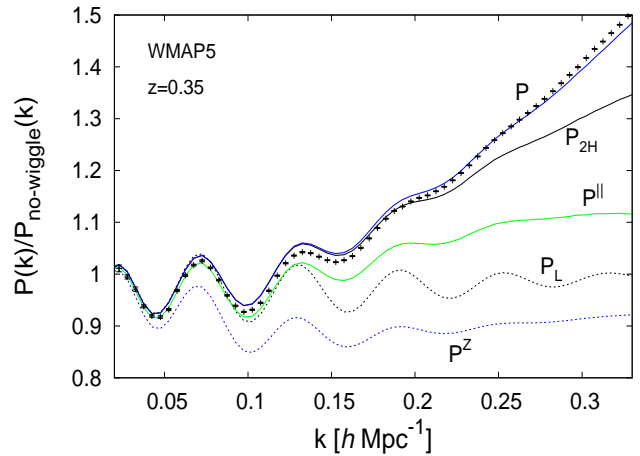}}
\end{center}
\caption{Ratio of the power spectrum $P(k)$ to a reference linear power spectrum
$P_{\rm no-wiggle}(k)$ without baryon oscillations, at $z=0.35$
We show the WMAP5 linear power $P_L$, the Zel'dovich power spectrum $P^{\rm Z}$ from
Eq.(\ref{PZ-def}), the non-Gaussian perturbative power spectrum $P^{\parallel}$ from
Eq.(\ref{Pa-1}), the two-halo contribution $P_{\rm 2H}$, and the full nonlinear power spectrum
$P(k)$.
The points are the results from numerical simulations.}
\label{fig-rPk_z0.35}
\end{figure}

\begin{figure}
\begin{center}
\epsfxsize=8.5 cm \epsfysize=6.5 cm {\epsfbox{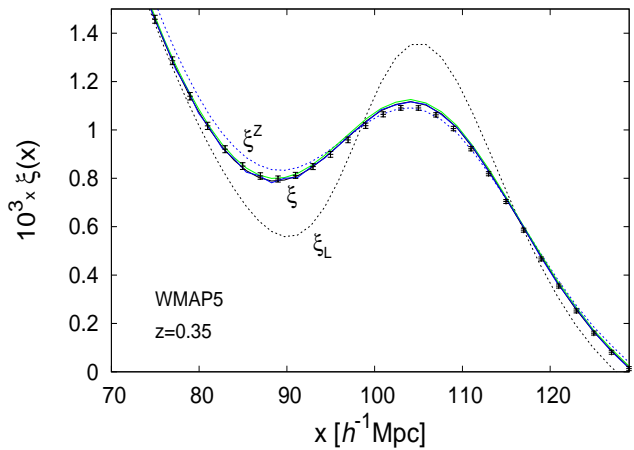}}
\end{center}
\caption{The matter density correlation function $\xi(x)$ at $z=0.35$, multiplied by a factor
$10^3$, on BAO scales. 
The points are the results from numerical simulations.
We show the linear correlation $\xi_L$ (black dotted line), the
correlation $\xi^{\rm Z}$ associated with the Zel'dovich power spectrum (blue dotted line
that is slightly above the simulation results at $x \sim 85 h^{-1}$Mpc), and the correlations
$\xi^{\parallel}$, $\xi_{\rm 2H}$, and $\xi$, which almost coincide with each other and
closely follow the simulations.}
\label{fig-xi_BAO_z0.35}
\end{figure}

In this section, we focus on BAO scales to investigate how the oscillations in $P(k)$
and the peak in $\xi(r)$ depend on the details of the model.
As usual, to emphasize the baryon acoustic oscillations, we show in Fig.~\ref{fig-rPk_z0.35}
the ratio of the nonlinear power spectrum $P(k)$ to a reference reference linear power
spectrum $P_{\rm no-wiggle}(k)$ without baryon oscillations, at $z=0.35$.
(This is the lowest redshift of this set of simulations, but it also corresponds to the range
probed by some surveys that use the baryon oscillations to probe cosmology
\cite{Eisenstein2005,Kazin2010}.)
Because of high-order mode couplings, all nonlinear results plotted in
Fig.~\ref{fig-rPk_z0.35} show a damping of high-$k$ oscillations, as compared with the
linear power spectrum, where we can still distinguish the oscillations at
$k\sim 0.25 h$Mpc$^{-1}$ and $k\sim 0.31 h$Mpc$^{-1}$.
This damping is common to most perturbative schemes that go beyond linear order,
because the contribution to the full nonlinear power spectrum of the linear part
decreases (e.g., because it is multiplied by decaying propagators \cite{Crocce2008}),
whereas the higher-order contributions mix different wave numbers (through 
high-order convolutions of the linear power) and smooth the final power.

From a physical point of view, this damping is due to the displacements of matter particles,
through large-scale bulk flows and small-scale virial motions \cite{Eisenstein2007}.
This broadens the peak in the real-space correlation function (as seen in
Fig.~\ref{fig-xi_BAO_z0.35}), over a width of a few Mpc set by the typical relative
displacements of particle pairs.
Then, this yields a damping of the oscillations in Fourier
space. As shown by Fig.~\ref{fig-rPk_z0.35}, this effect is already captured by the
Zel'dovich power spectrum $P^{\rm Z}$ (\ref{PZ-def}), which corresponds to a
Gaussian approximation for particle displacements. However, in agreement with the
analysis in Sec.~\ref{Zeldovich}, this also yields an overall damping of the power spectrum
that is too strong.
Then, the perturbative approximation $P^{\parallel}$ (\ref{Pa-1}) increases the power by
going beyond the Gaussian approximation (taking into account the exact one-loop
contribution through the skewness of the relative displacement, as in
Fig.~\ref{fig-Pkappa_z0.35}), and the
nonperturbative adhesion-like continuation of Sec.~\ref{shell-crossing}, which enters the
two-halo term $P_{\rm 2H}$ (\ref{Pk-2H-1}), provides a further increase (associated
with pancakes).
These two nonlinear corrections, which involve high-order mode couplings, do not keep
much of the initial linear oscillations, and this even further damps the relative importance
of the oscillations at high $k$. Finally, the one-halo contribution, which corresponds to the
difference $P(k)-P_{\rm 2H}(k)$, only becomes important for $k > 0.23 h$Mpc$^{-1}$
and is also very smooth.

We show the real-space correlation function associated with these different models
in Fig.~\ref{fig-xi_BAO_z0.35}. We clearly see how the damping of the oscillations
found in Fig.~\ref{fig-rPk_z0.35} corresponds to a broadening of the real-space peak,
for all nonlinear models. One striking result is that all nonlinear correlations are very close
to each other, despite the differences seen in Fig.~\ref{fig-rPk_z0.35}.
In particular, $\xi^{\parallel}$, $\xi_{\rm 2H}$, and $\xi$ cannot be distinguished in this
figure. This shows that small-scale virial motions are largely subdominant as
compared with large-scale bulk flows, with respect to the broadening of the acoustic peak.
Moreover, the Zel'dovich approximation $\xi^{\rm Z}$ already provides a remarkably good
description of the broadening, even though it shows a small but noticeable departure from
the simulations.
(Of course, these curves start to show significant deviations from each other on smaller
nonlinear scales, $x < 4 h^{-1}$Mpc.)
This shows that even though the baryon acoustic peak is significantly modified from
linear theory, all reasonable Lagrangian-space-based models are able to model this
nonlinear evolution (see also \cite{Matsubara2008,Carlson2012}). 
This also holds for the simplest one, i.e. the Zel'dovich approximation, although it shows
small departures from simulations, and higher-order models (i.e., that include one-loop 
corrections) provide a very good match.
This explains why reconstruction techniques \cite{Eisenstein2007a}, which are inspired
by the Zel'dovich approximation, perform very well.
For observational purposes, this confirms again that the baryon acoustic
peak is a very robust probe of cosmology \cite{Eisenstein2007}, and that the
real-space correlation may be more convenient than the Fourier-space power spectrum, 
because these nonlinearities may be more naturally understood as particle displacements
than resonant wave couplings \cite{Valageas2007a,Bernardeau2012}.

Redshift-space distortions also have a (smaller) impact on the acoustic peak of the correlation
function \cite{Eisenstein2007a}, but we leave the study of these effects for future works.

\subsection{Other cosmologies}
\label{Other-cosmologies}

To further test our model, we also compare our predictions with numerical simulations
for two other cosmologies.

\subsubsection{WMAP3 cosmology}
\label{WMAP3}

\begin{figure*}
\begin{center}
\epsfxsize=6.35 cm \epsfysize=6.5 cm {\epsfbox{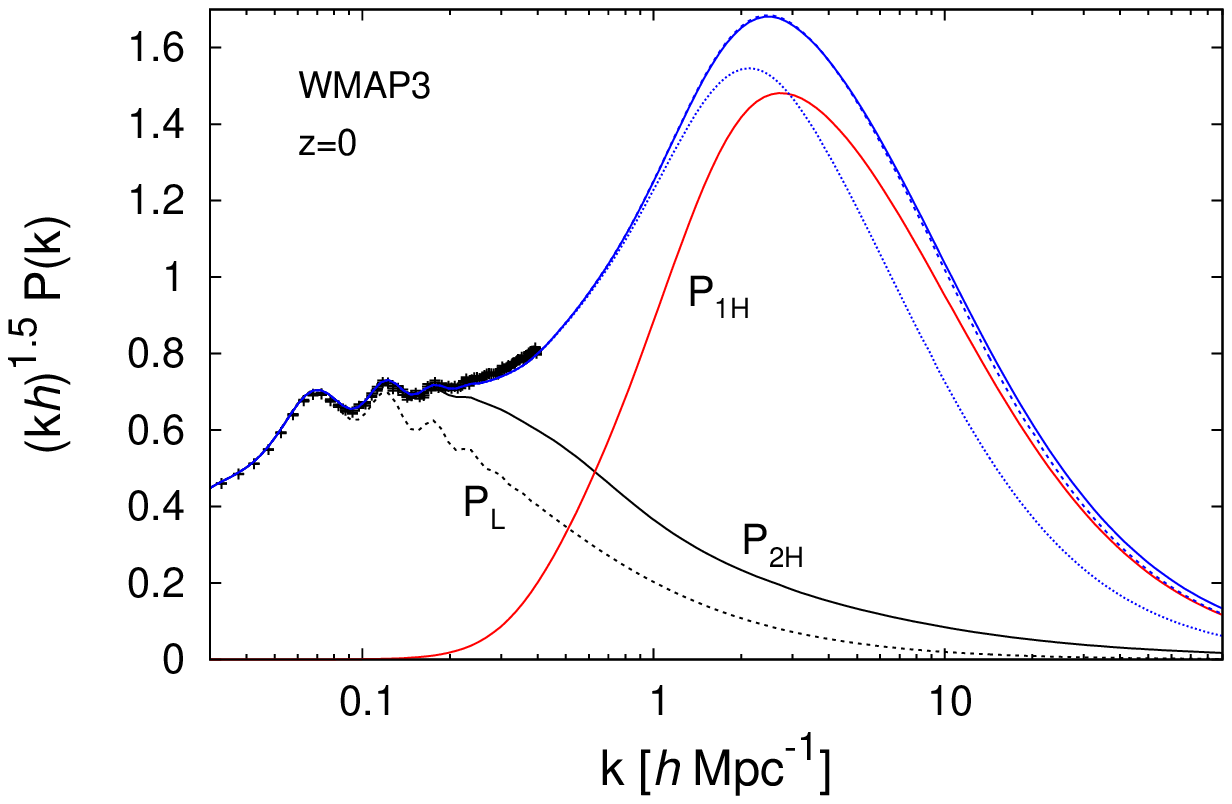}}
\epsfxsize=5.65 cm \epsfysize=6.5 cm {\epsfbox{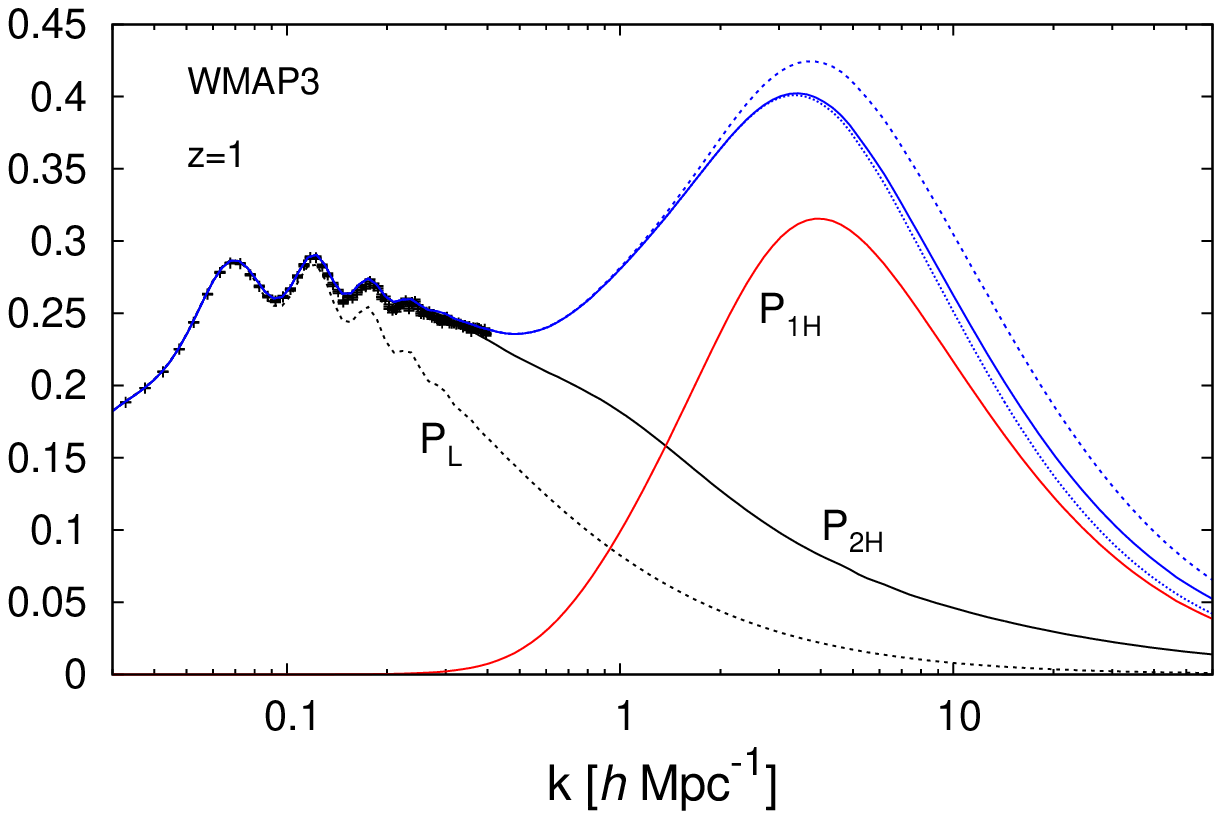}}
\epsfxsize=5.65 cm \epsfysize=6.5 cm {\epsfbox{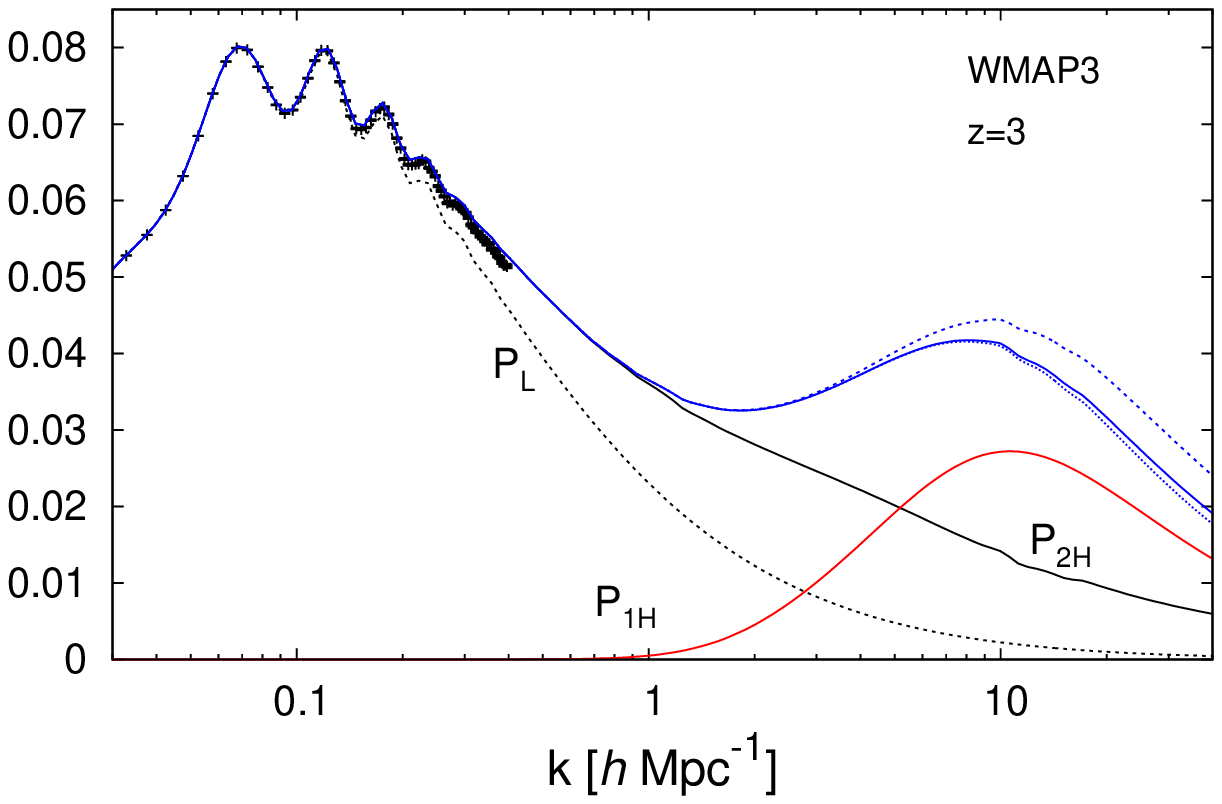}}
\end{center}
\caption{Matter density power spectrum $P(k)$ at $z=0$, $1$, and $3$,
from left to right, multiplied by a factor $(k h)^{1.5}$ as in Fig.~\ref{fig-Pk_Halo} but for
a different cosmology (WMAP3).
Again, in addition to the result obtained
with the mass-concentration relation (\ref{c-def}) (solid line) we also show the results
obtained using the mass-concentration relations from \cite{Duffy2008} and
\cite{Bhattacharya2013} (dashed lines).}
\label{fig-Pk_Halo_WMAP3}
\end{figure*}

\begin{figure}
\begin{center}
\epsfxsize=8.5 cm \epsfysize=6.5 cm {\epsfbox{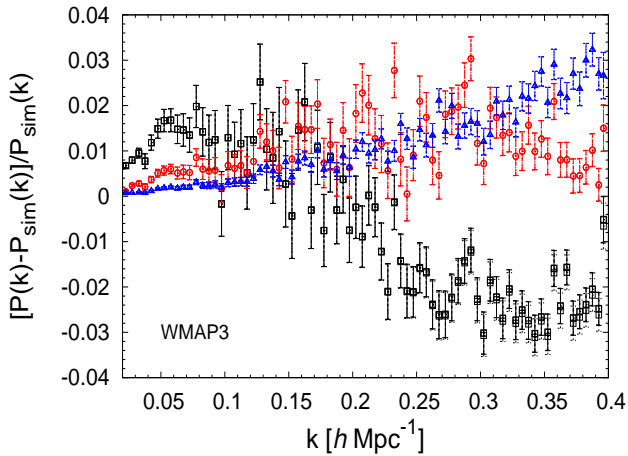}}
\end{center}
\caption{Relative deviation between the density power spectrum predicted by our model
and the result from numerical simulations, at $z=0$ (black squares), $1$ (red circles),
and $3$ (blue triangles), as in Fig.~\ref{fig-dPk} but for
a different cosmology (WMAP3).
We show the results (which almost coincide) obtained using the mass-concentration
relation (\ref{c-def}) and the ones from \cite{Duffy2008} and \cite{Bhattacharya2013}.}
\label{fig-dPk_WMAP3}
\end{figure}

We first use an older set of WMAP3 simulations, with $h=0.734$, $\Omega_{\rm m}=0.234$,
$\sigma_8=0.76$, $n_s=0.961$, and $w_{\rm de}=-1$, performed in \cite{Nishimichi2009}.
These simulations have a smaller box size $(1000\,h^{-1}$ Mpc$)$ and a lower resolution
($512^3$ particles); hence we only perform the comparison on large scales.
On the other hand, they allow us to go down to redshift $z=0$.

We show our results for the power spectrum in Figs.~\ref{fig-Pk_Halo_WMAP3} and
\ref{fig-dPk_WMAP3}.
These results are similar to those obtained in Figs.~\ref{fig-Pk_Halo} and
\ref{fig-dPk} for the WMAP5 cosmology, with larger error bars in Fig.~\ref{fig-dPk_WMAP3}
because of the smaller box size of the simulations.

\subsubsection{Quintessence cosmology}
\label{w-cosmology}

\begin{figure*}
\begin{center}
\epsfxsize=6.35 cm \epsfysize=6.5 cm {\epsfbox{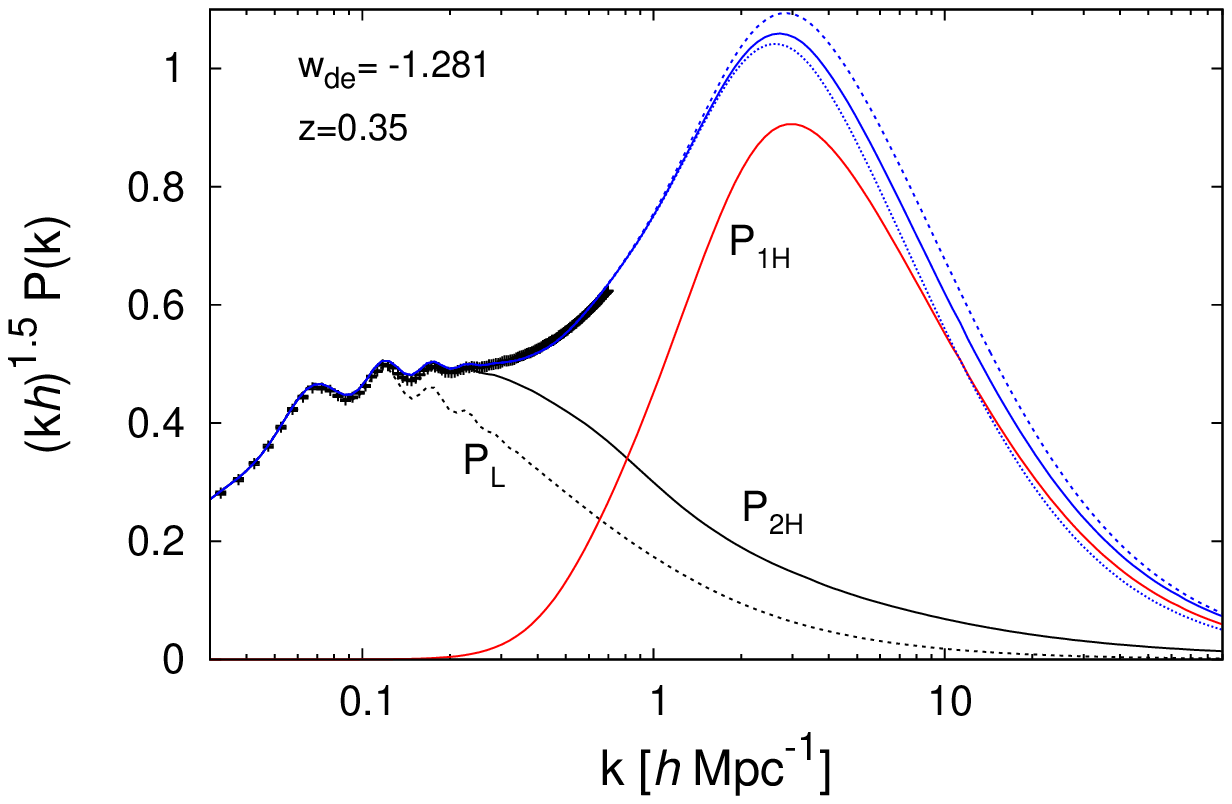}}
\epsfxsize=5.65 cm \epsfysize=6.5 cm {\epsfbox{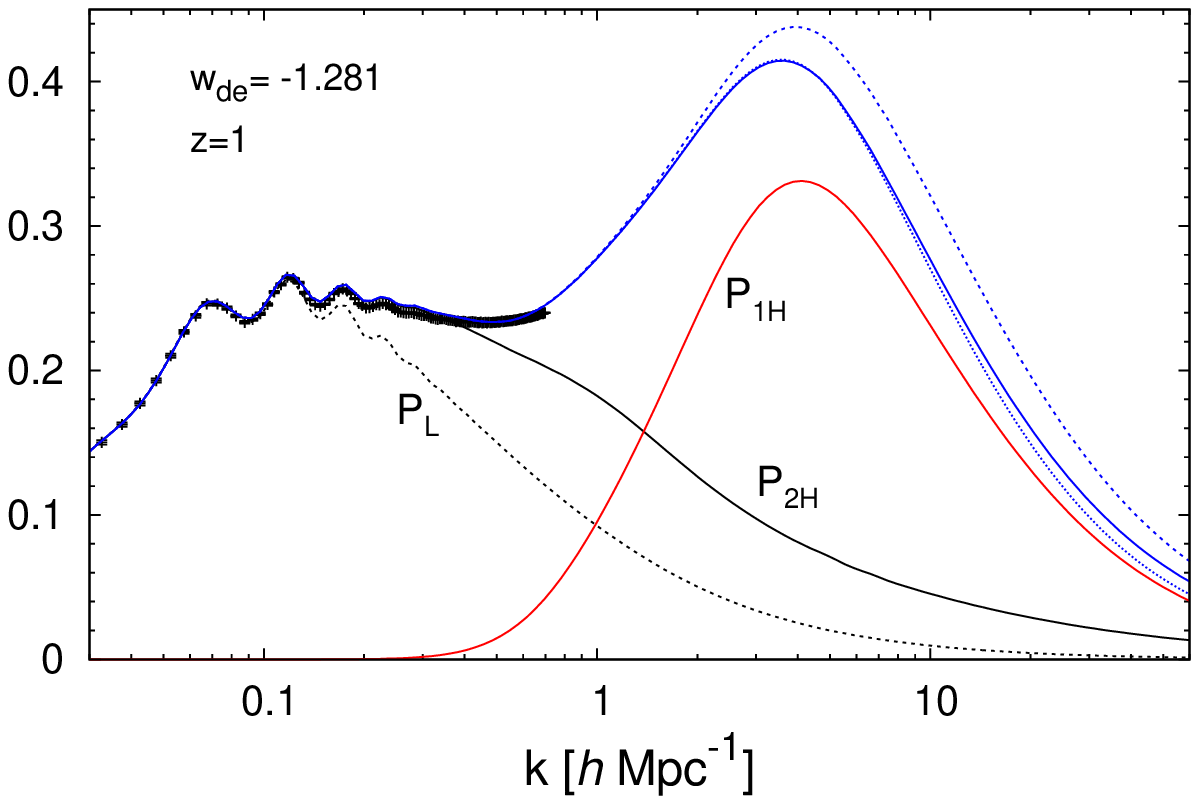}}
\epsfxsize=5.65 cm \epsfysize=6.5 cm {\epsfbox{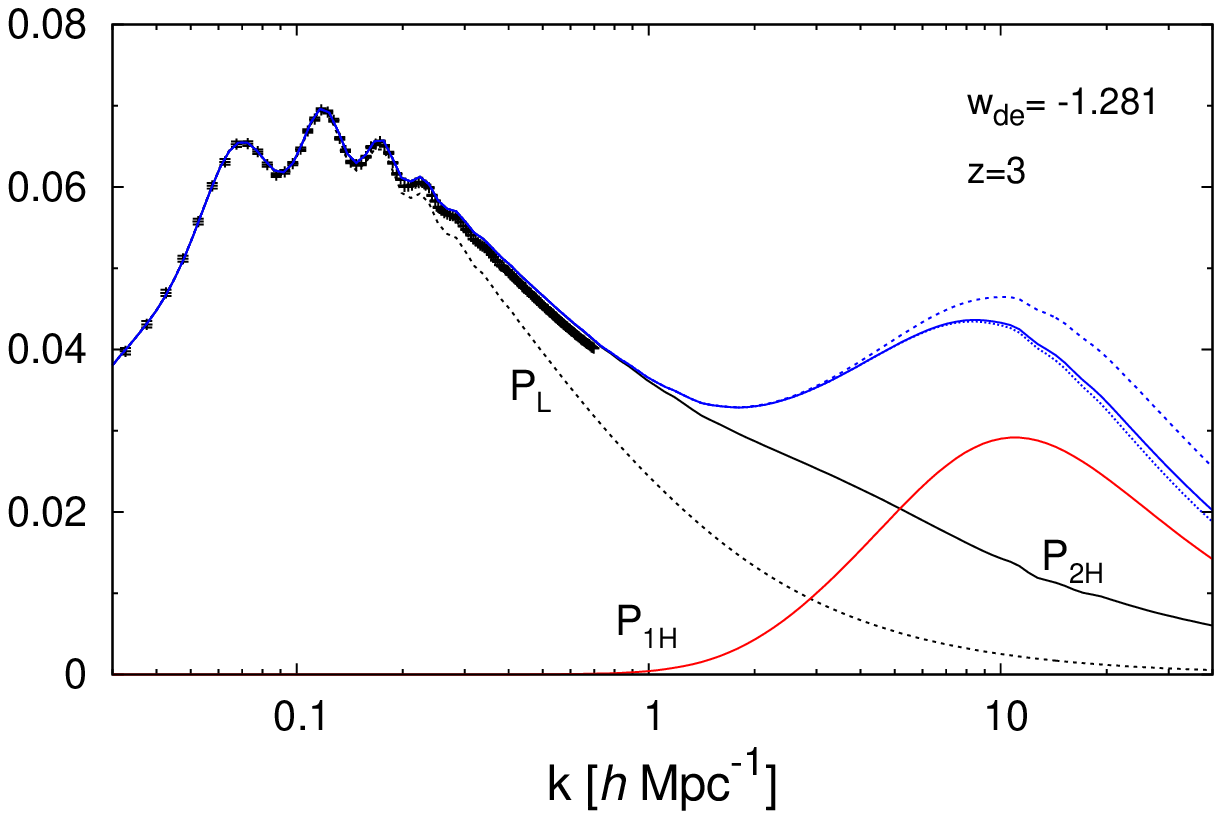}}
\end{center}
\caption{Matter density power spectrum $P(k)$ at $z=0.35$, $1$, and $3$,
from left to right, multiplied by a factor $(k h)^{1.5}$ as in Fig.~\ref{fig-Pk_Halo} but for
a different cosmology ($w_{\rm de}=-1.281$).
Again, in addition to the result obtained
with the mass-concentration relation (\ref{c-def}) (solid line) we also show the results
obtained using the mass-concentration relations from \cite{Duffy2008} and
\cite{Bhattacharya2013} (dashed lines).}
\label{fig-Pk_Halo_w}
\end{figure*}

\begin{figure}
\begin{center}
\epsfxsize=8.5 cm \epsfysize=6.5 cm {\epsfbox{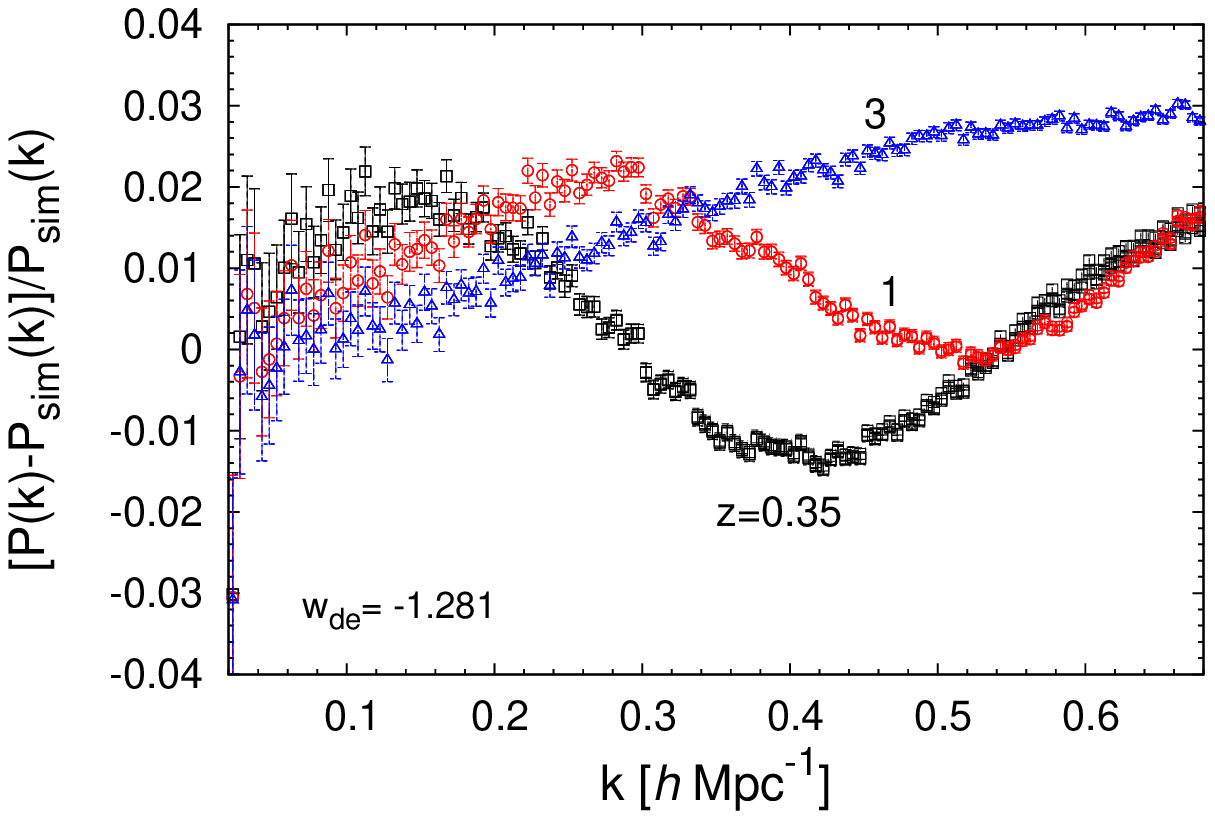}}
\end{center}
\caption{Relative deviation between the density power spectrum predicted by our model
and the result from numerical simulations, at $z=0.35$ (black squares), $1$ (red circles),
and $3$ (blue triangles), as in Fig.~\ref{fig-dPk} but for
a different cosmology ($w_{\rm de}=-1.281$).
We show the results (which almost coincide) obtained using the mass-concentration
relation (\ref{c-def}) and the ones from \cite{Duffy2008} and \cite{Bhattacharya2013}.}
\label{fig-dPk_w}
\end{figure}

\begin{figure*}
\begin{center}
\epsfxsize=6.35 cm \epsfysize=6.5 cm {\epsfbox{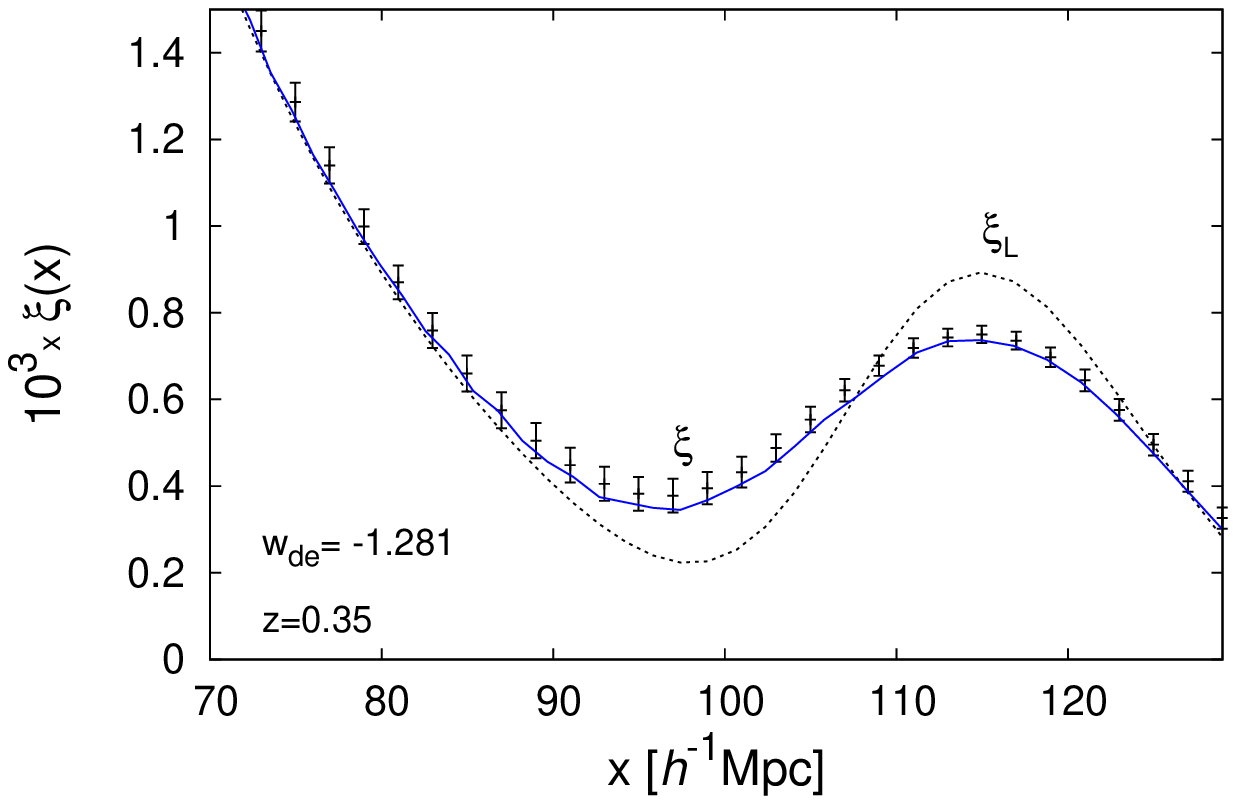}}
\epsfxsize=5.65 cm \epsfysize=6.5 cm {\epsfbox{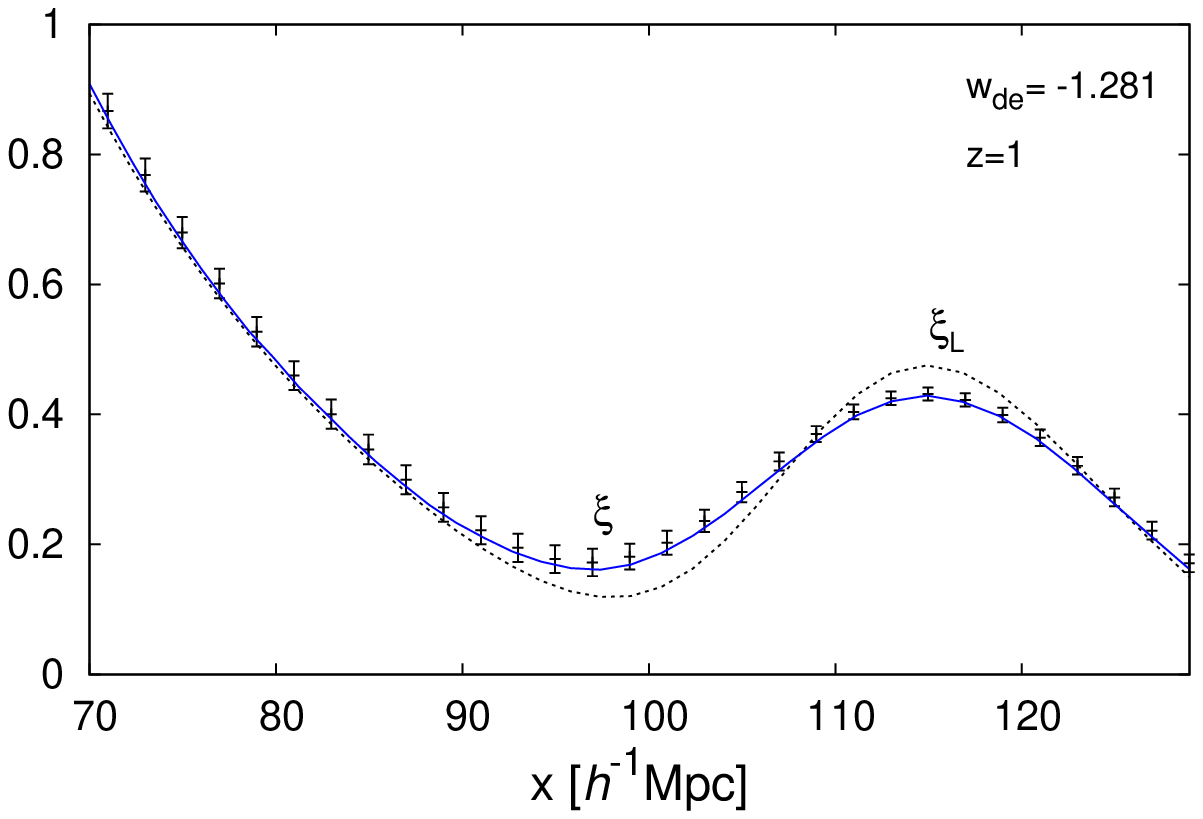}}
\epsfxsize=5.65 cm \epsfysize=6.5 cm {\epsfbox{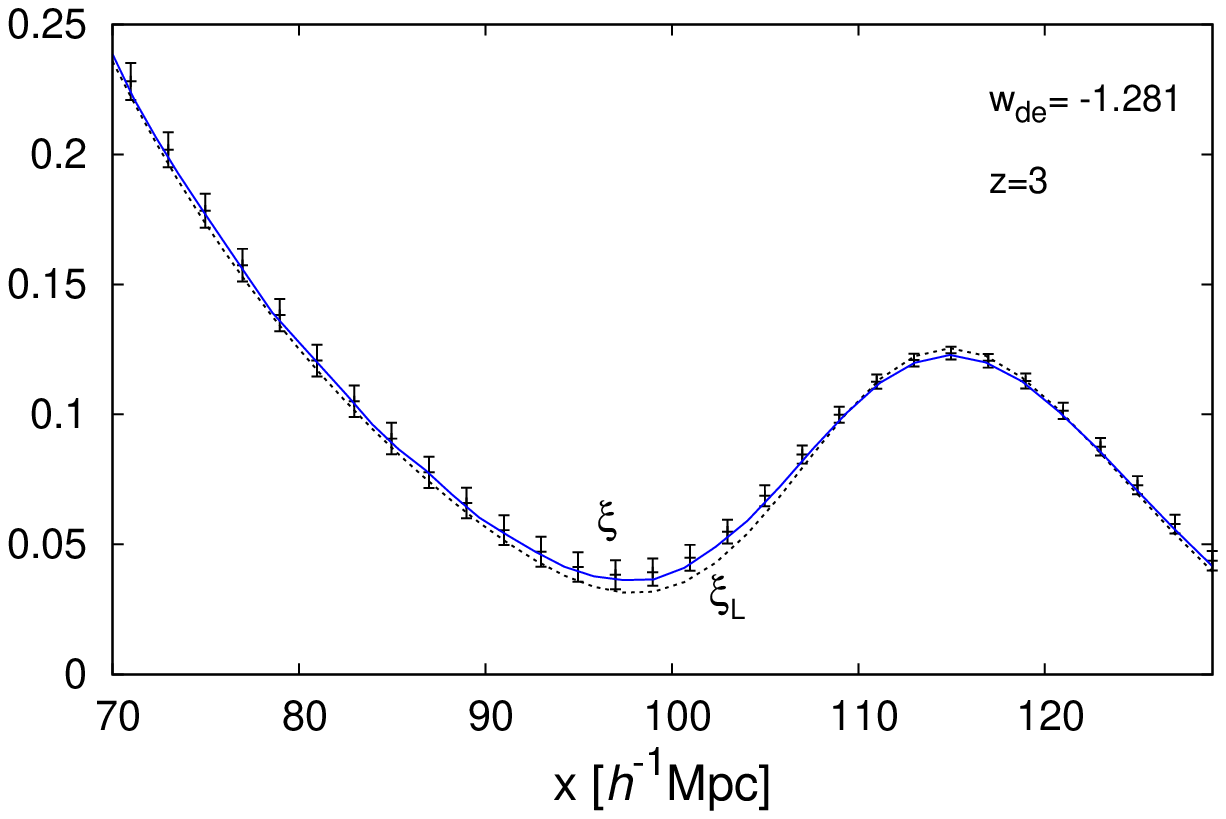}}
\end{center}
\caption{Matter density correlation function $\xi(x)$ at $z=0.35$, $1$, and $3$,
from left to right, multiplied by a factor $10^3$, as in Fig.~\ref{fig-xi_w} but for
a different cosmology ($w_{\rm de}=-1.281$).}
\label{fig-xi_w}
\end{figure*}

\begin{figure}
\begin{center}
\epsfxsize=8.5 cm \epsfysize=6.5 cm {\epsfbox{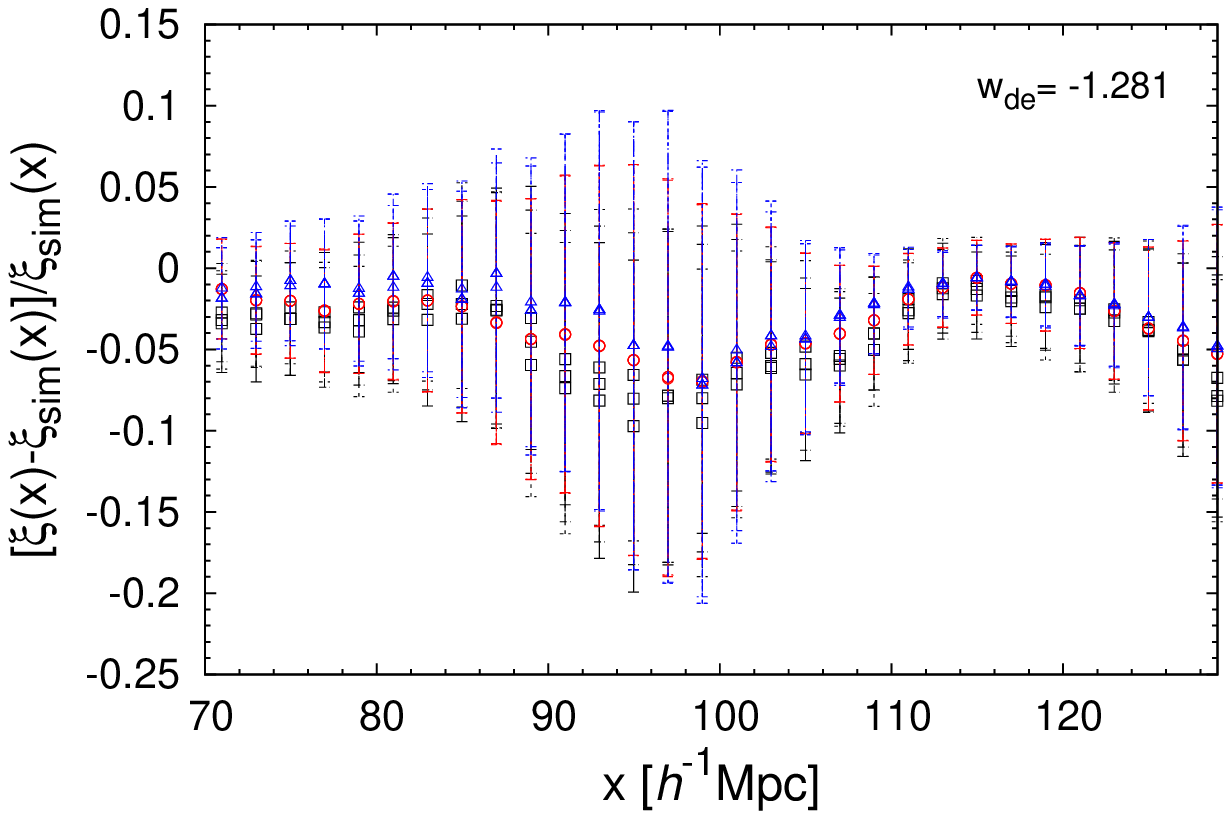}}
\end{center}
\caption{Relative deviation between the two-point correlation $\xi(x)$ predicted by our model
and the result from numerical simulations, at $z=0.35$ (squares), $1$ (circles), and $3$
(triangles), as in Fig.~\ref{fig-dxi} but for
a different cosmology ($w_{\rm de}=-1.281$).
We show the results (which almost coincide) obtained using the mass-concentration
relation (\ref{c-def}) and the ones from \cite{Duffy2008} and \cite{Bhattacharya2013}.
The error bars are the statistical error bars of the N-body simulations.}
\label{fig-dxi_w}
\end{figure}

We finally consider a less standard cosmology, where the dark energy does not behave as
a cosmological constant. Thus, we choose a dark energy equation-of-state parameter
$w_{\rm de}=-1.281$, and $h=0.7737$, $\Omega_{\rm m}=0.23638$,
$\sigma_8=0.7692$,  and $n_s=1.0177$.
We use the measurements from N-body simulations done in \cite{Taruya2012} for this
cosmological model. These simulations are done in $2048\,h^{-1}$~Mpc boxes with $1024^3$ particles,
and eight statistically independent realizations are available.

We show our results for the matter power spectrum in Figs.~\ref{fig-Pk_Halo_w} and 
\ref{fig-dPk_w}, and for the correlation function in Figs.~\ref{fig-xi_w} and \ref{fig-dxi_w}.
We again obtain similar behaviors to those found for the LCDM WMAP5 cosmology in
Sec.~\ref{WMAP5}.

This confirms that our method provides efficient predictions that can be used for a variety
of cosmologies.

\section{Conclusion}
\label{Conclusion}

We have developed in this paper a new approach to a systematic modeling of the
cosmological density field: instead of looking for explicit partial resummation schemes,
we embed
the standard perturbation theory within a realistic nonlinear ansatz.
This automatically ensures a reasonable behavior on small scales, while being consistent
with perturbative approaches up to the required order. Then, this allows us to obtain
a reasonable matching with the highly nonlinear regime.

We have shown how this can be achieved within a Lagrangian-space framework.
Because the lowest-order approximation that appears in this context is the Zel'dovich
approximation \cite{Zeldovich1970}, we first discussed the lack of power on
weakly nonlinear scales associated with this Gaussian approximation.
Then, we explained how this can be partially cured by including higher-order terms
(i.e., the skewness) while keeping a well-behaved ansatz. Because the Gaussian is the
only probability distribution with a finite set of nonzero cumulants, this requires
including nonzero cumulants at all orders and we describe a simple ansatz that provides
a well-behaved probability distribution for the longitudinal displacement field.
This provides a perturbative power spectrum that has the same qualitative properties as
the true gravitational dynamics power spectrum.
Moreover, it can be made consistent with the exact perturbative expansion up to the
required order, while keeping all these qualitative properties.
Here we only go up to order $P_L^2$, but higher orders could be taken into account
by explicitly choosing the kurtosis and higher-order cumulants of the displacement
field (at the price of a more complex ansatz for the cumulant-generating function).

Next, we argued that for a good description of intermediate scales it is important to
include some effects associated with the shell-crossing regime. In particular, we
pointed out that the small-scale behavior implied by the Zel'dovich power spectrum
and most previous Lagrangian-space schemes, associated with the escape of particles
 to infinity, is not adequate, and a more realistic picture is provided by the adhesion model.
 We presented a simplified sticking model that captures some features of this trapping of
 particles within potential wells and describes the first stages of pancake formation.
 In particular, it captures the (nonperturbative) increase and change of sign of the
 skewness of the displacement field on small scales. This gives a ``cosmic web'' power
 spectrum that shows a broader high-$k$ tail than the original Zel'dovich power spectrum,
 with a universal $k^{-2}$ decay instead of the steeper-than-$k^{-3}$ falloff. 
The final expression for this ``cosmic web'' power spectrum is 
summarized as follows [see Eq.~(\ref{Psc-def})]:
\beqa
P_{\rm c.w.}(k) & = & \int \frac{\dd\Delta\vq}{(2\pi)^3} \; \frac{1}{1+A_1}
\biggl \lbrace e^{-\varphi_{\parallel}(-\ii k\mu\Delta q\,\sigma_{\kappa_{\parallel}}^2)/\sigma_{\kappa_{\parallel}}^2}
+ A_1 \nonumber \\
&& \hspace{-1.3cm} + \int_{0^+-\ii\infty}^{0^++\ii\infty} \frac{\dd y}{2\pi\ii} \; 
e^{-\varphi_{\parallel}(y)/\sigma^2_{\kappa_{\parallel}}} 
\left[ \frac{1}{y} - \frac{1}{y\! + \! \ii k\mu\Delta q \, \sigma^2_{\kappa_{\parallel}}} \right]
\biggl \rbrace \nonumber \\
&& \hspace{-1.3cm} \times \; e^{-\frac{1}{2} k^2 (1-\mu^2) \sigma_{\perp}^2} ,
\nonumber
\eeqa
with the quantities $\sigma_\perp^2$, $\sigma_{\kappa_\parallel}^2=\sigma_\parallel^2/(\Delta q)^2$ 
and $A_1$ respectively given by 
Eqs.~(\ref{sig-par-I}), (\ref{sig-perp-I}), and (\ref{An-def}). For the cumulant-generating
function $\varphi_\parallel$, we adopt the simple ansatz of Eq.~(\ref{phi-alpha-def}), which is 
parameterized by the effective skewness (\ref{S3-1loop}) 
through the scale-dependent parameter $\alpha(\Delta q)$ [Eq.~(\ref{S3-alpha})].

This provides a well-behaved cosmic web power spectrum that can be combined with a
halo model to describe all regimes of the cosmological density field, from large linear
to small, highly nonlinear scales. 
The explicit expression for the power spectrum, given as the sum of the 
two contributions $P_{\rm 1H}(k)$ and $P_{\rm 2H}(k)$, is described in Sec.~\ref{Combining} 
[see Eqs.~(\ref{Pk-1H}) and (\ref{Pk-2H-1})].

A comparison with numerical simulations shows that our new proposition extends the
range of validity of theoretical predictions over previous models. We obtain an
accuracy better than $2\%$ for $k \leq 0.3 h$Mpc$^{-1}$ and
better than $5\%$ for $k \leq 3 h$Mpc$^{-1}$, at $z \geq 0.35$.
This also yields an accuracy of $2\%$ on BAO scales for the two-point correlation
function and of $5\%$ down to $0.5 h^{-1}$Mpc.
We checked that a similar accuracy is obtained for an alternative less standard cosmology,
with a dark energy equation-of-state parameter $w_{\rm de}=-1.28$.

Our approach should be generalized in two directions to provide a more direct comparison
with observations. First, we should take into account redshift-space distortions, which
should be possible within our Lagrangian-space framework.
Second, we could consider the statistics of biased tracers, such as galaxies of different
mass or luminosity. We leave these generalizations to further works.

The model developed in this paper can also be directly applied to several topics.
First, following \cite{Valageas2012,Valageas2012a},
we can use our predictions for the 3D density field power spectrum to obtain the
power spectrum of the weak-lensing shear (in the Born approximation), by integrating
along the line of sight.
Second, as in \cite{Koyama2009,Brax2012}, we can study less standard scenarios that
involve modifications of gravity on cosmological scales, or consider the possible impact
on the power spectrum of a warm dark matter component \cite{Valageas2012b}.
It should also be possible to study the effect of neutrinos \cite{Lesgourgues2009} or
of baryons \cite{Bernardeau2013} on the power spectrum.

\acknowledgments

This work has benefited from exchange visits supported by a 
bilateral grant from the
Minist\`ere  des Affaires Etrang\`eres et Europ\'eennes in France
and the Japan Society for the Promotion of Science (JSPS). 
P. V. is supported in part by the French Agence Nationale de la Recherche under Grant
ANR-12-BS05-0002.
A.T. is supported in part by a Grant-in-Aid for Scientific 
Research from the JSPS (No.~24540257). 
T. N. is supported by a Grant-in-Aid for JSPS Fellows (PD: 22-181) and 
by the World Premier International Research Center Initiative
(WPI Initiative), MEXT, Japan. Numerical computations for the present work 
have been
carried out in part on Cray XT4 at the Center for Computational Astrophysics, 
CfCA, of the National
Astronomical Observatory of Japan, and in part under the Interdisciplinary 
Computational 
Science Program in the Center for Computational Sciences, University of Tsukuba.

\appendix

\section{Halo model details}
\label{Halo-details}

\begin{figure}
\begin{center}
\epsfxsize=8.5 cm \epsfysize=6.5 cm {\epsfbox{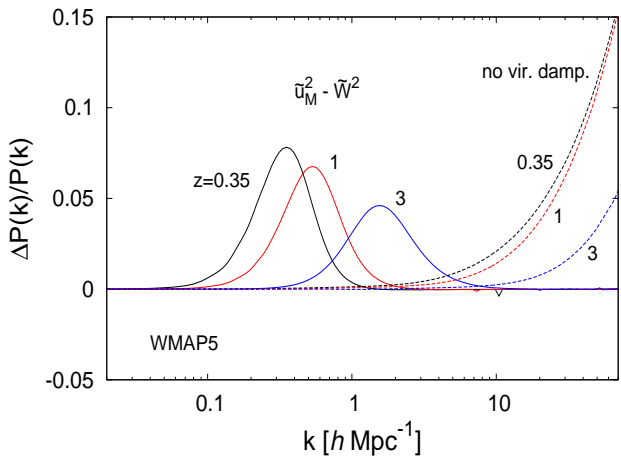}}
\end{center}
\caption{Relative deviation from our model (\ref{Pk-1H})-(\ref{Pk-2H-1}) of the power
spectra obtained with a factor
$(\tu_M^2-\tW^2)$ instead of $(\tu_M-\tW)^2$ in Eq.(\ref{Pk-1H})) (left solid lines), or
with a virial damping factor $\lag e^{\ii\vk\cdot\Delta\vx} \rag^{\rm vir}_{\Delta q}$
set to unity instead of Eq.(\ref{vir-1}) in Eq.(\ref{Pk-2H-1}) (right dashed lines).
We show our results for the WMAP5 cosmology of Sec.~\ref{WMAP5} at redshifts
$z=0.35, 1$, and $3$.}
\label{fig-dDk_k4}
\end{figure}

We show in Fig.~\ref{fig-dDk_k4} the impact on the matter power spectrum of some details of
the halo model.

The solid curves show the relative deviation from our prediction that would be obtained by using
a factor $(\tu_M^2-\tW^2)$, as in \cite{Valageas2011d}, instead of $(\tu_M-\tW)^2$
in Eq.(\ref{Pk-1H}). This only gives a $k^2$ decay at low $k$ instead of a $k^4$ tail
(i.e., it satisfies matter conservation but not momentum conservation \cite{Peebles1974}).
This would slightly overestimate the power spectrum on weakly nonlinear scales, where the
one-halo term starts being non-negligible while remaining sensitive to its asymptotic low-$k$
behavior.

The dashed curves show the relative deviation from our prediction that would be obtained by
neglecting the virial damping factor $\lag e^{\ii\vk\cdot\Delta\vx} \rag^{\rm vir}_{\Delta q}$
in Eq.(\ref{Pk-2H-1}). This would overestimate the power spectrum in the
highly nonlinear regime on small scales, which are sensitive to motions within halo cores.
However, on these scales our model suffers from other uncertainties, due to the limited
accuracy of halo profiles and concentration parameters, and to the approximate form
of our two-halo contribution.

\bibliography{ref2}   % name your BibTeX data base

\end{document}